\documentclass[12pt]{article}

\usepackage{subcaption}
\usepackage[font=small,labelfont=bf]{caption}
\usepackage{floatrow}
\floatstyle{plaintop}
\restylefloat{table}

\usepackage{hyperref}
\usepackage{float}
\usepackage{amsmath}
\usepackage{amsthm}
\usepackage{amssymb}
\usepackage{setspace}
\usepackage{placeins}
\usepackage{threeparttable}	
\usepackage{comment}
\usepackage{adjustbox}
\usepackage{booktabs}
\usepackage{multicol}
\usepackage[authoryear,round]{natbib}
\usepackage{afterpage}
\usepackage{changepage}
\usepackage{pdflscape}
\usepackage{microtype}
\usepackage{afterpage}
\usepackage{caption}

\newtheorem{prop}{Proposition}

\ifx\pdfoutput\@undefined\usepackage[usenames,dvips]{color}
\else\usepackage[usenames,dvipsnames]{color}
\usepackage[letterpaper,top=1in, bottom=1in, left=1in, right=1in]{geometry}
\usepackage{setspace}

\definecolor{oceanboatblue}{rgb}{0.0, 0.47, 0.75}
\definecolor{LightCyan}{rgb}{0.88,1,1}

\hypersetup{
colorlinks=true,	
citecolor=oceanboatblue,	
urlcolor=oceanboatblue,	
linkcolor=oceanboatblue	
}

\usepackage{pgfplotstable}
\pgfplotstableset{
    every head row/.style={
    before row=\toprule,after row=\midrule},
    every last row/.style={
    after row=\bottomrule},
    col sep = &,
    row sep=\\,
    string type,
}

\usepackage[splitrule,multiple]{footmisc}  

\usepackage{array,etoolbox}
\preto\tabular{\setcounter{magicrownumbers}{0}}
\newcounter{magicrownumbers}
\def\rownumber{}

\usepackage{longtable}
\usepackage{epigraph}

\usepackage{etoolbox}
\AtBeginEnvironment{quote}{\par\onehalfspacing\small}

\makeatletter

\date{October 19, 2022}
\newcolumntype{P}[1]{>{\centering\arraybackslash}p{#1}}

\title{Making the Elite: Top Jobs, Disparities, and Solutions}

\vspace{0.2cm}
\author{Soumitra Shukla\thanks{ \scriptsize Shukla: Economist, Board of Governors of the Federal Reserve System (\href{mailto:soumitra.shukla@frb.gov}{soumitra.shukla@frb.gov}) and Research Affiliate, Department of Economics, Yale University (\href{mailto:soumitra.shukla@yale.edu}{soumitra.shukla@yale.edu}). This paper was previously circulated under the title ``Between College and That First Job: Designing and Evaluating Policies for Hiring Diversity." I would especially like to thank many members of the educational institution I worked with for immense support, motivation, and encouragement. I would like to thank Joseph Altonji, Costas Meghir, and John Eric Humphries for serving on my committee and providing guidance and feedback. I would also like to thank Daniel Aaronson, Jason Abaluck, Peter Arcidiacono, Carolina Arteaga, Ian Ball, Prashant Bharadwaj, Barbara Biasi, Peter Blair, Shoumitro Chatterjee, Ashwini Deshpande, Alessandra Fenizia, Claudia Goldin, Peter Hull, Yujung Hwang, David Jenkins, Ezra Karger, Patrick Kline, Kala Krishna, Adam Kapor, Ro'ee Levy, Eliza McNay, Tatiana Mocanu, Jack Mountjoy, Dylan Moore, Karthik Muralidharan, Ahyan Panjwani, Brendan Price, Cheyenne Quijano, Evan Rose, Fran\c cois de Soyres, Basit Zafar, Seth Zimmerman, seminar participants at the Yale labor workshop, the Federal Reserve Bank of Chicago, the Board of Governors of the Federal Reserve System, Wharton (BEPP), Penn State, UCSD, and SOLE for helpful feedback and support. This paper was also selected to be presented at the Barcelona GSE Summer Forum 2022 (Structural Microeconometrics), which I was unable to attend due to COVID-related issues. I gratefully acknowledge financial support from the Cowles Foundation for Research in Economics and a Yale Dissertation Fellowship. The views in this paper are solely the responsibility of the author and should not be interpreted as reflecting  the  views  of  the  Board  of  Governors  of  the  Federal  Reserve  System or  of  any  other  person  associated  with  the  Federal Reserve System.}}

\makeatother

\begin{document}

\maketitle

\onehalfspacing


\vspace{-0.8cm}
\hfil \href{https://soumitrashukla.github.io/Shukla_JMP.pdf}{[\large Link to most current version]} \hfil 


\vspace{0.5cm}
\thispagestyle{empty}

How do socioeconomically unequal screening practices impact access to elite firms and what policies might reduce inequality? Using personnel data from elite U.S. and European multinational corporations recruiting from an elite Indian college, I show that caste disparities in hiring do not arise in many job search stages, including: applications, application reading, written aptitude tests, large group debates that assess socio-emotional skills, and job choices. Rather, disparities arise in the final round, comprising non-technical personal interviews that screen on family background, neighborhood, and ``cultural fit." These characteristics are plausibly weakly correlated with productivity (at the interview round) but strongly correlated with caste. Employer willingness to pay for an advantaged caste is as large as that for a full standard deviation increase in college GPA. A hiring subsidy that eliminates the caste penalty would be more cost-effective in diversifying elite hiring than equalizing the caste distribution of pre-college test scores or enforcing hiring quotas.

\vspace{0.2cm}

\hspace{-0.6cm}\textbf{JEL codes:} J15, J44, J70, J71, J78, M51 

\smallskip

\hspace{-0.6cm}\textbf{Keywords:} Discrimination, Inequality, Elite Labor Markets, Multinational Corporations, Organizational Practices, Personnel Screening, Personal Interviews, Cultural Fit, Hiring Discretion

\clearpage

\setcounter{page}{1}

Socioeconomically unequal screening practices commonly determine access to elite jobs and colleges \citep{stevens2009, riverapedigree}. Such screening practices are often outwardly blind to race, gender, ethnicity, or social class. However, they can implicitly penalize certain groups by screening on cultural fit, personal hobbies, legacy and athlete preferences, and subjective impressions of personality \citep{rivera_interviews, harvardversussffa, peter_legacy}. Therefore, such screening practices can create barriers to elite jobs and colleges (henceforth, ``elite attainment").

The pivotal socioeconomic role of elite attainment has led many governments, firms, and educational institutions around the world to advance policies to improve access to such opportunities \citep{forbes2021}. Despite progress, substantial disparities persist. Among all current CEOs of Fortune 500 firms, only 1\% are black \citep{forbeswahba}. In India, most elite educational institutions have implemented caste-based quotas since the early 1990s, assigning 50\% of their college seats to disadvantaged castes. However, stark caste disparities remain a feature of elite jobs: nearly 95\% of board members in India's top 1,000 businesses belong to advantaged castes, which make up about 30\% of the overall population \citep{ajitdonkersaxena2019}. 

This paper uses insights from the screening practices of elite foreign-based multinational corporations (MNCs) hiring in the Indian private sector to isolate the sources of and evaluate potential solutions to caste disparities. Elite private-sector jobs lack affirmative action policies for disadvantaged castes. There is also little empirical evidence on the sources of caste disparities in elite hiring \citep{madheswaran2008}. Moreover, hiring practices of elite firms are often non-transparent, making it challenging to understand why inequalities arise and persist \citep{riverapedigree}.

My paper proceeds in two parts. In the first part, I uncover the mechanisms driving the caste gap in earnings. I employ novel administrative data on each stage of the job placement process from an elite Indian college. I combine this data with evidence on common questions asked by Human Resources (HR) managers in personal interviews. My paper offers a representative window into how elite college graduates transition into ``elite entry-level jobs," defined as the top 1\% paying entry-level jobs in the Indian private sector. These jobs are primarily in U.S. and European MNCs that hire for their Indian locations. About 96\% of the jobs in my data are ``elite" (Section~\ref{sec:data1}). 

Despite the elite college assigning nearly 50\% of its seats to disadvantaged castes, there are large disparities in labor market outcomes. The unconditional caste gap in earnings among graduates from the elite college is 17\%. In the presence of detailed controls on pre-college skills, within-college academic performance, previous labor market experience, and other employer-relevant skills, the gap is reduced to 11\%. There are no caste differences in pay for a given job, so the earnings gap is due to differences in job composition across castes. 

I further decompose the remaining 11\% caste gap in earnings into components attributable to each successive stage of the job placement process. The stages are job applications, application reading, written aptitude tests (``technicals"), group debates, non-technical personal interviews, job offers, and job choices. Group debates are \emph{non-technical}. They typically comprise two separate teams with 10-15 students in each, last a few hours, and assess a wide array of socio-emotional skills, including: communication, mannerisms, consensus building, collegiality, confidence, and teamwork. In my setting, firms post uniform job-specific (not match-specific) wages that are non-negotiable over the course of job search. Therefore, the remaining 11\% earnings gap can be further decomposed by comparing the caste
difference in the composition of jobs at each stage of job search. The composition of job applications does not explain the remaining earnings gap. After job applications are submitted, the three pre-interview screening stages together contribute to only about one-tenth of the remaining 11\% earnings gap. The composition of job choices over offered jobs also does not contribute to the remaining earnings gap. Therefore, almost 90\% of the remaining 11\% earnings gap emerges between non-technical personal interviews and job offers, suggesting that policies informing applicants about job opportunities, modifying student preferences, or improving performance at university are unlikely to mitigate disparities \emph{for this population}.  

Personal interviews, also called HR interviews, are a widely employed screening practice by elite firms in the Indian private sector. Importantly, they are \emph{not} technical or case-study-based \citep{deshpande2011, jodhka2017, fernandez2018}. Based on evidence on the content of these interviews, I argue that caste disparities in earnings are primarily due to employers screening on (researcher-unobserved) \emph{background characteristics}, such as educational qualifications of family members, neighborhood of residence, family background, father's job, cosmopolitan attitudes, upbringing, personal hobbies, and subjective impressions of ``cultural fit" \citep{newmananddeshpande2007, rivera_interviews, riverapedigree,jodhka2017}.\footnote{\emph{Naukri.com}, India's leading job search portal with a market share over 60\%, suggests that the ``best way to answer this common interview question [when asked by recruiters to introduce oneself] is to tell the hiring manager about your education and family background'' \citep{naukri2019}.} These characteristics are plausibly weakly correlated with productivity (at the personal interview round) but are strongly correlated with caste.

Unstructured interviews that screen on conversations related to shared experiences, aspects of cultural fit, and personal hobbies are not unique to India. Such practices are also common in the screening practices of elite U.S. colleges and corporate America \citep{stevens2009, riverapedigree}. For example, \citet{riverapedigree} shows that managers in the U.S. offices of 120 elite firms prioritized cultural fit in interviews largely because they believed that, unlike job skills, ``fit" could \emph{not} be easily taught. Nearly 80\% of the surveyed professionals endorsed the use of an interview evaluation heuristic called the ``airplane test": basically, would I like to be stranded in an airport with the candidate?

The emergence of earnings disparities closely parallels caste revelation. Caste in \emph{elite, urban-educated India} is almost perfectly signaled through background characteristics, similar to those screened on during personal interviews. However, surface-level cues observed by employers during the application reading, written aptitude test, and group debate rounds---such as last names, skin color, facial features, accents, and dialects---are \emph{highly noisy} signals of caste in elite, urban-educated India, given large regional variation in these cues (Section~\ref{sec:caste reveal}).\footnote{Surnames like ``Singh,'' ``Sinha,'' ``Verma,'' ``Chaudhary,'' ``Mishra,'' and ``Das'' are shared across castes \citep{asi2009}. Relatedly, in a recent audit study based on firms in the New Delhi area, \citet{banerjee2009} state that the ``enormous regional variations [in last names] mean that the precise coding of a particular last name is unlikely to be familiar to people from a different linguistic region of India.'' Naming conventions also differ significantly across regions. For example, for South Indians, personal (first) names often perform the role of traditional ``surnames" \citep{jayaraman2005}.}\footnote{Scholars have argued that there is no association between skin color and caste, especially since Indian skin color is influenced mostly by geographic location rather than caste status \citep{mishra1994, parmeswaran2009}.}\footnote{Perception of accent variation among young, English-speaking university graduates in India is linked to broad regional factors instead of caste status \citep{wiltshire2020}.}\footnote{It is difficult to ``fake" caste status because caste networks are typically quite deep \citep{beteille1965, mamidi2011}.} Thus, personal interviews likely reveal caste leading to direct discrimination and, even if employers do not value caste \emph{per se}, worsen disparities due to indirect discrimination on background characteristics.

The paper's descriptive facts advance the literature on the detection and measurement of labor market disparities.  Discrimination based on socioeconomic cues in elite, urban-educated settings is likely to become more salient as the world becomes increasingly multi-ethnic and diverse and standard characteristics by which to differentiate groups become less perceptible \citep{loury2002,freeman2011,gaddis2017}. I provide an important example of the kinds of data future researchers may have to collect in such settings to better detect disparities from outwardly neutral screening practices. Recent works studying the role of hiring discretion have either exclusively focused on low-skilled jobs or provided correlational evidence between callback disparities and HR practices being more subjective \citep{hoffmankahn2015,klinerosewalters2021}. By collecting data on all steps of the placement process, my paper is the first to decompose the earnings gap at successive stages of job search and quantify the role of a widely employed subjective screening practice---non-technical personal interviews---in determining access to elite jobs.

Other alternative explanations are unlikely to explain the remaining earnings gap of 11\%. These include differences in socio-emotional skills, outside options, negotiation abilities, the possibility that employers may be ``playing along'' at earlier rounds due to government audits or internal institutional pressure, competition from the government sector, employers casting a wider net before interviews, and student preference for living in a metropolitan city (Section~\ref{sec:alternative explanations}).

In the second part of the paper, I propose policies and provide new evidence on their effectiveness to diversify hiring in elite entry-level jobs. Such an analysis could also be useful in other contexts, as elite MNCs have global footprints. To evaluate potential policies, I calculate employer willingness to pay for key characteristics, such as pre-college test scores and caste. These estimates are obtained from a novel model of the hiring process. Using the model estimates, I propose and evaluate three counterfactual policies to diversify elite hiring. The model incorporates the economically relevant stages of job search in my setting: firm hiring and final job choices. This is also the first empirically estimated model of the job placement process from an elite college and could serve as a prototype for the standardized job placement processes of other elite engineering, business, and law schools in India (Section~\ref{sec:data1}). While the details vary, it is an open question as to how best to efficiently match students to firms while also addressing concerns about equity. 

The employer willingness to pay for caste is calculated through a reduced-form caste coefficient in the employer's utility function. The ``caste penalty" is identified under the plausible assumption that---conditional on resume screening and performance in both technical tests and non-technical group debates---parental background, neighborhood, and subjective impressions of cultural fit are plausibly weakly correlated with productivity. Moreover, since jobs set wages \emph{nationally}, offering about the same job-specific wage across all locations in India, wage can enter employer utility as an exogenous regressor, with its corresponding coefficient identified from within-job time variation in job offer rates and wages. Thus, the caste penalty can be converted into dollar units. 

As mentioned previously, background characteristics are almost perfectly predictive of caste in elite, urban-educated India. Therefore, the reduced-form coefficient representing the caste penalty could capture discrimination due to employers directly valuing caste \emph{in addition to} indirect, researcher-unobserved characteristics (family background, neighborhood of residence, upbringing, and cultural fit) revealed during personal interviews. Caste disparities could stem from either taste-based or statistical discrimination and the reduced-form caste coefficient \emph{embeds a mechanism for both}. In other words, the magnitude of the caste penalty, and therefore, the employer willingness to pay for caste is \emph{invariant} to the underlying motivations for caste disparities.

My empirical approach to model the caste penalty through a reduced-form caste coefficient that captures both direct and indirect sources of disparities helps advance recent research that argues for a constructivist understanding of group identities, instead of treating them as immutable facts \citep{bhi2022,rose2022,sarsons2022}. Such an approach is crucial to better understand ``caste," classifications of which are rooted in the economic, political, and material history of India  \citep{beteille1965,beteille1969}. In addition, perceptions of caste in elite, urban-educated India are guided by a myriad of socioeconomic cues, paralleling the impressions of social class in other contexts, especially Britain \citep{deshpande2011, mamidi2011, savage2015, jodhka2017}.

I show that firms discount the value of disadvantaged castes at the equivalent of 4.8\% of average annual salary, holding other student attributes constant. Employer willingness to pay for an advantaged caste is as large as that for a full standard deviation increase in college GPA. In addition, eliminating the caste gap in each pre-college test score quantile closes only about 10\% of the model-implied caste penalty, suggesting the need for policies that directly mitigate caste disparities.

Potential policies need to be consistent with India's legal environment that enforces explicit caste-based compensatory policies, but \emph{only in} public jobs and colleges. Therefore, there is no federal ombudsman to regulate private-sector hiring. Moreover, while explicit caste-based discrimination is illegal, the Indian legal system does not recognize ``disparate impact." Neither is there a systematic legal provision (anywhere) to penalize employers for judging ``cultural fit" based on myriad characteristics correlated with protected status \citep{jodhka2017, lang2020}.

Motivated by the need for policies to directly mitigate caste disparities, I first consider a hiring subsidy that eliminates the caste penalty by making elite firms indifferent between observably identical applicants across castes. Therefore, the subsidy is equivalent to the amount employers discount the value of disadvantaged castes---i.e., 4.8\% of average annual salary. This amount is a \emph{one-time common payment} to each elite entry-level job, per disadvantaged caste hired, and is similar in spirit to the incentive-based Diversity Index proposed by the Ministry of Minority Affairs \citep{sachar2006, diversityindex2008}. Note that, in principle, the hiring subsidy reimburses the employer for a \emph{stream of costs} incurred in the future and \emph{not just} the cost of hiring a disadvantaged caste over a single year. In the second policy, I consider a ``pre-college intervention'' that equalizes the distribution of pre-college skills (college entrance exam scores) across castes. In the third, and final, counterfactual policy, I consider a hiring quota that requires firms to hire an equal proportion of applicants from advantaged and disadvantaged castes: a policy that mirrors caste-based quota policies in government jobs \citep{madheswaran2008}.

Note that subsidies are unlikely to \emph{further} stigmatize beneficiaries \emph{from this population} for three reasons. First, while theoretical works have suggested that stigma could be worsened due to affirmative action policies, empirical research has found slim evidence in support of this contention \citep{coateandloury1993, bok1998, deshpanderedtape}. Second, since employers are unlikely to know caste until the final round HR interviews, potential beneficiaries are likely as capable as non-beneficiaries in technical skills judged by written tests and socio-emotional skills judged by group debates. These skills are judged \emph{before} HR interviews and are plausibly more strongly correlated with productivity than subjective impressions of ``fit." Third, even granting the purported worsening of stigma, compensatory policies could still be efficiency enhancing, as disadvantaged groups likely benefit the most from elite attainment, whereas displaced advantaged groups are likely not much worse off \citep{blackdemmingrothstein2020}. This reason is also why it could still be meaningful to intervene through policies to address disparities from accurate statistical discrimination.

I evaluate the cost-effectiveness of counterfactual policies in improving both the absolute and relative caste hires at elite firms in a partial equilibrium framework. I show that omitting aspects such as wage setting, reallocation to elite entry-level jobs, caste share of applicants to elite jobs, firm entry, and information-based policies is not a major limitation for my analysis (Section~\ref{sec:comments_modeling_choices}). 

To evaluate the cost-effectiveness of subsidies and pre-college interventions, I compare the model-implied subsidy equivalent of the pre-college intervention policy to the direct costs of changing test scores. My estimates show that elite firms put modest weights on pre-college test scores, suggesting that the efficiency gains from pre-college interventions (which my cost-effectiveness analysis omits) are likely small. The model-implied subsidy equivalent of the pre-college intervention policy is about 0.6\% of average annual salary, which is only about 10\% of the caste penalty. To calculate the direct costs of improving pre-college test scores, I use estimates from a meta-analysis of education-focused impact evaluations that documents the costs of changing test scores of primary and secondary school students in India \citep{asim2015}. Even under extremely conservative assumptions to extrapolate the direct costs of test score changes, subsidies to hire applicants from disadvantaged castes are twice as cost-effective in diversifying \emph{elite hiring}. 

Finally, I discuss results from a hiring quota policy that requires firms to hire an equal proportion of applicants across castes, mirroring the caste shares in the elite college. While more disadvantaged castes are hired, the caste penalty is large enough to eventually make the average marginal utility of filling two slots lower than the average marginal cost. This happens well before firms can achieve baseline levels of hiring. Firms counteract the quota policy by making fewer job offers and decrease overall recruitment from the university by 7\%. Empirical evaluations of hiring quotas in elite public-sector jobs in India and elite private-sector jobs in other contexts have found analogous effects \citep{payresearchunit2018, nitaqat2021}.

\section{Caste and Affirmative Action in India}
\label{sec:historical detour}

Consistent with the practice of affirmative action policies in India, \emph{this paper focuses on two caste groups}: advantaged (``upper") and disadvantaged (``lower") castes. This section provides a brief history of caste-based affirmative action policies in India and its present limitations. I will also emphasize that categorizations of ``caste" have constantly interacted with social class and have been forged over political and historical processes spanning decades. 

The first provisions for uplifting ``depressed'' or socioeconomically disadvantaged classes of Indian society were made possible after the Government of India Act of 1919 established self-governing institutions (i.e., provisional assemblies and central legislative assemblies), which introduced limited self-government to a majority British-controlled India. The Government of India Act of 1935 replaced the words ``depressed classes'' with ``Scheduled Castes'' \citep{bayly2008}. Many articles of the Constitution of India, ratified in 1949, formalized reservation-based affirmative action policies in legislatures, higher-educational institutions, and government jobs for the so-called ``backward" classes. Backward classes were intended to include not only members of Scheduled Castes (SCs) and Scheduled Tribes (STs), but also those from the Other Backward Classes (OBCs). These provisions begged an obvious question: what determines ``backwardness''?

In 1979, the Mandal Commission was set up with a mandate to ``identify the socially and educationally backward classes in India'' \citep{mandal1979}. The Mandal Commission recommended caste as the basis for reservation. In particular, it recommended a 27\% reservation (quota) in central and state services, public undertakings, and educational institutions for OBCs. Given the already existing 22.5\% reservation for SCs and STs, the fraction of reserved seats for \emph{disadvantaged castes} (SCs, STs, and OBCs) was brought up to 49.5\%. The recommendations of the Mandal Commission were formally implemented in 1990. However, \emph{none} of the current constitutional provisions extend to advancing compensatory hiring policies for disadvantaged castes in \emph{private-sector jobs}. One of the goals of this paper is to assess the potential of such policies to diversify \emph{elite hiring} in the Indian private sector. I do not focus on other private-sector jobs.

\section{Institutional Setting and Key Definitions}
\label{sec:data1}

In this section, I define key terms and elaborate on important features of my institutional setting.

\medskip

\textbf{1) Defining elite colleges and elite entry-level jobs.} An ``elite entry-level job" is defined as an entry-level job in the Indian private sector paying among the top 1\% of entry-level salaries \citep{stateofinequality2022}. I define ``elite colleges" as those consistently ranked in the top 10 in their respective fields---such as science, engineering, humanities, commerce, or law---by \emph{India Today}, which is the Indian equivalent of \emph{U.S. News \& World Report}. Nearly all elite Indian colleges are \emph{public institutions} \citep{altbach,datta2017}.

\medskip

\textbf{2) Elite colleges have similar job placement processes.} Elite Indian colleges have similar job placement processes primarily due to an extended process of historical imitation. Post independence, the earliest elite Indian colleges were built in the early 1950s and were closely modeled on elite U.S. universities, particularly MIT and Stanford. The earliest elite Indian colleges served as role models for elite Indian colleges built later, which closely imitated key aspects such as academic calendars, faculty-to-student ratios, and job placement processes. Hence, almost all elite Indian colleges today have similar mechanisms for selectively placing graduates into elite entry-level jobs \citep{altbach,datta2017}. Examples of elite Indian colleges sharing common placement processes include 23 Indian Institute of Technologies, 20 Indian Institute of Managements, and several colleges under the ambit of the prestigious Delhi University.\footnote{See \href{https://www.iitbhilai.ac.in/index.php?pid=AIPC\_COVID19}{All IITs Placement Committee Brochure} and \href{http://placement.du.ac.in/}{Central Placement Cell (Delhi University).}}

\medskip

\textbf{3) Elite college graduates primarily work in elite entry-level jobs.} Almost 96\% of elite college graduates work in elite entry-level jobs in the Indian private sector \citep{newman2009, jodkhanaudet2019, subramanian2019}.\footnote{There are about 30 million jobs in India's organized private sector of which about 50,000 jobs are \emph{elite entry-level jobs} \citep{qes2021}, per the definition in the first point of Section~\ref{sec:data1}. Similarly, per the definition in the first point of Section~\ref{sec:data1}, about 60 Indian colleges are ``elite," each admitting about 800 students. Therefore, another way to convey the third point of Section~\ref{sec:data1} is that ``most elite college graduates in India work in (top jobs) in the private sector."}

\medskip

\textbf{4) Elite entry-level jobs almost exclusively hire from elite colleges.} Graduates from elite Indian colleges account for more than 95\% of the hires of elite entry-level jobs in India \citep{newman2009, jodkhanaudet2019, subramanian2019}.

\medskip

\textbf{5) Small proportions of students skip the job placement processes of elite colleges.} These proportions range from 5-8\% of graduating classes \citep{IndiaToday2015}.  

\medskip

\textbf{6) Elite entry-level companies recruiting from this elite college recruit representatively.} Scraping data from the placement websites of colleges, I find that about 94\% of the firms that recruit students from this elite college also visit other elite colleges for their on-campus job fairs.

\medskip

Together, the facts presented in points 2), 3), 4), 5), and 6) above imply that the job placement process from the college I examine in the paper offers a representative window into how elite college graduates transition into elite entry-level jobs in the Indian private sector.

\medskip

\textbf{7) Elite entry-level jobs are in the Indian offices of foreign-based MNCs.} About 97\% of the firms that recruit from this elite college are foreign-based (\emph{mostly U.S. and European}) MNCs. They also \emph{hire primarily for their Indian offices} (Online Appendix Table~\ref{tab:foreign_based_versus_india_based}). Despite a recent rise in India-based startups, a negligible proportion of such firms recruit from elite Indian colleges. Such firms comprise less than 0.5\% of elite entry-level firms in India, pay much lower than foreign-based MNCs, and are considered by placement officers as highly risky to invite, especially after many instances of startups reneging on job offers \citep{timesofindia2016, economicsurvey2022}.

\medskip

\textbf{8) Most of the offices and entry-level labor force of foreign-based MNCs are outside of India.} Almost 97\% of the offices and 96\% of the entry-level labor force of foreign-based MNCs in my setting are outside of India (Online Appendix Table~\ref{tab:proportion_offices_outside_india}). 

\medskip

\textbf{9) Job placement process of an elite public college.} The process takes place \emph{entirely on campus} and can be divided into two broad phases: 1) pre-placement phase, and 2) placement phase.

\begin{itemize}
\item[1.] \textbf{Pre-Placement phase.} The pre-placement phase can be further subdivided into the following steps. First, the placement office invites firms prior to June. Second, between June to mid-August, firms visit the college campus and conduct pre-placement talks to advertise job profiles and gauge student interest. Third, firms make return offers from summer internships by late August. These offers are also called pre-placement offers (PPOs) and have late August deadlines. Students who accept their PPOs are disallowed from participating in the formal placement process for full-time jobs. Fourth, students register for the formal on-campus job placement process by late August. Fifth, by early September, firms submit \emph{employer registration forms} to the placement office; these forms list details including job positions, compensation packages, and the probable number of slots (vacancies) firms want to fill from the college that year (Online Appendix Section~\ref{sec:employer_registration_form}). Sixth, after these forms are submitted, advertised job profiles are considered ``locked.'' They cannot be changed by firms during the course of the placement cycle. Moreover, students are prohibited by the placement office from bargaining over compensation bundles. The placement office verifies advertised compensation bundles by requiring students to submit copies of their job offer letters.

\item[2.] \textbf{Placement phase.} The placement phase can be further subdivided into the following steps. First, students start applying for jobs in mid-September. Second, firms make the ``first cut'' after skimming through applications and invite students for additional screening. Third, firms conduct written and verbal tests to determine eligibility for on-campus interviews. Fourth, firms conduct interviews. Fifth, firms make job offers. Sixth, students make final job choices and the placement process concludes around early January. The following are some key rules of the placement process set by the elite college's placement office:

\hspace{0.6cm}\textbf{a) Interview day allotment.} Each firm is allotted one interview day by the college's placement office to conduct personal interviews on campus. Unlike job recruitment at U.S. colleges, there are \emph{no further onsite interviews}. A particular rule of the job placement process is that conditional on getting a job offer on a given interview day, a student can no longer participate in interviews on future interview days. At best, a student can receive multiple job offers within a given interview day. If a student does not get any job offers on a particular interview day, he can participate in interviews on future interview days.\footnote{I defer the discussion of how the process of interview day allotment affects strategic behaviors of firms (if at all) to Section~\ref{sec:model}, where I discuss the model of the job placement process.}

\hspace{0.6cm}\textbf{b) Students cannot ``reject" firms midway or accept offers ``early."} Having applied, students cannot skip any of the pre-interview screening rounds or the sequence of scheduled interviews, typically spread over multiple days. Neither can they accept offers ``early" in the process (e.g., by negotiating offers midway before they are officially announced for others). 

\hspace{0.6cm}\textbf{c) All job offers are announced at the end of the interview day.} Per placement office rules, all job offers are announced within a short interval of time at the end of the interview day to prevent firms that are allotted the \emph{same interview day} from coordinating on offers.

\end{itemize}

\section{Data Overview}
\label{sec:data overview}

\subsection{Students}
\label{sec:data overview students}

The administrative data belong to the job placement cycles corresponding to four years: e.g., 2012, 2013, 2014, and 2015 (exact years omitted to preserve anonymity). Online Appendix Table~\ref{tab:descriptive statistics students appendix} shows the total number of students belonging to each caste for each college degree. There are 4207 students in the sample. Male students comprise about 90\% of the sample.\footnote{The fraction of males in the data is typical of those observed in elite technical colleges in India \citep{datta2017}.} Caste in the data is self-declared by students and is used as the basis for quota-based policies in admissions.\footnote{Indian colleges implement quotas under the labels ``General" (advantaged) and ``Non-General" (disadvantaged).} These policies equalize the share of both disadvantaged and advantaged castes \emph{within each college major}. Therefore, nearly 50\% of the students are from disadvantaged castes for three of the four college degrees in the sample (Section~\ref{sec:historical detour}).\footnote{The Master of Science (M.S.) degree has a slightly larger proportion of advantaged castes. \citet{krishna2015} also document similar patterns.}  

The four college degrees in the sample are the Bachelor of Technology (B.Tech.), Dual, Master of Technology (M.Tech.), and Master of Science (M.S.) degrees. A Dual degree integrates undergraduate and post-graduate studies. I omit students pursuing a different Master of Science (M.Sc.) degree, as they comprise less than 2\% of the student population.\footnote{There are two different Master of Science degrees offered at the college. M.S., which is included in the sample, has an industry focus. M.Sc., which is not included in the sample, is a much smaller program with a research focus.} 

\smallskip

\textbf{College admissions criteria.} Students are admitted to the elite college through entrance exams based on caste-major-specific cutoffs that comprise the \emph{only criterion for college admission}. 

\subsection{Differences in Baseline Characteristics across Castes}
\label{sec:differences in baseline characteristics}

\medskip

\hspace{0.6cm}\textbf{Differences in pre-college skills across castes.} Caste differences in college entrance exam scores are large, averaging about 0.6 standard deviations across college degrees (Online Appendix Table \ref{tab:differences pre-college skills across castes}). Entrance exam scores in the data are \emph{originally exam ranks}, which have been renormalized so that larger numbers are better. Caste differences in 10th and 12th grade national examination scores are modest and not statistically significant, likely because such tests do not distinguish at the top of the test-taking ability distribution. Importantly, there are students from both castes (common support) within each entrance exam score decile (Online Appendix Figure \ref{fig:common support}). 

\medskip

\textbf{Differences in college GPA across castes.} As with entrance exam scores, there are substantial differences in college GPA across castes but students from both castes within each GPA decile (Online Appendix Table~\ref{tab:differences GPA across castes}; Online Appendix Figure \ref{fig:common support}).

\medskip

\textbf{Differences in previous labor market experience across castes.} I find only modest differences in previous labor market experience across castes. Previous labor market experience comprises detailed information on both summer and winter internships, including duration of internship employment, duration of part-time or full-time employment, total pay during internships, total pay during part-time or full-time employment, sector of internship employment, and employment in startups. Internship descriptions typically include application eligibility criterion, desired skills, and expectations on the job (Online Appendix Table \ref{tab:differences pre-college skills across castes previous labor}).

\medskip

\textbf{Differences in other employer-relevant skills across castes.} Admissions quotas coupled with fairly rigid engineering curricula lead to almost no caste differences on many measures of other employer-relevant skills. These measures include college major, college degree, and coursework. The dataset can also proxy for other employer-relevant skills by including indicators for getting past the various stages of job search, as will be elaborated upon in  Section~\ref{sec:data overview firms} below.

\subsection{Firms}
\label{sec:data overview firms}

Below, I provide some key definitions and an overview of firms recruiting from the elite college.

\medskip

\textbf{Definition of a job, locations of advertised jobs, and the number of years each firm recruits from the elite college.} In the sample, a ``job'' means a job designation within a firm. For example, Google can hire a product manager and a software engineer. These are two different jobs. Almost all  these jobs (i.e., about 97\%) are  located in the Indian offices of U.S. and European MNCs (see point 7 in Section~\ref{sec:data1}). Each job included in my sample---spanning four placement cycles---arrives on the elite college's campus for \emph{each} of the four years to conduct recruitment. 

\medskip

\textbf{Omitting public-sector jobs.} I omit such jobs for four reasons. First, such jobs comprise less than 4\% of all jobs available to students in the degree programs included in the sample. Second, public-sector jobs are quite different from their private-sector counterparts, especially in areas like salary structure and job stability.\footnote{See the report of the \href{https://doe.gov.in/seventh-cpc-pay-commission}{Seventh Central Pay Commission, 2016}.} Third, the probability of transitioning from elite public-sector jobs to elite private-sector jobs is less than 2.5\% and vice versa (Online Appendix Table~\ref{tab:firm_transition_matrix}; Section~\ref{sec:comments_modeling_choices}). Fourth, public-sector jobs already have strong government-mandated hiring quotas for disadvantaged castes \citep{madheswaran2008}.

\medskip

\textbf{Distribution of salaries and firms by sector and job types.} Online Appendix Table~\ref{tab:descriptive statistics firms main} shows the distribution of firms by sector and the average salary across all jobs by sector: 52\% of all firms belong to the technology sector, 20\% belong to the consulting sector, and 28\% belong to the manufacturing sector. Non-client-facing jobs comprise almost 85\% of all available jobs.\footnote{Detailed job descriptions (particularly, job titles and job functions) were used to categorize jobs as client-facing versus non-client-facing. Typically, a software engineering role would be considered non-client-facing, whereas a consulting or managerial role would be considered client-facing.} 

Conditional on college degree, job salaries do \emph{not} vary across major, caste, or gender. See the employer registration form in Online Appendix Section~\ref{sec:employer_registration_form} for how firms declare salaries for job profiles. Average salaries across all jobs in the technology, consulting, and manufacturing sectors are \$67,302.64 (PPP), \$63,544.02 (PPP), and \$43,525.25 (PPP), respectively.

\medskip

\textbf{Declaration of job descriptions, non-pecuniary amenities, and vacancies per job.} Firms declare job details in the \emph{employer registration forms} made available to them by the placement office. These details include information on various job characteristics, including job descriptions, job designations, sector, salaries, non-pecuniary amenities, desired skills, expectations on the job, the expected number of slots (vacancies) a job wants to fill from the college, and even job application eligibility criterion. The number of advertised non-pecuniary amenities is high and ranges between 40-50 \emph{per job} (see ``Pre-Placement Phase" in Section~\ref{sec:data1}; Online Appendix Section~\ref{sec:employer_registration_form}, Online Appendix Table~\ref{tab:non pecuniary amenities full 1}).
 
\medskip

\textbf{Return offers from internships.} These usually have late August deadlines. Students who accept them are not allowed to participate in the regular placement cycle for full-time jobs (see ``Pre-Placement Phase" in Section~\ref{sec:data1}). Therefore, a typical student who applies to a firm during the regular placement season would \emph{not} have completed a summer internship in his junior year at the same firm. I discuss selection induced by the placement rule regarding return offers in Section~\ref{sec: earnings gap}.

\medskip

\textbf{Data on multiple stages of job search.} The data comprise \emph{job-level} information regarding the number of students who applied, qualified for each round of screening, received job offers, and decided to accept jobs. \emph{All jobs conduct four rounds of screening before making job offers}. Students apply for jobs through a centralized job application portal, like Job Openings for Economists. Applying to a job only involves clicking on the name of the job in the application portal and does not require additional cover letters or other statements. Employers only request student resumes that are automatically made available to them when the student ``clicks'' on the name of a company to apply. Resumes are written in a standardized format prescribed by the placement office. I do not have access to student resumes. Application eligibility depends upon a combination of major, degree, and GPA---information that the dataset contains.

The first screening round is the application reading stage. This ``first cut'' is typically made on a GPA cutoff to select students for the next round. A simple cutoff rule is followed at this stage plausibly because the average job recruiting from campus \emph{receives over 600 applications per year}. The second round is a written aptitude test. These tests, also called ``technicals," are mostly conducted online. Technicals differ by job roles. They typically include coding tests for manufacturing or technology firms and case-study-based tests for consulting firms. The third round is a large group debate, usually comprising 25 to 30 students and lasts a few hours. In this round, employers come face-to-face with students for the first time but do \emph{not} extensively interact with them. Recruiters divide students into two teams and make them discuss a general or non-technical topic (e.g., ``Is Technology a Necessary Evil?"), with one team speaking for the motion and vice versa. Meanwhile, recruiters behave like passive observers, their only active role consisting of duties such as starting and ending the discussion on time and ensuring decorum. Discussion moderators organically ``emerge'' from the group of students participating in the debate. Overall, group debates are used to assess a wide array of socio-emotional skills, including: communication, mannerisms, consensus building, collegiality, confidence, and teamwork. The fourth round is a personal interview featuring the only extensive interactions between students and recruiters during the job search process. The \emph{same} group of recruiters conduct group debates as well as personal interviews.

\medskip

\textbf{More on personal interviews.} Personal interviews, also called HR interviews, are \emph{not} technical or case-study-based. Such interviews commonly screen on \emph{background characteristics} (e.g., educational qualifications of family members, father’s job, neighborhood of residence, preference for living in a cosmopolitan city, desire for traveling, and ``cultural fit") and are commonly used in recruitment practices of elite private-sector MNCs hiring both in India as well as in the U.S.

In a detailed survey of HR managers in the Indian offices of elite MNCs, \citet{newman2009} documented opinions regarding the role of personal interviews in the hiring process. These elite MNCs employ nearly 2 million workers worldwide. The authors found unanimous agreement on the importance of asking questions pertaining to background characteristics in personal interviews. Interestingly, HR managers saw little contradiction in judging a candidate's individual merit through background characteristics, arguing instead that trustworthiness to potential clients, attrition, and cultural fit can be identified through such inquiries. One HR manager at a large MNC expressed what he looks for in personal interviews:

\begin{quote}
We also ask a lot of questions related to family background. Questions like how many family members are there, how many are educated, etc. The basic assumption behind these questions is that a good person comes from a good and educated family. If parents have good education, the children also have good education. Some questions about their schooling $\hdots$ and the locality where they [grew up].
 \end{quote}

Another HR manager expressed the signaling value of background characteristics by mentioning that conversations about family background and upbringing are useful in forming impressions about trustworthiness to potential clients, cultural fit, attrition, and long-term professional behavior:

\begin{quote}
{As personal traits are developed with the kind of interaction
 you have with society $\hdots$ Where you have been brought up, the
 kind of environment you had in your family, home, colony and
 village, these things shape the personal attributes of people $\hdots$
 This determines his behavior, and working in a group with
 different kinds of people. We have some projects abroad, and
 if a person doesn't behave properly with them, there is a loss
 for the company. Here the family comes in, whether the person
 behaves well and expresses himself in a professional way, for a
 longer term and not for a short term $\hdots$ This is beneficial.}
\end{quote}

Recent quantitative research supports the above qualitative evidence. Using Likert scale ratings, \citet{mamgain2019} shows that HR managers in elite firms value family background almost as much as candidate experience, quality of the institution of education, aptitude, or technical skills. 

These prevailing personal interview conventions have also influenced the advice provided to candidates by job portals. \emph{Naukri.com}, India's largest job search portal with over 60\% market share, suggests that the ``best way to answer this common interview question [when asked to introduce oneself] is to tell the manager about your education and family background'' \citep{naukri2019}. 

Unstructured interviews that screen on cultural fit and ``background" are \emph{not} unique to India \citep{rivera_interviews, riverapedigree}. For example, \citet{rivera_interviews} shows that managers in the U.S. offices of 120 elite firms prioritized cultural fit in interviews largely because they believed that, unlike job skills, ``fit" could \emph{not} be easily taught. One evaluator categorically desired biographic similarities: 

\begin{quote}
    I usually start an interview by saying, "Tell me about yourself." When I get asked that, I talk about where I'm from, where I was raised, and then my background $\hdots$ I want to hear your life story. Hopefully there's something more interesting about your life than deciding to go to school $\hdots$ When they tell me about their background, it's easier to find things in common $\hdots$ Maybe $\hdots$ they're from Seattle and I've been to Seattle. We can talk about that and develop a connection.
\end{quote}

Nearly 80\% of the professionals surveyed by \citet{rivera_interviews} also endorsed the use of an interview evaluation heuristic called the ``airplane test." One manager said:

\begin{quote}
One of my main criteria is what I call the ``stranded in the airport test." Would I want to be stuck in an airport in Minneapolis in a snowstorm with them? And if I'm on a business trip for two days and I have to have dinner with them, is it the kind of person I enjoy hanging with? And you also have to have some basic criteria, skills and smarts or whatever, but you know, but if they meet that test, it's most important for me.
\end{quote}

\section{Descriptive Facts}
\label{sec:descriptive facts}

I document large earnings disparities across castes, discuss selection, argue that the earnings gap is conservative, and show that disparities arise mostly due to non-technical personal interviews.

\subsection{Large Earnings Gap across Castes}
\label{sec: earnings gap}

The unconditional earnings gap across castes is -0.174 (0.016) log points, or about 17\%, where the number in parentheses denotes the standard error. In the presence of detailed controls for pre-college skills, within-college academic performance, previous labor market experience, and other employer-relevant skills, the remaining gap is -0.113 (0.014) log points, or about 11\%. These results are robust to many different specifications (Online Appendix Table~\ref{tab:earnings gap main}).

There is \emph{modest} heterogeneity in the earnings gaps across job sectors and job types. For example, the earnings gap among firms in the consulting sector is -0.119 (0.032) log points, whereas it is -0.080 (0.022) log points among firms in the technology sector and -0.084 (0.022) log points among firms in the manufacturing sector (Online Appendix Table~\ref{tab:earnings gap main}). Non-client-facing and client-facing jobs also have modestly different earnings gaps. The earnings gap is -0.080 (0.016) for non-client-facing jobs, which comprise 85\% of all advertised jobs, and rises to -0.117 (0.029) log points for client-facing jobs (Online Appendix Table~\ref{tab:earnings gap main}).

\medskip

\textbf{Matching with Exit Data.} ``Exit data" is a college-administered survey of students,  conducted two months after graduation, that includes their job designations and responses to whether offer terms were negotiated between the conclusion of the placement process (i.e., around early January) and the rollout of the exit survey (i.e., around late July). Most jobs have \emph{starting dates around mid-July}. The data also include specifics of the negotiated terms and the reasons for negotiation. Note that students are restricted from bargaining over advertised compensation bundles only \emph{during} the course of the placement process and not after its conclusion (see ``Pre-Placement Phase" in Section~\ref{sec:data1}). The placement office administers exit surveys to validate its employment data.

Exit data is filled by nearly 98\% of the graduating cohort, possibly since passing along information regarding their current roles and contact details is relevant for students to take advantage of the elite college's alumni network. Comparable response rates are found in similar exit surveys collected by the career offices of elite U.S. MBA programs \citep{yalesomemployment201718}. 

Reassuringly, exit data show that nearly 99\% of job getters began in the same jobs they obtained through the placement process and \emph{did not negotiate} salaries or non-pecuniary amenities, even two months after graduation or several months after the conclusion of the placement process. Among non-job getters, almost all students indicated that they were still ``unemployed" two months after graduation, a catch-all category that may include self-employed students or those taking gap years. 

\medskip

\textbf{Selection and Implications for the Earnings Gap.} I now discuss differential selection by caste among those omitted from the earnings regressions.

\begin{itemize}
\item[1.]\textbf{About 6\% of the student sample deregisters from the job placement process.} ``Deregistered" students are those who either do not sign up to participate in the job placement process for full-time jobs or accept their return offers from summer internships, and therefore, \emph{cannot} participate in the job placement process for full-time jobs (see ``Pre-Placement Phase" in Section~\ref{sec:data1}). Similar proportions of students skip the placement processes of other elite Indian colleges suggesting that, unlike in the U.S., internships pursued by students from elite Indian colleges are exploratory (see Section~\ref{sec:data1}; \citealp{IndiaToday2015}). 

\item[2.]\textbf{The earnings gap is conservative.} \emph{In addition} to belonging to the ``deregistered" category, students omitted from the earnings regressions could also belong to the ``registered" category. Omitted students from the registered category comprise those who participate in the full-time job placement process but either do not get full-time jobs or reject all of their job offers. 

While job outcomes for students omitted from the earnings regressions are not consistently observed, many of their other characteristics (e.g., pre-college skills, college GPA, and internships completed) are. Comparing characteristics of students omitted from the earnings regressions, I find that disadvantaged castes are much \emph{more negatively selected} on college GPA and entrance exam scores (Online Appendix Table~\ref{tab:comparison gpa overall and without jobs}). Since average earnings are increasing in college GPA and entrance exam scores (not shown), the reported earnings gap is conservative.

\end{itemize}

\subsection{Almost All of the Earnings Gap Is at the Offer Stage}
\label{sec:hiring stage}

I lay out one of the key contributions of the paper. I quantify the role of a widely employed subjective screening practice---non-technical personal
interviews---in determining access to elite jobs.

\begin{figure}[h]
\centering
\caption{\footnotesize Earnings Gap across Castes at Each Job Search Stage}
\includegraphics[width=0.9\textwidth]{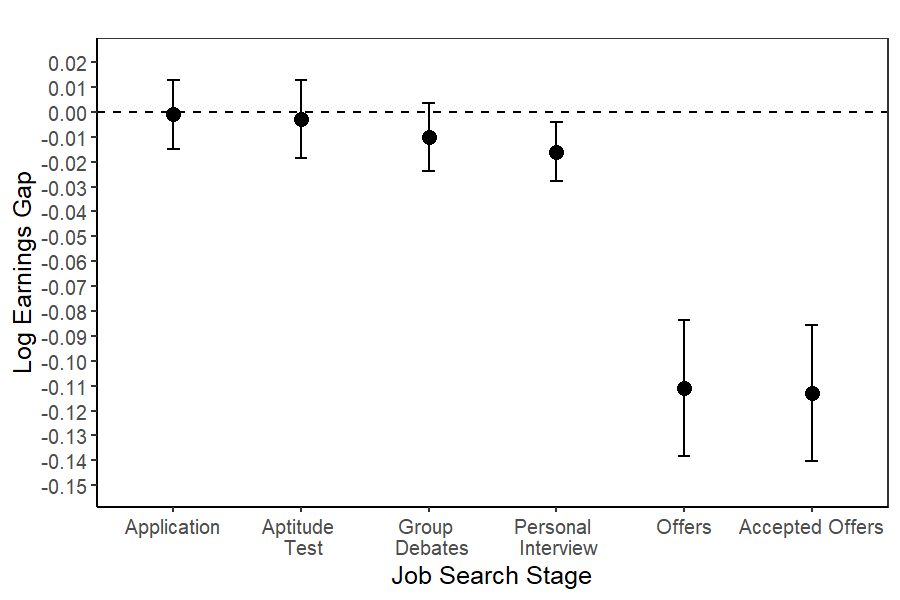}
\label{fig:earnings gap hiring stage}
    \floatfoot{\scriptsize Notes: Figure \ref{fig:earnings gap hiring stage} shows the log earnings gap across castes at successive stages of job search. Each coefficient in the figure is represented by a black dot and reports the percentage difference in the average salary at each job search stage between advantaged and disadvantaged castes. The vertical bars are 95\% confidence intervals. These regressions include controls.  Note that firms post uniform job-specific (not match-specific) wages that are non-negotiable over the course of job search. Thus, the earnings gap can be decomposed by comparing the group
difference in the composition of jobs at each stage of job search.}
\end{figure}

Recall, firms post job-specific (not match-specific) wages that are non-negotiable over the course of job search (see ``Pre-Placement Phase" in Section~\ref{sec:data1}). Thus, the \emph{remaining} (conditional) earnings gap of 11\% reported in Secton~\ref{sec: earnings gap} can be \emph{further} decomposed by comparing the caste difference in the composition of jobs that remain in contention at each stage of the placement process.

I show that almost all of the remaining earnings gap of 11\% occurs between personal or HR interviews and job offers. To do so, I run different specifications of the following regression: 
\begin{align}
\label{eq:app behavior main}
\text{log}(\text{Avg. Job Salary}^{\text{Search Stage}}_{i}) = \alpha + \beta \times \text{Disadv. Caste}_{i} + \text{Controls}_{i} + \epsilon_{i}, 
\end{align}
where \text{Search Stage} $\in$ \{\text{Application}, \text{Aptitude Tests}, \text{Group Debates (GD)}, \text{Personal Interviews}, \text{Offers}, \\ \text{Accepted Offers}\}.
The coefficient of interest is $\beta$, shown in Figure~\ref{fig:earnings gap hiring stage} for each job search stage. 

Figure \ref{fig:earnings gap hiring stage} shows that the earnings disparities primarily occur due to personal or HR interviews. The composition of job applications does not contribute to the remaining earnings gap plausibly because the streamlined job application process (similar to Job Openings for Economists) makes the marginal cost of an application effectively zero. After applications are submitted, firms conduct three pre-interview screening stages: application reading, written aptitude tests (``technicals"), and large group debates that test for socio-emotional skills. Together, these stages contribute to only about one-tenth of the remaining 11\% earnings gap. The composition of job choices over offered jobs also does not contribute to the remaining earnings gap. Therefore, almost 90\% of the remaining 11\% earnings gap emerges between non-technical personal interviews and job offers. Interestingly, the emergence of earnings disparities is \emph{even more concentrated} for technology, manufacturing, and non-client-facing jobs. The entire earnings gap among these jobs occurs between personal interviews and job offers (Online Appendix Figures~\ref{fig:earnings gap hiring stage sectors and job types}).

\medskip

\textbf{There is a substantial winnowing down in the number of jobs that remain in contention at each successive stage of job search.} The number of jobs available to each student reduces by about 35\% between any two stages, except between interviews and offers, where the reduction is much sharper due to placement office rules regarding job offer acceptance (see ``Placement Phase" in Section~\ref{sec:data1}). Therefore, meaningful cuts are made by employers at each screening stage.

\section{Interpretation, Contributions, and Alternative Explanations}
\label{sec:caste_penalty_intepretation}

This section offers a leading explanation for the earnings disparities shown in Figure~\ref{fig:earnings gap hiring stage}. I also discuss key contributions and alternative explanations to the emergence of earnings disparities.

\subsection{When Does Caste Get Revealed to Employers?}
\label{sec:caste reveal}

I argue that caste is plausibly revealed or strongly signaled during HR interviews, whereas prior job search stages most likely offer noisy signals that obfuscate caste identification.

The average job recruiting from campus receives over 600 applications per year (see Section~\ref{sec:data overview firms}). It is unlikely that caste status is known during the application reading stage (first round) given enormous regional variation in last names, different naming conventions, migration, and other factors. For example, surnames like ``Singh,'' ``Sinha,'' ``Verma,'' ``Chaudhary,'' ``Mishra,'' and ``Das'' are shared across castes \citep{asi2009}.\footnote{In a recent audit study based on firms in the New Delhi area, \citet{banerjee2009} state that the ``enormous regional variations [in last names] mean that the precise coding of a particular last name is unlikely to be familiar to people from a different linguistic region of India.''} Naming conventions also differ significantly across regions. For example, South Indians typically do not have conventional last names. Rather, for them, personal (first) names often perform the roles of traditional ``surnames" \citep{jayaraman2005}. It is also unlikely that caste status is known during written aptitude tests or ``technicals" (second round), as these tests are typically conducted online.

Caste identification is also unlikely to be reliable during group debates (third round). Recall, these debates are conducted among large groups comprising 25 to 30 students, who either argue for or against a given topic. In the data, students at group debates are about evenly split between castes. At this round, employers finally get to observe but do \emph{not} extensively interact with students after an initial setup. They also become privy to additional information like facial features, skin tones, accents, dialects, and demeanor. Scholars have argued that there is no association between skin color and caste, especially since Indian skin color is influenced mostly by geographic location rather than caste status \citep{mishra1994, parmeswaran2009}. Among educated elites, like those in my sample, English has emerged as a caste-neutral language with no remnants of caste dialects that are prevalent in most Indian languages \citep{kothari2013, ransubhe2018}. Rather, perception of accent variation among young, English-speaking university graduates in India is linked to broad regional factors \citep{wiltshire2020}.\footnote{The overwhelming influence of regionality in common parlance is perhaps most clearly expressed by \citet{gumperz1957}, who states that a ``high-caste villager may speak the same form of urban Hindi as his untouchable neighbor.''}

Conversations during personal or HR interviews plausibly reveal caste. These interviews primarily screen on parental background, neighborhood, and cultural fit. Such \emph{background characteristics} are strongly correlated with caste in \emph{elite, urban-educated India} (see Section~\ref{sec:data overview firms}; \citealp{deshpande2011, jodhka2017}). Recent experimental studies among college educated students in urban India have also shown that background characteristics convey almost perfect signals of caste status relative to surface-level attributes (last names, facial features, and dialects) that are highly evenly shared across castes, and therefore, obfuscate caste identification \citep{mamidi2011}.\footnote{It is difficult to ``fake" caste status because caste networks are typically quite deep \citep{beteille1965, mamidi2011}.}

\subsection{Nature of Disparities and Key Contributions}
\label{sec:taste_stat_client_test}

Earnings disparities are plausibly due to non-technical personal interviews that screen on parental background, neighborhood, and cultural fit. These characteristics are plausibly weakly correlated with productivity (at the interview round) but are strongly correlated with caste. Such interviews are more likely to reveal caste leading to direct discrimination and, even if HR managers do not value caste \emph{per se}, worsen disparities due to indirect discrimination on background characteristics.

There are \emph{three other key takeaways} from the descriptive facts. First, since the earnings disparities primarily occur due to non-technical personal interviews, policies informing applicants about job opportunities, modifying student preferences, or improving performance at university are unlikely to close the earnings gap \emph{for this population}. Second, the paper's descriptive facts advance the literature on the detection and measurement of labor market disparities. Disparities due to discretionary screening practices that value background characteristics are unlikely to be detected through correspondence studies. In addition, recent works studying the role of hiring discretion have either exclusively focused on low-skilled jobs or provided correlational evidence between callback disparities and HR practices being more subjective \citep{hoffmankahn2015,klinerosewalters2021}. By collecting data on all steps of the placement process, my paper is the first to decompose the earnings gap at successive stages of job search and quantify the role of a widely employed subjective screening practice---non-technical personal interviews---in determining access to elite jobs. 

Finally, these descriptive facts provide an important example of the kinds of data future researchers may have to collect in \emph{elite, urban-educated} settings to better detect disparities from outwardly neutral screening practices. Discrimination based on socioeconomic cues is likely to become more salient in elite, urban-educated settings, as they increasingly become more multi-ethnic and diverse and standard characteristics by which to differentiate groups become less perceptible \citep{loury2002,freeman2011,gaddis2017}. 

\subsection{Alternative Explanations}
\label{sec:alternative explanations}

\begin{enumerate}
\item \textbf{Differences in socio-emotional skills across castes.} Group debates that test for socio-emotional skills do not account for caste disparities. Moreover, as nearly 85\% of the jobs are non-client-facing, firms may not have a large preference for students at the right tail of the socio-emotional skills distribution compared to those at the mean (Section~\ref{sec:data overview firms}).

\item \textbf{Better outside options for advantaged castes procured ``offline."} Students must decide on internship offers \emph{before} registering for the full-time job placement process (Section~\ref{sec:data overview firms}). Registered students are prohibited by the elite college from searching ``offline" (i.e., outside of the centralized placement process) and risk being debarred from the services of the placement office if discovered doing so. These rules are shared across the placement processes of elite Indian colleges.\footnote{See \href{https://www.iitbhilai.ac.in/index.php?pid=AIPC\_COVID19}{All IITs Placement Committee Brochure} and \href{http://placement.du.ac.in/}{Central Placement Cell (Delhi University)}.} Thus, registered students do not simultaneously search offline. 

\item \textbf{Advantaged castes can bargain better over salaries and amenities.} Salaries and non-pecuniary amenities posted by jobs are non-negotiable during the course of the placement process. Compensation bundles are also are verified by the placement office at the conclusion of the placement process. Furthermore, ``exit data" confirms that job getters in my setting start in the same jobs and receive the same compensation bundles, months after the placement process has concluded (see ``Pre-Placement Phase" in Section~\ref{sec:data1} and Section~\ref{sec: earnings gap}). 

\item \textbf{Employers are ``playing along'' at rounds prior to personal interviews due to the possibility of government audits or internal institutional pressure.} Government pressure to promote caste diversity has been weak in the absence of a formal regulatory agency to oversee hiring practices in the Indian private sector \citep{jodhka2007}. Overtly expressed attitudes by private-sector employers also suggest a lack of support for compensatory hiring policies, \emph{even absent government pressure} \citep{newmanandjodhka2007,jodhka2007}. Perhaps unsurprisingly, as recently as 2018, only 3 of the top 100 firms listed on India's premier stock exchange claimed to maintain caste data for internal HR purposes \citep{businessline2018}. 

Moreover, diversity practices in foreign-based MNCs employing Indians are overwhelmingly influenced by the historical priorities of the West, where caste is not a protected category, suggesting that internal institutional pressure to rectify caste disparities has either been sluggish or even non-existent \citep{newyorktimes2021}. Further, even granting internal institutional pressure, the difficulty of observing caste \emph{before} HR interviews makes it unlikely that firms could easily ``play along" in prior screening rounds (Section~\ref{sec:caste reveal}).

\item \textbf{Competition from the government sector.} While government jobs have caste-based quotas, they comprise less than 4\% of jobs in my sample (Sections~\ref{sec:historical detour} and \ref{sec:data overview firms}). Moreover, private-sector firms are \emph{themselves} aware of the meagre presence of public-sector firms in the on-campus job fair of the elite college (the list of firms is available on the online job portal).

\item \textbf{Employers are advancing disadvantaged castes before interviews due to uncertainty about ``fit."}  Caste is difficult to discern before personal interviews (Section~\ref{sec:caste reveal}).\footnote{Note that while employers observe GPA (entrance exam scores) at the application reading stage, it is a very noisy signal of caste, especially outside the tails of the distribution (Online Appendix Figure~\ref{fig:common support}).} Moreover, even if employers decide to advance slightly weaker performers in earlier rounds---thereby possibly (but not always) advancing more disadvantaged castes---such considerations are likely to be relevant only for marginal candidates and should not meaningfully skew the caste difference in the composition of jobs that remain in contention prior to HR interviews.

\item \textbf{Caste differences in preferences over job characteristics.} A model of the placement process developed later confirms that caste differences in preferences over job characteristics do not drive caste disparities in job offers. These job characteristics include stock options, signing bonuses, job sectors, and whether the job is located in a metropolitan city (Section~\ref{sec:student preferences over student characteristics}).

\end{enumerate}

\section{A Model of the Job Placement Process}
\label{sec:model}

In the second part of my paper, I propose policies and provide new evidence on their effectiveness to diversify hiring in \emph{elite entry-level jobs} in the Indian private sector. Such an analysis could also be useful in other contexts, especially since elite MNCs have global footprints. To evaluate policies, I calculate \emph{employer willingness to pay} for key characteristics, such as pre-college test scores and caste. These estimates are obtained from a novel model of the hiring process. As will be emphasized later, employer willingness to pay for caste is \emph{invariant} to the underlying motivations for caste disparities, be it taste-based or statistical discrimination (see ``Role of the caste coefficient" in Section~\ref{sec:joboffers}). Therefore, the model \emph{does not} distinguish between the sources of caste disparities. 

This is also the first empirically estimated model of the job placement process from an elite college and could serve as a prototype for the standardized placement processes of other elite engineering, business, and law schools in India (Section~\ref{sec:data1}). While the details vary, it is an open question as to how best to efficiently match students to firms while also addressing concerns about equity. 

\subsection{Key Unmodeled Features}
\label{sec:key_unmodeled}

I begin by discussing some key features of the job placement process my model takes as exogenous.

\medskip

\textbf{National Wage Setting.} The model takes wage setting as exogenous. This assumption is plausible because firms set job-specific wages \emph{nationally} and do not pay meaningfully different salaries across their Indian locations to new hires from other universities. Using data from a major salary reporting platform in the U.S., I find that there is only a 2\% average difference in salaries offered by firms for the same job (job designation within a firm) across their Indian locations (Online Appendix Table~\ref{tab:Salaries for Select Firms in the Same Location}).\footnote{Note that salaries on the platform are self-reported by employees.} National wage setting is closely tied to features of the job placement process, typically shared across elite Indian colleges (Section~\ref{sec:data1}). Firms cannot change advertised wages or non-pecuniary amenities during the course of the placement processes of elite Indian colleges. Therefore, elite firms typically provide similar job details on the employer registration form equivalents required by the placement offices of other elite universities, as they recruit new hires from those universities during a single on-campus placement cycle lasting from mid-September to early January (see ``Pre-Placement Phase" in Section~\ref{sec:data1}, Online Appendix Section~\ref{sec:employer_registration_form}).\footnote{The phenomenon of firms setting wages nationally has also been documented in the U.S. labor market for elite entry-level jobs \citep{sarsonshazellpattersontasaka2021}.}

\medskip

\textbf{Interview Day Allotment.} The model also takes interview day allotment as exogenous. Recall that each firm is allotted an interview day to conduct on-campus interviews (see ``Placement Phase" in Section~\ref{sec:data1}). Past interview day allocations and job characteristics are almost perfectly predictive of current interview day allocations (Online Appendix Table~\ref{tab:hedonic regression firm characteristics only 1}). Among these job characteristics, job salaries are the \emph{only} significant determinants of interview day assignments. A one standard deviation increase in salary increases the probability of getting assigned the first interview day by 8\%. However, since job salaries are nationally set, it is plausible to assume interview day allocations as exogenous (see ``Pre-Placement Phase" in Section~\ref{sec:data1}). 

\medskip

\textbf{Student Application Behavior.} I do not model student application behavior because students effectively apply to all eligible jobs (Section~\ref{sec: earnings gap}). Omitting application behavior of elite Indian college students from the model is not necessarily a limitation. Job placement processes of elite Indian colleges typically feature streamlined and centralized application systems, such as Job Openings for Economists, thereby making the marginal cost of an additional application effectively zero \citep{mamgain2019}. Still, I show how one could extend the model to incorporate application behavior, as it may be important in settings besides elite Indian colleges (Online Appendix Section~\ref{sec:appendixmodelingapplications}).

\subsection{Modeled Features}

I model the two economically significant stages of job search in my setting: firm hiring and final job choices. 
The model is solved backwards starting from final job choices followed by job offers. For the remainder of the section, note that a ``job" means a job designation within a firm. 

\subsubsection{Stage 2: Job Choice by Students}
\label{sec:finalchoice}

At the job choice stage, students know their job offers and have no uncertainty about their preferences. The set of job options for student $i$ is denoted by $\mathcal{O}(Z_{i})$ defined as
\begin{equation}
\mathcal{O}(Z_{i}) = \{0\} \cup \{j: Z_{ij} = 1\},
\end{equation}
where the outside option, which is indistinguishable from unemployment, is denoted by $j = 0$. The vector $Z_{i} = (Z_{i1}, \dots, Z_{iJ})$ collects all job offers for student $i$, where $Z_{ij}$ is an indicator variable that takes the value 1 if student $i$ receives an offer from job $j$ and 0 otherwise. 

Let $U_{ij}$ be the utility of student $i$ from job $j$. $U_{ij}$ depends upon student and job characteristics, a random effect $q_i$ that is unobserved by the econometrician, and a job offer acceptance shock, $\epsilon_{ij}$, realized after job offers are known but before final job choices are made. Mathematically,
\begin{equation}
\label{eq:finalstage}
U_{ij} = X_{ij}'{\beta} + \text{NP}_{j}'{\Psi}  + w_{j} \tau + q_i + q_i \times \sum_{m = 1}^{M} \gamma_{m} \text{NP}_{jm} + \epsilon_{ij},
\end{equation}
where the vector $X_{ij}$ includes student and job characteristics, particularly interactions between caste and non-pecuniary amenities. The vector $\text{NP}_{j} = (\text{NP}_{j1}, \dots, \text{NP}_{jM })$ is a vector of over 50 unique non-pecuniary amenities for job $j$, and $w_j$ is the (log) salary offered by job $j$. Note that I categorize some fringe benefits as ``non-pecuniary'' amenities because I do not have information on direct cash-equivalents of such benefits for a substantial portion of the sample. 

For identification, the econometrician-unobserved $q_i$ does not enter the utility for the outside option---i.e., $q_i$ shifts the value of all jobs uniformly relative to the value of unemployment. Furthermore, interacting $q_i$ with non-pecuniary amenities like stocks, signing bonuses, and relocation allowances allows for random marginal effects of non-pecuniary amenities and drives preferential selection over job offers. I will assume that $q_i \sim \mathcal{N}(0, \sigma_{q}^2)$. Additionally, each element in the vector of idiosyncratic student preference shocks over jobs, denoted by $\epsilon_{i} = (\{\epsilon_{ij}\}_{j \in J}, \epsilon_{i0})$, is drawn from an independent, identically distributed Type-1 extreme value distribution. I normalize the value of the outside option.\footnote{Despite being more negatively selected (Section~\ref{sec: earnings gap}), caste differences in average welfare may be modest within the sample of non-job getters \emph{in the baseline} because elite disadvantaged caste students may sort into other jobs (e.g., relatively lower-paying elite public- or private-sector jobs with better amenities). Finally, even if we grant that outside options are \emph{worse} for disadvantaged castes, my paper likely underestimates caste disparities in initial placements.} This value is given by
\begin{equation}
\label{eq:normalize}
U_{i0} = \epsilon_{i0}.
\end{equation}

Student $i$'s optimal choice of job $j$ given his set of job offers $\mathcal{O}(Z_{i})$ solves:
\begin{equation}
\label{eq:beforeshock}
C^{*}_{i} = \text{arg}\max_{j \in \mathcal{O}(Z_{i})} U_{ij} - U_{i0}.
\end{equation}

\subsubsection{Stage 1: Student Choice by Jobs}
\label{sec:joboffers}

In this section, I introduce a model of labor demand with limited job-level heterogeneity. I assume that a firm allotted interview day $k$ makes job offers independently of any other firm allotted the \emph{same} interview day. This assumption is plausible, as the placement office requires firms to announce job offers within a very short interval of time at the end of the interview day---typically late in the evening---to prevent firms conducting interviews on the \emph{same interview day} from coordinating on whom to hire (see ``Placement Phase" in Section \ref{sec:data1}). 

Recall, a ``job'' means a job designation within a firm. Let the binary variable $A_{ij}$ indicate whether student $i$ applies to job $j$. The vector $A_{i} = (A_{i1}, \dots, A_{iJ})$ collects these indicators for all jobs. Taking student applications as given, job $j$ accepts student $i$ on interview day $k$ with probability $\pi^{i}_{j}$ that depends on both student and job characteristics. Let $f(Z_i | A_i)$ denote the probability of realizing a job offer vector $Z_i$ given an application vector $A_i$. The formula for $f(Z_i | A_i)$ is shown in Online Appendix Section~\ref{sec:appendix1}.

I now describe how jobs choose students in more detail. Motivated by the earnings decomposition shown in Figure~\ref{fig:earnings gap hiring stage}, I model firm hiring as a one-stage process. Each job chooses an incoming cohort of students to maximize expected utility. In Proposition~\ref{prop} in Online Appendix Section~\ref{sec:appendix_prop}, I show that each job $j$ follows a job-specific cutoff hiring rule. Thus, each job $j$ hires a student $i$ iff
\begin{align}
\label{eq:8}
V_{ij} &= S_{ij}'{\alpha} + \text{Disadv. Caste}_i \times  \eta  - w_{j}\phi  + q_{i} \delta  + \mu_{ij} > \underline{k^{*}_j}. \nonumber  \\ 
&= S_{ij}'{\alpha} + \text{Disadv. Caste}_i \times  \eta  - w_{j}\phi  + q_{i} \delta  + \mu_{ij} - \underline{k^{*}_j} > 0.
\end{align}
where $V_{ij}$ is the utility that job $j$ gets from student $i$, $\underline{k^{*}_j}\equiv\underline{k^{*}_j}(w_j, \overline{X_j})$ is a job-specific cutoff that is estimated for each job $j$ with the vector $\overline{X_j}$ denoting features of the job besides wage, $S_{ij}$ includes student and job characteristics, $w_{j}$ denotes the (log) salary offered by job $j$, $q_i \sim \mathcal{N}(0, \sigma_{q}^2)$ denotes the econometrician-unobserved random effect, and $\mu_{ij}$ is an idiosyncratic match term, which is unobservable to student $i$ but observable to job $j$. Each $\mu_{ij}$ follows a standard logistic distribution and is independent across all students and jobs. Job salaries entering Equation~\ref{eq:8} are taken as exogenous based on evidence that suggests firms set entry-level salaries \emph{nationally} (Section~\ref{sec:key_unmodeled}). 

\medskip

\textbf{Job-specific cutoff.} The hiring cutoff result proven in Online Appendix Section~\ref{sec:appendix_prop} relies on the assumption that the information observed by job $j$ is sufficient for its valuation of $V_{ij}$. In other words, observing decisions of other jobs does not affect job $j$'s best estimate of $V_{ij}$. Note also that $\underline{k^{*}_{j}}$ is \emph{not a structural parameter} and will be allowed to change in counterfactual exercises.

The inclusion of the job-specific cutoff implies that parameters entering Equation~\ref{eq:8} are identified from \emph{within-job} variation. Recall, each firm arrives on campus for a total of four years to conduct recruitment and typically offers \emph{different} job-specific salaries and non-pecuniary amenities across years (Section~\ref{sec:data overview firms}). Overall, $\underline{k^{*}_j}$ acts as a fixed effect that controls for job-level (firm-job-designation-level) heterogeneity. In practice, $\underline{k_{j}}$ is modeled as a job-specific constant.

\medskip

\textbf{Student and job characteristics.} Student characteristics entering $S_{ij}$ include controls for pre-college skills, within-college academic performance, previous labor market experience, and other employer-relevant skills, including indicators for whether the student qualified for various stages of job search. Job characteristics entering $S_{ij}$ could
potentially include the entire set of over 50 non-pecuniary amenities, as applicable. Importantly, while firm hiring is modeled as a one stage process, \emph{I retain information about prior screening stages}. I do so by adding indicators in $S_{ij}$ for students passing the application reading, written test, or group debate stage. Note also that \emph{employers observe pre-college skills} (e.g., college entrance exam scores) on the online job portal.\footnote{Note that while employers observe GPA (entrance exam scores) at the application reading stage, it is a very noisy signal of caste, especially outside the tails of the distribution (Online Appendix Figure~\ref{fig:common support}).}

\medskip

\textbf{Role of the caste coefficient.} The ``caste penalty" is captured by $\eta$ in Equation~\ref{eq:8}. Together, the coefficients on wages and caste in Equation~\ref{eq:8} allow us to calculate employer willingness to pay for caste. This estimate is crucial to assess the effectiveness of new policies to diversify elite hiring. 

As mentioned previously, background characteristics are almost perfectly predictive of caste in \emph{elite, urban-educated India}. Therefore, the reduced-form coefficient representing the caste penalty could capture discrimination due to employers directly valuing caste \emph{in addition to} indirect, researcher-unobserved characteristics (e.g., family background, neighborhood of residence, upbringing, and cultural fit) revealed during personal interviews. Caste disparities could stem from either taste-based or statistical discrimination and the reduced-form caste coefficient \emph{embeds a mechanism for both}. In other words, the magnitude of the caste penalty, and therefore, the employer willingness to pay for caste is \emph{invariant} to the underlying motivations for caste disparities.  

My empirical approach to model the caste penalty through a reduced-form caste coefficient that captures both direct and indirect sources of disparities helps advance recent research that argues for a constructivist understanding of group identities, instead of treating them as immutable facts \citep{bhi2022,rose2022,sarsons2022}. Such an approach is crucial to better understand ``caste," classifications of which are rooted in the economic, political, and material history of India  \citep{beteille1965,beteille1969}. In addition, perceptions of caste in elite, urban-educated India are guided by a myriad of socioeconomic cues, paralleling the impressions of social class in other contexts, especially Britain \citep{deshpande2011, mamidi2011, savage2015, jodhka2017}. 

\medskip

\textbf{Incoming hires.} I now complete the firm's problem. Let $C(j)$ denote the set of applicants who accept an offer from job $j$. I will assume that the utility of job $j$ from cohort $C(j)$ is given by
\begin{equation}
\label{eq:9}
\overline{V}_{j}(C(j)) = \sum_{i \in C(j)} V_{ij}.
\end{equation}
Equation~\ref{eq:9} says that jobs do not focus on team building during initial hiring. This is plausible as the college comprises a small fraction of a job's overall incoming cohort. Plus, a job hiring from this college has only one day to decide its cohort after interviews and has to do so at every other elite Indian college it hires from, making it difficult to coordinate on team building (Section~\ref{sec:data1}). Moreover, jobs select students based on screening tests that are general in scope (Section~\ref{sec:data overview firms}).

In Equation~\ref{eq:9} above, the utility of job $j$ is defined for a \emph{given} cohort $C(j)$. Note that $C(j)$ is random from the perspective of job $j$ when it is deciding which students to extend offers to. Accepting an offer from job $j$ depends upon employer-unobserved idiosyncratic student preferences for job $j$ as well as those for other jobs (through $\epsilon_{i}$ in Equation \ref{eq:finalstage}), while getting other jobs depends upon idiosyncratic match terms not observed by job $j$ (through $\mu_{ij'}$ in Equation \ref{eq:8}). Although job $j$ does not observe $\mu_{ij'}$ for $j' \ne j$, it observes $(S_{ij}, w_j, q_{i}, \mu_{ij})$ for each student $i$. Job $j$ solves
\begin{align}
Z^{*}(j) &= \text{arg} \max_{Z(j) \in \{0, 1\}^{|A(j)|}} \mathbb{E} \bigg[\overline{V}_{j}(C(j))\bigg], \\ 
\label{eq:eleven}
&\text{s.t.} \hspace{0.2cm} \mathbb{E}(|C(j)|) \le \overline{M_{j}} \\ \nonumber 
&= \hspace{-1.3cm}\sum_{\hphantom{kvvvvvvvv}i: V_{ij} > \underline{k^{*}_{j}}, \hspace{0.1cm} j \in A_i} \text{Pr}(C^{*}_{i} = j) \le \overline{M_{j}}.
\end{align}

where the above expectation is taken over unknowns from the perspective of job $j$, $A(j)$ is the set of applicants to job $j$, $Z(j)$ is the set of applicants who receive offers from job $j$, and Equation~\ref{eq:eleven} is the ex-ante hiring constraint faced by job $j$. The left-hand side of Equation~\ref{eq:eleven} is the expected size of the incoming cohort $C(j)$ for job $j$, where $V_{ij}$ is the utility to job $j$ from student $i$, $\underline{k^{*}_{j}}$ is the job-specific hiring cutoff, $A_i$ is the application vector of student $i$, and $C^{*}_i$ is the optimal job choice by student $i$ at the job choice stage. The right-hand side of Equation~\ref{eq:eleven} is the ex-ante hiring cap of each job $j$, denoted by $\overline{M_{j}}$ (which is not a parameter).

Notice that job salaries also enter the ex-ante hiring constraint since they enter $V_{ij}$ through Equation~\ref{eq:8}, rationalizing the fact that a  job (job designation within a firm) may make offers in proportion to their wages. For example, a job paying a higher wage in a given year may make fewer offers, all else being equal, and vice versa.

\medskip

\textbf{Interpreting the econometrician-unobserved random effect \boldmath{$q$}.} Notice that the random effect $q$ enters the utility functions of both students and jobs. An economic interpretation of such a specification is that jobs may choose students either because they like high $q$ students (Equation \ref{eq:8}) or because high $q$ students are more likely to accept an offer conditional on getting one (Equations \ref{eq:finalstage} and \ref{eq:eleven}). Hence, $q$ acts as a productivity term while also affecting preferences over jobs. See \citet{howell2010} for a similar treatment of unobserved heterogeneity. An example of $q$ could be ``student interest level," which employers could learn during non-technical personal interviews.

\medskip

\textbf{Comment regarding the specification of the random effect.} One might wonder if instead of the same $q$ entering the utilities of students and jobs, it would be more reasonable to allow for two different, but correlated, sources of unobserved heterogeneity: one that affects how students value jobs and vice versa. For example, if we consider such a correlation to represent the ``quality'' of the private information observed by the student about his employer-observed $q$, then the ideal data should have observably identical students with better signals applying more ``aggressively.'' However, with little to no variation in student application behavior, conditional on observables, such a correlation is infeasible to identify in my setting. 
     
More precisely, the lack of variation in application behavior, conditional on observables, restricts the modeling choice on the student side to essentially one that just incorporates final job choice behavior. Recall that modeling application behavior of elite Indian college students is not necessarily economically interesting, as streamlined and centralized application systems effectively make students apply for all eligible jobs (see ``Student Application Behavior" in Section~\ref{sec:key_unmodeled}). However, modeling only final job choices of students restricts how flexible one can be with random effects that enter students' decisions over job offers \emph{and} are correlated with the random effect entering firms' hiring decisions. The basic constraint is that students can only accept one job offer.

Note also, however, that the main result on the student side---there are no average caste differences in preferences over non-pecuniary amenities---holds with or without the inclusion of the random effect $q$ in student utility (Section~\ref{sec:student preferences over student characteristics}). Additionally, employer willingness to pay estimates (crucial for counterfactuals) do not critically depend upon $q$ entering \emph{student} utility.

\subsection{Equilibrium}
\label{sec:equilibrium}
An equilibrium is a tuple
\begin{align*}
\big\{\underline{k^{*}_{j}}, C^{*}_{i} \big\}_{i = 1, \dots, I, j = 1, \dots, J}
\end{align*}
where $i \in \{1, \dots, I\}$ indexes the student and $j \in \{1, \dots, J\}$ indexes the job such that:
\begin{enumerate}
\item At the final stage, student $i$'s optimal choice of job $j$ given his set of job offers $\mathcal{O}(Z_{i})$ solves 
\begin{align}
\label{eq:equilibrium1}
C^{*}_{i} = \text{arg}\max_{j \in \mathcal{O}(Z_{i})} U_{ij} - U_{i0},
\end{align}
where $U_{ij}$ and $U_{i0}$ are given by Equations \ref{eq:finalstage} and \ref{eq:normalize}, respectively.
\item Given the application vector $A_{i}$ of student $i$, each job $j$ solves 
\begin{align}
\label{eq:equilibrium2}
Z^{*}(j) &= \text{arg} \max_{Z(j) \in \{0, 1\}^{|A(j)|}} \mathbb{E}
\bigg[\overline{V}_{j}(C(j))\bigg], \\ 
\label{eq:equilibrium3}
& \text{s.t.} \hspace{0.2cm} \mathbb{E}(|C(j)|) \le \overline{M_{j}},
\end{align}
where the expectation above is taken over unknowns from the perspective of job $j$,  $C(j)$ is the incoming cohort for job $j$, $A(j)$ is the set of applicants to job $j$, $Z(j)$ is the set of applicants who receive offers from job $j$, Equation~\ref{eq:equilibrium3} is the ex-ante hiring constraint faced by job $j$, and $\overline{M_{j}}$ is the ex-ante hiring cap for job $j$. 

\end{enumerate}

\section{Identification and Estimation}
\label{sec:identification and estimation}

This section describes the identification of key model parameters: student preferences over job characteristics, wage effects entering the employer's utility function, the econometrician-unobserved random effect $q$, and the ``caste penalty.'' Estimation is done via maximum simulated likelihood (Online Appendix Section~\ref{sec:appendix2}).

Recall that a ``job'' is a job designation within a firm. I assume that characteristics like caste, salaries, and non-pecuniary amenities entering the utility functions of students are exogenous. Similarly, exogenous characteristics entering the utility functions of jobs include salaries, sector, caste, major, and degree.

\begin{itemize}
\item[1.]\textbf{Student preferences over job characteristics.} Identification of student preference parameters comes from variation in job characteristics of both accepted and rejected job offers, which also lead to variation in job choices between students.

\item[2.]\textbf{Wage effects.} Identification of parameters entering the employer's utility function comes from \emph{within-job} variation since the job-specific cutoff, $\underline{k^{*}_j}$, basically acts like a fixed effect controlling for job-level heterogeneity. For example, identification of the wage effects in Equation~\ref{eq:8} comes from within-job time variation in wages and job offer rates. Note that each firm arrives on campus for a total of four years to conduct recruitment. Across years, a firm typically offers \emph{different} salaries and non-pecuniary amenities for the same job designation.

I assume that wages are causal primarily because evidence suggests that rookie salaries are set \emph{nationally}, so it is unlikely that local conditions or shocks influence them (Section~\ref{sec:key_unmodeled}). Note also that this college comprises only a tiny fraction of a given firm's total hiring pool. 

\item[3.]\textbf{Econometrician-unobserved random effect.} Econometrician-unobserved $q_i \sim \mathcal{N}(0, \sigma_q^2)$ is identified from correlation in offer probabilities across jobs \emph{within a student's application portfolio}. Conditional on observables, highly correlated job offer outcomes within a student's job application portfolio imply that econometrician-unobserved $q$ plays an important role in job hiring. Basically, identification comes from the fact that, conditional on observables, getting an offer from Facebook is correlated with getting an offer from Microsoft.

\item[4.]\textbf{Caste penalty.} Identification occurs under the plausible assumption that---conditional on resume screening and performance in both technical tests and non-technical group debates that assess a wide array of socio-emotional skills---parental background, neighborhood, and impressions of ``cultural fit" are plausibly weakly correlated with productivity. To address concerns regarding potential differences in unobserved ability by caste, Equation~\ref{eq:8} includes detailed measures of pre-college skills, within-college academic performance, previous labor market experience, and other employer-relevant skills, \emph{including} indicators for whether the student got past the application reading, written test or group debate stage. I also assume that the econometrician-unobserved $q$ in Equation~\ref{eq:8} is \emph{uncorrelated} with caste and that there is no other error term capturing unobserved ability of applicants. 

\end{itemize}

\section{Parameter Estimates}
\label{sec:parameter estimates}

In this section, I quantify student and employer willingness to pay for key characteristics. To do so, I scale the coefficients of interest by the coefficient on wage in either student or firm utility.

\subsection{Student Preferences over Job Characteristics}
\label{sec:student preferences over student characteristics}

Table \ref{tab:parameter estimates student utility main} shows select parameter estimates entering the utility functions of students. Unless otherwise stated, all compensation measures are interpreted for a student with mean $q$ and dollar amounts are in PPP terms.
 
\textbf{Main takeaways.} About 50\% of the students who participate in the placement process get multiple job offers.\footnote{Multiple offers are from firms interviewing candidates on the \emph{same} interview day (Section~\ref{sec:data1}).} While non-pecuniary amenities are valuable to students on average, there are no caste differences in preferences over them. Recall, non-pecuniary amenities comprise a list of almost 50 unique items \emph{per job}, including stock options, signing bonuses, relocation allowances, medical insurance, performance bonuses, whether the job is located in a metropolitan city, job sectors, and so on.\footnote{Recall, I categorize some fringe benefits as ``non-pecuniary'' amenities because, for a substantial portion of the sample, I do not have information on direct cash-equivalents of such benefits (Online Appendix Table~\ref{tab:non pecuniary amenities full 1}).}

\subsection{Job Preferences over Student Characteristics}
\label{sec:job preferences over student characteristics}

Table \ref{tab:select parameter estimates firm utility main} shows select parameter estimates entering the utility functions of jobs.

\textbf{Main takeaways.}  Firms discount the value of disadvantaged castes at the equivalent of 4.8\% of average annual salary (\$2721), holding other student attributes constant. Employer willingness to pay for an advantaged caste is large. All else being equal, an increase in college GPA of about one standard deviation equalizes hiring probabilities across castes. In addition, closing the caste gap in each pre-college test score quantile closes only about 10\% of the model-implied caste penalty.

\subsection{Modeled Unobservables}
\label{sec:modelled unobservables}

Table~\ref{tab:random marginal effects for non-pecuniary amenities main} reports the standard deviation of the econometrician-unobserved random effect $q$.

\textbf{Main takeaways.} Econometrician-unobserved  $q$ plays only a modest role in student and firm utility. Consider a job that does not offer any non-pecuniary amenities. To get the same utility from that job as a student with one standard deviation higher $q$, a student with mean $q$ needs to be compensated about 1.7\% (\$969) of average salary. Similarly, a firm needs to be subsidized 1.1\% of average salary (\$623) to offset a one standard deviation decrease in $q$.

\subsection{Model Fit and Job Cutoffs}
\label{sec:modelfit_and_jobcutoffs}

The model does a good job fitting many moments, including job offers, job choices, unemployment, and the earnings gap (Online Appendix Table~\ref{tab:model fit job choice and job offers main 1}). As expected, the highest paying firms have the highest hiring cutoffs and vice versa (Online Appendix Table~\ref{tab:firm cutoffs main}). 

{\eject \pdfpagewidth=10in \pdfpageheight=10in
\begin{landscape}
\thispagestyle{empty}
 \begin{table}[!htbp]
 \begin{adjustwidth}{-0.3cm}{}
    \centering
        \small
    \caption{Select Parameter Estimates (Student Utility)}
    \label{tab:parameter estimates student utility main}
    \begin{threeparttable}
    \begin{tabular}{c c c c c c c} 
        \\[-1.8ex]
        \hline 
        \hline 
        
        \\[-1.8ex] 
         \\[-1.8ex] Parameter & Estimate & Std. Error &  Compensation (\$) & Std. Error (\$) & Compensation (\%) & Std. Error (\%)\\
         \hline 
         \\[-1.8ex] 
         
         Salary (log), $\tau$ & $2.482^{***}$ & 0.008 & --- & ---  & --- & ---  \\
         
         \\

        Signing Bonus & $0.156^{***}$ & 0.005 &  $+3683.111^{***}$ & 120.058 & $+6.489^{***}$ & 0.211 \\
        
        \\
        
        Performance Bonus & $0.049^{***}$ & 0.008 & $+1132.033^{***}$ & 199.491 & $+1.994^{***}$ &  0.351\\

        \\

        Medical Insurance & $0.046^{***}$ & 0.010 &  $+1062.080^{***}$ &  233.872 & $+1.871^{***}$ & 0.412\\

        \\

        Relocation Allowance & $0.078^{***}$ & 0.010 & $+1812.616^{***}$ & 246.859 & $+3.193^{***}$ &  0.434\\

        \\

        Restricted Stock Units & $0.124^{***}$ & 0.002 & $+2908.609^{***}$ & 50.599 & $+5.123^{***}$ & 0.089 \\

        \\

        Getting a Job in Technology & $0.078^{***}$ &  0.005 & $+1812.616^{***}$ & 115.655 & $+3.193^{***}$ & 0.204\\
        
        \\
        
        Getting a Job in Consulting & $0.087^{***}$ & 0.006 & $+2025.454^{***}$ & 143.100 & $+3.567^{***}$ &  0.252\\

        \\
        
        Metro City & $0.045^{***}$ & 0.009 & $+1038.842^{***}$ & 213.458  & $+1.830^{***}$ &  0.357\\

        \\

         Disadv. Caste $\times$ Salary (log) &$-0.013$ & 0.099 & --- & ---  & --- & ---   \\
         
         \\

        Disadv. Caste $\times$ Signing Bonus  & $-0.026$ & 0.061 &  $-591.654$ & 1380.824 & $-1.042$ & 2.432\\        
        \\

        Disadv. Caste $\times$ Performance Bonus  & $-0.011$ & 0.117 & $-251.072$ & 2664.572 & $-0.442$ & 4.693\\
        
        \\

        Disadv. Caste $\times$ Medical Insurance  & $-0.013$ & 0.134 & $-296.602$ & 3049.280 & $-0.522$ & 5.371\\
        
        \\

        Disadv. Caste $\times$ Relocation Allowance  & $-0.039$ & 0.131 &  $-885.165$ & 2949.910 & $-1.559$ & 5.196\\
        
        \\

        Disadv. Caste $\times$ Restricted Stock Units  & $-0.012$ & 0.127 & $-273.842$ &  2891.160 & $-0.482$ & 5.093\\
        
        \\
        
        Disadv. Caste $\times$ Technology  & $-0.046$ & 0.065 & $-1042.574$ & 1459.487 & $-1.836$ & 2.571\\
        
        \\

        Disadv. Caste $\times$ Consulting  & $\phantom{-}0.016$ & 0.079 & $+367.188$ & 1818.833 & $+0.647$ & 3.204\\
        
        \\
        
        Disadv. Caste $\times$ Metro City & $\phantom{-}0.015$ & 0.080 & $+ 344.010$ & 1834.969  & $+0.606$ &  3.261\\

        \hline 
        \hline \\[-1.8ex] 
        \end{tabular}
          \begin{tablenotes}
            \item Average Salary = \$56,767.29 (PPP), $N$ = 4207 (no. of students), $J$ = 644 (no. of jobs). PPP stands for purchasing power parity.
            \item  Notes: Table \ref{tab:parameter estimates student utility main} includes estimates for select student preference parameters over job characteristics. The compensation terms are calculated in units of dollars (PPP) for a person with mean econometrician-unobserved $q$. ``Metro City" is a dummy for whether a job was located in a traditional metropolitan city, like Delhi, Mumbai, Kolkata, or Chennai. Full estimation tables are available upon request.\\ * significant at 10\%, ** significant at 5\%, *** significant at 1\%.
          \end{tablenotes}
\end{threeparttable}
\end{adjustwidth}
\end{table}
\end{landscape}}

\onehalfspacing

{\eject \pdfpagewidth=12in \pdfpageheight=12in
\begin{landscape}
\eject
\thispagestyle{empty}
\begin{table}[!htbp]
\begin{adjustwidth}{-3.5cm}{}
    \centering
    \small
    \caption{Select Parameter Estimates (Job Utility)}
       \label{tab:select parameter estimates firm utility main}
  \begin{threeparttable}
  \begin{tabular}{c c c c c c c} 
\\[-1.8ex]\hline 
\hline \\[-1.8ex] 
        
        \\[-1.8ex] Parameter & Estimate & Std. Error & Employer Subsidy (\$) & Std. Error (\$) & Employer Subsidy (\%) & Std. Error (\%) \\
        
        \hline \\[-1.8ex] 

        Salary (log), $\phi$ & $\phantom{-}1.893^{***}$ & 0.074 & --- & --- & --- & --- \\
        
        \\
        
       Disadv. Caste, $\eta$ & $-0.093^{***}$ & 0.030 & $+2721.486^{***}$ & 863.231 & $+4.794^{***}$ & 1.521 \\ 
       
      \\[-1.8ex]\hline
        \\[-1.8ex]  &  & & B.Tech. Degree & & &  \\
        
        \hline \\[-1.8ex] 
       
      College GPA  & $\phantom{-}0.077^{***}$ & 0.023 & $+2262.744^{***}$ & 667.570 & $+3.986^{***}$ & 1.175 \\ 
       
       \\

       College GPA $\times$ Consulting & $\phantom{-}0.018^{**{\phantom{*}}}$ & 0.010 & $+537.226^{**\phantom{*}}$ & 299.516 & $+0.946^{**\phantom{*}}$ & 0.522 \\
       
       \\

      College GPA $\times$ Technology & $\phantom{-}0.028^{**{\phantom{*}}}$ & 0.012 & $+833.485^{**\phantom{*}}$ & 357.073 & $+1.468^{**\phantom{*}}$ & 0.630 \\
      
      \\

      Entrance Exam Score & $\phantom{-}0.022^{**{\phantom{*}}}$ & 0.011 & $+655.917^{**\phantom{*}}$ & 326.920 & $+1.155^{**\phantom{*}}$ & 0.576\\

    \\[-1.8ex]\hline
        \\[-1.8ex]  &  & & Dual Degree & & &  \\
        
        \hline \\[-1.8ex]

      College Degree  & $\phantom{-}0.039\phantom{^{***}}$ & 0.033 & $+1157.567\phantom{^{***}}$ & 972.072 & $+2.039\phantom{^{***}}$ & 1.712 \\ 
      
      \\
       
       College GPA &  $\phantom{-}0.121^{***}$ & 0.021 & $+3515.013^{***}$ & 604.677 & $+6.192^{***}$ & 1.065 \\

       \\

       College GPA $\times$ Consulting & $\phantom{-}0.012\phantom{^{***}}$ & 0.076 & $+358.718\phantom{^{***}}$ & 2264.842 & $+0.632\phantom{^{***}}$ & 3.990 \\

       \\

      College GPA $\times$ Technology & $\phantom{-}0.014\phantom{^{***}}$ & 0.052 & $+418.283\phantom{^{***}}$ & 1548.101 & $+0.737\phantom{^{***}}$ & 2.727 \\
      
      \\

      Entrance Exam Score & $\phantom{-}0.019^{**\phantom{*}}$ & 0.010 & $+566.922^{**\phantom{*}}$ & 297.577 & $+0.998^{**\phantom{*}}$ & 0.524\\

          \\[-1.8ex]\hline
        \\[-1.8ex]  &  & & M.Tech. Degree & & &  \\
        
        \hline \\[-1.8ex] 
       
      College Degree  & $\phantom{-}0.203^{***}$ & 0.041 & $+5772.520^{***}$ & 1130.359 & $+10.169^{***}$ & 1.991 \\ 
      
      \\

       College GPA  & $\phantom{-}0.123^{***}$ & 0.028 & $+3571.245^{***}$ & 796.479 & $+6.291^{***}$ & 1.403 \\ 
       
       \\

       College GPA $\times$ Consulting & $\phantom{-}0.038^{**\phantom{*}}$ & 0.017 & $+1128.183^{**\phantom{*}}$ & 503.132 & $+1.987^{**\phantom{*}}$ & 0.886 \\
       
       \\

      College GPA $\times$ Technology & $\phantom{-}0.048\phantom{^{***}}$ & 0.052  & $+1421.328\phantom{^{***}}$ & 1521.945 & $+2.504\phantom{^{***}}$ & 2.681 \\
      
      \\

     Entrance Exam Score & $\phantom{-}0.003^{***}$ & 0.001 & $+89.893^{***}$ & 29.988 & $+0.158^{***}$ & 0.053 \\

    \\[-1.8ex]\hline
        \\[-1.8ex]  &  & & M.S. Degree & & &  \\
        
        \hline \\[-1.8ex] 
       
      College Degree  & $\phantom{-}0.182^{***}$ & 0.063 & $+5203.660^{***}$ & 1727.431 & $+9.167^{***}$ & 3.043 \\ 
      
      \\

       College GPA  & $\phantom{-}0.090^{***}$ & 0.022 & $+2635.767^{***}$ & 636.632 & $+4.643^{***}$ & 1.121 \\ 
       
       \\

       College GPA $\times$ Consulting & $\phantom{-}0.023\phantom{^{***}}$ & 0.057 & $+685.550\phantom{^{***}}$ & 1689.161 & $+1.207\phantom{^{***}}$ & 2.976 \\
       
       \\

      College GPA $\times$ Technology & $\phantom{-}0.078\phantom{^{***}}$ & 0.051 & $+2291.530\phantom{^{***}}$ & 1472.316 & $+4.036\phantom{^{***}}$ & 2.593 \\
      
      \\

      Entrance Exam Score & $\phantom{-}0.003^{***}$ & 0.001 & $+89.893^{***}$ & 29.998 & $+0.158^{***}$ & 0.053 \\ 

 \hline 
\hline \\[-1.8ex] 
       \end{tabular}
          \begin{tablenotes}
            \item Average Salary = \$56,767.29 (PPP), $N$ = 4207 (no. of students), $J$ = 644 (no. of jobs). PPP stands for purchasing power parity.
            \item Notes: Table \ref{tab:select parameter estimates firm utility main} includes estimates for the preference parameters of jobs over student characteristics. Employer subsidy measures for entrance exam scores (GPA) are calculated for a unit standard deviation decrease in entrance exam score (GPA).  College entrance exam scores are originally ranks, which have been renormalized so that higher numbers are better. The standard errors for the employer subsidy terms are calculated through the delta method. Degree fixed effects are shown relative to the bachelor's degree. College GPA and sector interactions have been reparametrized to reflect differences relative to the manufacturing sector. Full estimation tables are available upon request.  * significant at 10\%, ** significant at 5\%, *** significant at 1\%.
          \end{tablenotes}
\end{threeparttable}
\end{adjustwidth}
\end{table}
\end{landscape}}

\begin{table}[!htbp]
    \centering
    \normalsize
    \caption{Modeled Unobservables}
    \label{tab:random marginal effects for non-pecuniary amenities main}
  \begin{threeparttable}
    \begin{tabular}{ c c c} 
\\[-1.8ex]\hline 
\hline \\[-1.8ex]

        \\[-1.8ex] Parameter & Estimate & Std. Error \\
        
        \hline \\[-1.8ex] 
                 
        Standard deviation of $q$, $\sigma_{q}$  & $0.042^{***}$ & 0.004 \\
        
        \\
        
        Parameter on $\sigma_q$, $\delta$ & $0.512^{***}$  & 0.024  \\
        
        \\

        $\gamma_{\text{Signing Bonus}}$ & $0.217^{***}$ & 0.053 \\ 
        
        \\

        $\gamma_{\text{Performance Bonus}}$ & $0.526^{***}$ & 0.049 \\ 
        
        \\

        $\gamma_{\text{Medical Insurance}}$ & $0.017\phantom{^{***}}$ & 0.079 \\ 

        \\

        $\gamma_{\text{Relocation Allowance}}$ & $0.286^{***}$ & 0.051 \\ 

        \\

       $\gamma_{\text{Restricted Stock Units}}$ & $0.487^{***}$ & 0.104 \\ 

\hline 
\hline \\[-1.8ex] 
        \end{tabular}
          \begin{tablenotes}
            \item \scriptsize Notes: Table \ref{tab:random marginal effects for non-pecuniary amenities main} includes estimates of the standard deviation of econometrician-unobserved $q$, the factor loading $\delta$ in Equation \ref{eq:8}, and factor loadings ($\gamma_m$) in Equation \ref{eq:finalstage}, where $m$ indexes non-pecuniary amenities or fringe benefits. Full estimation tables are available upon request. 
            * significant at 10\%, ** significant at 5\%, *** significant at 1\%.
          \end{tablenotes}
\end{threeparttable}
\end{table}

\section{Counterfactuals}
\label{sec:counterfactuals}

Using \emph{employer willingness to pay} for key characteristics, such as caste and pre-college test scores, I evaluate three counterfactual policies to improve both the absolute and relative caste hires at \emph{elite private-sector firms} in a partial equilibrium framework. These policies are:

\textbf{1. Hiring subsidies.} As mentioned in Section~\ref{sec:parameter estimates}, model estimates show that eliminating the gap in each pre-college test score quantile closes only about 10\% of the model-implied caste penalty, suggesting the need for policies that directly mitigate caste disparities. In the first counterfactual, I consider one such policy: a hiring subsidy that eliminates the caste penalty by making elite firms indifferent between observably identical applicants across castes. The subsidy is equivalent to the amount employers discount the value of disadvantaged castes---i.e., 4.8\% of average annual salary. This amount is a \emph{one-time common payment} to each elite entry-level job, per disadvantaged caste hired, and is similar in spirit to the incentive-based Diversity Index proposed by the Ministry of
Minority Affairs \citep{sachar2006, diversityindex2008}. Note that, in principle, the hiring subsidy reimburses the employer for a \emph{stream of costs} incurred in the future and \emph{not just} the cost of hiring a disadvantaged caste over a single year. 

Given the structure of employee salaries in elite entry-level jobs in the Indian private sector, a subsidy is also a natural policy to diversify \emph{elite hiring}. Recall that firms cannot not change advertised wages \emph{during} the placement processes of elite Indian colleges (see ``Pre-Placement Phase" in Section~\ref{sec:data1}). Therefore, a subsidy can be conceptually thought of as firms ``adjusting'' wages for disadvantaged castes, which they could in a less restrained entry-level market, with the difference being made up by the government.

\textbf{2. Pre-college interventions.} Next, I consider a ``pre-college intervention'' that equalizes the distribution of pre-college skills (college entrance exam scores) across castes. The pre-college intervention policy encompasses different interventions---usually Randomized Controlled Trials (RCTs)---including hiring tutors, bonuses to teachers, and redesigning school curricula that are evaluated through their impact on educational outcomes, especially test scores \citep{asim2015}. 

\textbf{3. Hiring quotas.} Finally, I consider a hiring quota that requires
firms to hire an equal proportion of applicants from advantaged and disadvantaged castes: a policy that mirrors caste-based quota policies in government jobs \citep{madheswaran2008}.

\subsection{Counterfactual Results: Hiring Subsidies and Pre-College Intervention}
\label{sec:subsidies and pre-college intervention}

The counterfactual analysis uses \emph{employer willingness to pay} estimates to evaluate changes in the absolute and relative caste hires at elite firms in a partial equilibrium framework. Potentially relevant channels that are considered fixed in the counterfactual analysis include wage changes in elite entry-level jobs, reallocation of workers to elite entry-level jobs, caste share of applicants to such jobs (i.e., labor supply shares), and so on. In Section~\ref{sec:comments_modeling_choices}, I argue that omitting these channels does not necessarily limit the scope of my analysis for this population. 

In this section and in Section~\ref{sec:hiring quotas}, I discuss results from the counterfactual analysis. I begin by discussing how the model can bound absolute displacement effects under hiring subsidies and the pre-college intervention policy. I then compare absolute effects and the cost-effectiveness of these two policies in diversifying \emph{elite hiring} in the Indian private sector.

\medskip

\textbf{Bounding displacement effects.} First, note that both hiring subsidies and pre-college test score improvements explicitly improve employers' valuation of disadvantaged castes (Equation~\ref{eq:8}).

\smallskip

\textbf{1. Labor demand is perfectly elastic.} When labor demand is perfectly elastic, jobs do not adjust cutoffs and hire everyone who qualifies. Disadvantaged caste hires are at least as large as in the baseline and there is no displacement of advantaged castes (Online Appendix Figure~\ref{fig:elastic_inelastic_labor_demand}).

\smallskip

\textbf{2. Labor demand is perfectly inelastic.} When labor demand is perfectly inelastic, jobs do not relax their employment targets, raise their cutoffs, and displace advantaged castes in favor of disadvantaged castes (Online Appendix Figure~\ref{fig:elastic_inelastic_labor_demand}). 

Under both scenarios, more disadvantaged castes are hired relative to the baseline. However, disadvantaged caste hires are the highest (lowest) when labor demand is perfectly elastic (inelastic). Job displacements of advantaged castes are the highest (lowest) when labor demand is perfectly inelastic (elastic). This viewpoint is a natural way to bound plausible responses under policies that explicitly improve employers' valuation of disadvantaged castes. In such scenarios, firms would typically do a combination of increasing the hiring threshold a little and hiring a few more workers. 

\medskip

\textbf{Comparing absolute effects.} The model-implied subsidy equivalent of the pre-college intervention policy is about 0.6\% of average annual salary, \emph{which is only about 10\% of the employer willingness to pay for an advantaged caste}. In fact, employer willingness to pay for pre-college test scores is so small that even the upper bound of the earnings gap (in absolute value) under hiring subsidies is smaller than the lower bound of the earnings gap (in absolute value) under the pre-college intervention policy. Specifically, under the two alternative assumptions about labor demand, the earnings gap reduces from 11\% (in absolute value) in the baseline to between 6 to 8 percent under hiring subsidies and between 9 to 10 percent under the pre-college intervention policy (Online Appendix Table~\ref{tab:earnings gap baseline and cf}). Analogous comparisons hold for the reduction in job displacements of disadvantaged castes under both policies (Online Appendix Table~\ref{tab:unemployment baseline and cf}). 

\medskip

\textbf{Comparing cost-effectiveness.} To evaluate cost-effectiveness, I again use the employer willingness to pay estimates and compare the model-implied subsidy equivalent of the pre-college intervention policy to the direct costs of changing test scores.  To calculate the latter, I use estimates from a meta-analysis of education-focused impact evaluations that documents the costs of changing test scores of primary and secondary school students in India \citep{asim2015}. To extrapolate the direct cost of the pre-college intervention policy, I make three extremely conservative assumptions: 1) costs scale linearly with test score changes, 2) students can be perfectly targeted (i.e., the test score of a given student can be changed by any desired amount), and 3) there is no fade out (i.e., test score changes achieved through early
interventions persist over the lifetime). Even under these extremely conservative assumptions, subsidies to hire applicants from disadvantaged castes are twice as cost-effective in diversifying \emph{elite hiring} than the pre-college intervention policy. 

\subsection{Counterfactual Results: Hiring Quotas}
\label{sec:hiring quotas}

In this section, I evaluate the university-level displacement effects of a government-mandated quota that equalizes the caste share of hires within each elite entry-level job.

\medskip

\textbf{Implementation.} My model of the job placement process can readily accommodate hiring quotas. In contrast to responses to other hiring policies considered in this paper, jobs now \emph{explicitly} decide on two hiring thresholds: one for advantaged castes and vice versa.\footnote{I use the word ``explicitly'' because jobs implicitly solved for two hiring thresholds under previous counterfactual policies. The cutoffs for disadvantaged castes were shifted up by the ``caste penalty'' term, $\eta$, in Equation~\ref{eq:8}.}

Solving for two cutoffs per job instead of just one introduces additional computational complexity. The computational challenge can be overcome by leveraging a key institutional feature of the job placement process. Recall that patterns borne out by the data allow us to assume interview day allocation as exogenous (Section~\ref{sec:key_unmodeled}). Additionally, students are prevented from attending interviews on future interview days conditional on receiving job offers on the current interview day (Section~\ref{sec:data1}). Hence, firms allotted the first interview day can ignore firms allotted the second interview day onward as legitimate competition. Firms allotted the second interview day can, therefore, take the decisions of firms allotted the first interview day as given and ignore firms allotted the third interview day onward as legitimate competition, and so on.

\medskip
 
\textbf{Results.} Unlike hiring subsidies or the pre-college intervention policy, a quota policy that equalizes the caste-share of hires in elite entry-level jobs leads to a substantial \emph{decrease} in overall recruitment from the university, as firms counteract the policy by making fewer job offers in total.\footnote{In my model, job salaries are taken as given because they are \emph{nationally} set. As discussed in point 3 of Section~\ref{sec:comments_modeling_choices}, omitting wage setting behavior from the counterfactual analysis is not a major limitation for this population.}

The following elucidates the economic reasoning driving the result above. Under the quota policy, a firm needs to balance hiring from both castes. While more disadvantaged castes are hired under quotas, the caste penalty is large enough to eventually make the average marginal utility of filling two slots lower than the average marginal cost. This happens well before firms can achieve baseline levels of hiring. Therefore, firms counteract the quota policy by making fewer job offers and decrease overall recruitment from the university. Note that employers in my model do \emph{not} have a hard constraint on their hiring size.\footnote{In my model, a job's hiring cap is denoted by $\overline{M_j}$ and is \emph{not} treated as a structural parameter (Equation~\ref{eq:eleven}).}  Thus, quotas may either increase or decrease the total number of students recruited from the university. In other words, the results from the quota policy are not mechanical: they crucially depend upon the \emph{magnitude of the employer willingness to pay} for an advantaged caste. 

While more disadvantaged castes find jobs under quotas, the displacement effects on advantaged castes are severe. The proportion of unemployed disadvantaged castes falls from 36\% to 31\%. However, nearly two advantaged castes become unemployed for a newly employed disadvantaged caste. The proportion of unemployed advantaged castes increases from 25\% to 35\%. Overall, quotas reduce recruitment from the elite college by 7\% (Online Appendix Table~\ref{tab:unemployment baseline and cf}). Empirical evaluations of hiring quotas in elite public-sector jobs in India and elite private-sector jobs in other contexts have found analogous effects \citep{payresearchunit2018, nitaqat2021}.

Taken together, my findings in Sections~\ref{sec:subsidies and pre-college intervention} and \ref{sec:hiring quotas} suggest that a subsidy for hiring disadvantaged castes would be a cost-effective method to diversify \emph{elite hiring} in the Indian private sector. In Section~\ref{sec:comments_modeling_choices} below, I discuss my modeling choices and argue why omitting some aspects (e.g., wage setting, application shares) from my analysis is not a major limitation for this population. 

\subsection{Modeling Choices and Their Implications}
\label{sec:comments_modeling_choices}

In Sections~\ref{sec:subsidies and pre-college intervention} and \ref{sec:hiring quotas}, I used employer willingness to pay estimates to evaluate the performance of counterfactual policies in improving both the absolute and relative caste hires at elite firms. In doing so, I omitted other aspects, including wage changes in elite entry-level jobs, reallocation to elite entry-level jobs, caste share of applicants to such jobs (i.e., labor supply shares), information-based policies, and so on. Below, I argue that omitting these channels does not necessarily limit the scope of my analysis for this population. I also comment on some other modeling decisions below.
 
\begin{enumerate}

\item \textbf{Focusing only on one elite college and elite entry-level jobs in the private sector.} Elite entry-level jobs are important to focus on as they can shape not just an individual's economic trajectory but also broader societal inequalities \citep{riverapedigree}. Additionally, the job placement process of this elite college---the institutional setting of the paper---offers a representative window into how elite college graduates transition into elite entry-level jobs in the Indian private sector (see points 2, 3, 4, 5, and 6 in Section~\ref{sec:data1}). 

\item \textbf{No reallocation to elite entry-level jobs in the private sector from other jobs following compensatory policies and vice versa.} Separate samples from the Periodic Labour Force Survey, which is collected by the National Sample Survey Office (NSSO), and the India Human Development Survey show that the probability of transitioning into elite entry-level jobs in the private sector from ``other" jobs (elite entry-level public-sector jobs, other entry-level private-sector jobs, unemployment, and so on) is less than 2.5\% and vice versa (Section~\ref{sec:data overview firms}). Note that salaries in even elite entry-level jobs in the public sector are about 50\% of those in elite entry-level jobs in the private sector.\footnote{See the report of the \href{https://doe.gov.in/seventh-cpc-pay-commission}{Seventh Central Pay Commission, 2016}.} 

These transition probabilities suggest that elite entry-level workers in the private sector tend to stay there and transitioning into such jobs is challenging, especially since 95\% of the hires of elite jobs in India are from elite colleges (Section~\ref{sec:data1}). High concentration along the diagonals of job transition matrices is common in formal labor markets in India \citep{sarkar2017,jyotirmoy2021}.\footnote{Relatedly, the NSSO defines unemployment as a situation in which all those who owing to lack of work are not working, but seek work through employment exchanges, intermediaries, friends or relatives \citep{nssconcepts}. Therefore, being unemployed in the data collected by the NSSO is closer to being actually out of work (e.g., it does not include self-employment). On the other hand, students who are ``unemployed" through job the placement process I study could include those who are self-employed or taking gap years (Section~\ref{sec: earnings gap}).} Thus, omitting talent reallocation either from or to elite entry-level jobs in the private sector following compensatory policies is not a major limitation. 

\item  \textbf{No wage changes in counterfactuals.} Note that I take wages as exogenous because firms set them nationally (Section~\ref{sec:key_unmodeled}). I also argue that omitting wage changes from the counterfactual analysis is not a major limitation for two reasons. First, recall, my data collection shows that almost all firms recruiting from this elite college also visit other elite Indian colleges and are foreign-based MNCs that hire overwhelmingly for their Indian offices (Section~\ref{sec:data1}). However, about 97\% of the offices and 96\% of the entry-level labor force of these foreign-based MNCs are outside of India (Section~\ref{sec:data1}). Therefore, elite Indian college graduates comprise a small fraction of the global entry-level labor demand of foreign-based MNCs.

Second, I show that there is only a 3\% average difference between real job-specific salaries offered at establishments  in Indian locations versus those in countries similar to the ``MNC headquarters region" (typically locations in North America and Europe), which hire more than 90\% of the firm's entry-level labor force (Online Appendix Table~\ref{tab:differences in nominal salaries}). This finding corroborates recent research on the wage anchoring behavior of elite MNCs. Such firms may care about minimizing job-specific pay inequality across countries to facilitate the international movement of employees \citep{manelici2021, sarsonshjortli}. Relatedly, \citet{bloom2012} argue that elite MNCs typically follow firm-wide wage setting procedures \emph{internationally}, but decentralize hiring decisions or make them \emph{locally}. These behaviors are consistent with my model allowing elite foreign-based MNCs to locally (i.e., in India) adjust on the extensive margin by changing both absolute and relative caste hires due to compensatory policies, while keeping wages fixed due to strong wage anchoring behavior.

\item \textbf{Fixed share and composition of castes admitted to elite colleges following counterfactual policies.} Most elite colleges in India are \emph{public institutions} and explicitly assign 50\% of their seats \emph{within each major} to disadvantaged castes. Political establishments across India are also highly reluctant to modify the caste share of reserved seats in such colleges. Moreover, adding new seats in elite colleges is a long-drawn process due to bureaucratic red tape \citep{newman2009, datta2017}. Given that the number of elite colleges is likely fixed in the short-run, it is reasonable to assume that the share and composition of castes admitted to such colleges remain fixed following counterfactual policies.  

\item \textbf{Fixed share and composition of castes applying to elite jobs following  counterfactual policies.} These are not major limitations for the following reasons. First, almost 96\% of elite Indian college graduates work in elite entry-level jobs in the Indian private sector. Moreover, graduates from elite Indian colleges account for more than 95\% of the hires of elite jobs in India (Section~\ref{sec:data1}). Second, job placement processes of elite colleges typically feature streamlined and centralized application systems, effectively making students apply for all eligible jobs recruiting from campus (Section~\ref{sec: earnings gap}; \citealp{mamgain2019}). Third, within-major caste shares and the total number of seats in elite colleges are likely stable in the short-run (see point 4 above). Finally, the probability of transitioning into elite entry-level jobs in the private sector from other jobs is negligible and vice versa (see point 2 above). Given these facts, it is reasonable to assume that the caste share of students applying to elite entry-level jobs following compensatory policies is fixed in the short- to medium-run. 

Assuming that the composition of castes applying to elite jobs remains fixed following counterfactual policies is also not a major limitation. If advantaged castes prefer to go elsewhere (e.g., abroad) as a consequence of losing their ``unfair" advantage in elite entry-level jobs, my paper likely underestimates the effects on absolute and relative disadvantaged caste hires due to counterfactual policies. Similarly, assuming a fixed composition of disadvantaged caste applicants to elite jobs is likely to underestimate the effects on absolute and relative disadvantaged caste hires due to counterfactual policies.

\item  \textbf{Caste penalty fixed in the counterfactual analysis.} My counterfactual policies consider employer weights on various sources of caste disparities as policy invariant. These include weights on caste per se as well as on indirect sources facilitating caste identification (family background, father's job, cosmopolitan attitudes, upbringing, neighborhood of residence, personal hobbies, and desire for traveling) that are likely revealed during non-technical personal interviews. These weights are captured as a whole by the reduced-form caste coefficient in the employer's utility function used to calculate employer willingness to pay for caste (Section~\ref{sec:joboffers}). The main goal of the policy exercise is to use this estimate to motivate and evaluate potential solutions to caste disparities in elite jobs. Capturing the full equilibrium effects of counterfactual policies on the weights employers put on various direct (i.e., caste) and indirect characteristics that lead to disparities is outside the scope of this paper.

\item \textbf{Information-based policies.} Information-based policies to correct employers' potentially biased beliefs about the correlation between background characteristics and productivity are likely to be ineffective, especially given evidence that (unconditional) correlates of socioeconomic status are highly predictive of career success in elite firms \citep{Eschleman2014,clark2020,correa2019}. Moreover, ``valid" stereotypes regarding the correlation between ``cultural fit" and career success could have been reproduced through hiring heuristics (e.g., the ``airplane test" cited in Section~\ref{sec:data overview firms}) colored by ambient bias, further rendering information-based policies less effective \citep{riverapedigree}. Finally, even trying to statistically measure ``cultural fit" may be challenging or even infeasible. 

\item \textbf{Stigmatization and potentially addressing ``accurate" statistical discrimination:} Subsidies are unlikely to \emph{further} stigmatize beneficiaries \emph{from this population} for three reasons. First, while theoretical works have suggested that stigma could be worsened due to affirmative action policies, empirical research has found slim evidence in support of this contention \citep{coateandloury1993, bok1998, deshpanderedtape}. Second, since employers are unlikely to know caste until the final round HR interviews, potential beneficiaries are likely as capable as non-beneficiaries in technical skills judged by written tests and socio-emotional skills judged by group debates. These skills are judged \emph{before} HR interviews and are plausibly more strongly correlated with productivity than subjective impressions of ``cultural fit." Third, even granting the purported worsening of stigma, compensatory policies could still be efficiency enhancing, as disadvantaged groups likely benefit the most from elite attainment, whereas displaced advantaged groups are likely not much worse off \citep{blackdemmingrothstein2020}. This reason is also why it could still be meaningful to intervene through policies to address disparities from accurate statistical discrimination.

\item  \textbf{Firms not changing their recruitment practices.} Following counterfactual policies, elite firms might reevaluate potential trade-offs regarding the number of schools to visit, which screening steps to keep in place, how many candidates to interview, and how many resources to invest in hiring. Modeling such dynamics is beyond the scope of my paper. Moreover, it is also unlikely for such changes to materialize in the short-run, given that job placement processes of elite Indian colleges are standardized, dictated by universities, and closely modeled after those organized by elite Indian colleges established in the early 1950s (Section~\ref{sec:data1}). 

\item  \textbf{Not modeling GPA.} I do not model GPA since GPA and entrance exam scores are slightly negatively correlated in the data, \emph{conditional on caste and other observables} (Online Appendix Table~\ref{tab:gpa_and_test_score_negative_correlation}). Therefore, my model \emph{overestimates} the impact of the counterfactual policy exercise that increases college entrance exam scores of disadvantaged castes.\footnote{Recall that 10th and 12th standard  grades are not statistically different across castes (Section~\ref{sec:data overview students}).} One possible explanation for this (conditional) correlational pattern could be random variation in college entrance exam scores, conditional on ability. Students at the top of the distribution are more likely to have positive error in their entrance exam scores. Since the pool of students at the elite college is truncated at relatively high entrance exam scores, the correlation between GPA and entrance exam scores is plausibly dominated by the top of the distribution.

Omitting GPA effects from the counterfactual analysis is not necessarily a broader limitation. This is because of two reasons. First, \citet{krishna2015} also find slightly negative correlation between college GPA and entrance exam scores among disadvantaged caste students belonging to a different elite Indian college. As in my data, these effects are stronger within the most selective majors. Second, \emph{among elite Indian college students}, even high school grades and college performance are weakly correlated, largely because such students are already highly selected on the former \citep{krishna2015}.

\item  \textbf{Not modeling internship choices.} I omit modeling internship choices from my model as they are slightly negatively correlated with pre-college skills (Section~\ref{sec:differences in baseline characteristics}). The weak (and sometimes negative) correlation between internship outcomes and pre-college skills among elite Indian college students is a common pattern plausibly because, unlike in the U.S., internships are viewed by students as exploratory. This view is supported by the fact that only modest proportions ($\sim$ 5-8\%) of students accept return offers from summer internships, thereby not skipping the regular placement processes of elite colleges (\citealp{singh2018}; Section~\ref{sec: earnings gap}).
     
\item  \textbf{No bargaining by workers over wages and non-pecuniary amenities.} Placement processes of elite colleges prohibit bargaining over compensation bundles during the course of the placement cycle (see ``Pre-Placement Phase" in Section~\ref{sec:data1}). As mentioned previously, ``exit surveys" confirm that job getters in my setting start in the same jobs and receive the same compensation bundles, months after the placement process has concluded (Section~\ref{sec: earnings gap}). Therefore, my data offers an accurate description of the compensation bundles offered to students at the start of their new jobs. Moreover, under the plausible assumption that bargaining over salaries, non-pecuniary amenities, and promotions later in workers' careers could favor advantaged castes more, my paper likely underestimates caste disparities in initial placements. 

\item \textbf{Financing of subsidies.} The model does not consider financing of policies such as hiring subsidies. However, hiring subsidies may raise tax receipts by increasing employment and also reduce expenditure on unemployment assistance. Recent studies of hiring subsidies in Germany and France have even found them to be self-financing, even after accounting for bureaucratic costs of enforcement and roll-out \citep{brown2011, cahuc2014}.  
\end{enumerate}

In Online Appendix Section~\ref{sec:additional_channels_appendix}, I discuss other modeling choices and argue that they do not necessarily limit the scope of my analysis for this population. These include: 1) omitting firm entry, 2) fixing student preferences in the counterfactuals, 3) not modeling either multiple job screening stages or job applications, 4) not considering either wages, performance, hiring by firms of workers from other universities or job changing, 5) having the same random effect, $q$, enter both student and firm utility, 6) not omitting the random effect from student utility, 7) not modeling the equity-efficiency tradeoff, and 8) not incorporating legal challenges to subjective, personal interviews.

\section{Conclusion}
\label{sec:conclusion}

Discrimination based on socioeconomic cues in elite, urban-educated settings is likely to become more salient as the world becomes increasingly multi-ethnic and diverse and standard characteristics by which to differentiate groups become less perceptible \citep{loury2002,freeman2011,gaddis2017}. I provide an important example of the kinds of data future researchers may have to collect in such settings to better detect disparities from outwardly neutral screening practices. In doing so, my paper is the first to quantify the role of a widely employed subjective screening practice---non-technical personal interviews---in determining access to elite jobs. Additionally, by connecting how perceptions of socioeconomic cues determine barriers to elite attainment, this paper also helps advance how to conceptualize, quantify, and address racial, class, or caste disparities in such opportunities, most of which are situated in a rapidly diversifying urban landscape.
 
While this paper attempts to do many things, no paper is exhaustive.  Future research could collect similar data to detect less visible forms of discrimination in other parts of the world. Other works could also examine the \emph{evolution} of the caste penalty beyond the first job. Experimental follow-ups studying different \emph{firm-level} policies such as standardized interview questions, decision review, and interviewer representation are also promising areas for future exploration.

\setcitestyle{numbers}

\bibliographystyle{aer}
\bibliography{notes_refs}

\clearpage

\appendix

\setcounter{section}{0}

\setcounter{equation}{0}
\renewcommand{\theequation}{OA.\arabic{equation}}

\setcounter{table}{0}
\renewcommand{\thetable}{OA.\arabic{table}}

\setcounter{page}{1}
\renewcommand{\thepage}{OA.\arabic{page}}

\setcounter{figure}{0}
\renewcommand{\thefigure}{OA.\arabic{figure}}

\setcounter{footnote}{0}
\renewcommand{\thefootnote}{OA.\arabic{footnote}}

\setcitestyle{authoryear, round, comma, aysep={;}, yysep={,}, notesep={, }}

\hspace{-0.6cm}{\bf{\LARGE \underline{ONLINE APPENDIX}}}

\singlespacing

\section{Employer Registration Form}
\label{sec:employer_registration_form}

\vspace{1cm}
\begin{itemize}
    \item[\textbf{Step 1:}] After the invitation, companies should register basic details using the online portal. 
    
    \item[\textbf{Step 2:}] Register basic details on online portal:
    
    \vspace{0.2cm}
    
    \textbf{\underline{A) Company Details:}}
    \begin{itemize}
        \item Company Name* :
        \item Password* :
        \item Confirm Password* :
        \item Website* :

    \end{itemize}
    
    \textbf{\underline{B) Contact Details:}}
    
        \begin{itemize}
        \item Name* :
        \item Designation* :
        \item Contact Number* :
        \item Address* :
  
    \end{itemize}
    
        \item[\textbf{Step 3:}] After registering basic details, companies should enter job details, and select majors who qualify to apply.
        
        \vspace{0.2cm}
        
        \textbf{\underline{A) Job Details:}}
        
        \begin{itemize}
        \item Job Designation* :
        \item Offer Types* : Domestic $\square$ International $\square$
        \item Startup* : Yes $\square$ No $\square$
        \item Job Description* : Job\_Details.pdf [details of non-pecuniary amenities here]
        \item Probable number of slots per job*: 

        \end{itemize}
        
       \textbf{\underline{B) Select the Majors you wish to recruit from:}}
       
       \begin{itemize}
           \item \textbf{Bachelor of Technology:} \\
           All $\square$  Electrical Eng. $\square$ Aerospace Eng. $\square$ Mechanical Eng. $\square$	\\
            Metallurgical Eng. $\square$ Civil Eng. $\square$	 	 
            Material Eng. $\square$ \\
           Ocean Eng. $\square$ Computer Science $\square$
           
           \item \textbf{Dual Degree:} \\
            All $\square$  Electrical Eng. $\square$ Aerospace Eng. $\square$ Mechanical Eng. $\square$	\\
            Metallurgical Eng. $\square$	Civil Eng. $\square$	 
            Material Eng. $\square$ \\
           Ocean Eng. $\square$ Computer Science $\square$

           \item \textbf{Master of Technology:} \\
            All $\square$  Electrical Eng. $\square$ Aerospace Eng. $\square$ Mechanical Eng. $\square$	\\
            Metallurgical Eng. $\square$ Civil Eng. $\square$
            Material Eng. $\square$ \\
           Ocean Eng. $\square$ Computer Science $\square$
           
           \item \textbf{Master of Science:} \\
            All $\square$  Physics $\square$ Chemistry $\square$ Mathematics $\square$	\\
        \end{itemize}
        
         \textbf{\underline{C) Salary Details:}}
         
         \begin{table}[!htbp]
\begin{adjustwidth}{-0.3cm}{}
    \centering
    \caption*{}
        \label{tab:erf_salary}

  \begin{threeparttable}

        \begin{tabular}{@{\extracolsep{0pt}} c c c c}
                               
        \hline \\[-1.8ex]                 
                               
        \textbf{Degree} & \textbf{Gross Annual Pay} & \textbf{Gross Monthly Pay} & \textbf{Additional Comments}\\

        \hline \\[-1.8ex]
        Bachelor of Technology & & & \\
        \hline \\[-1.8ex]
        Dual Degree & & & \\
       \hline \\[-1.8ex]
       Master of Technology & & & \\
       \hline \\[-1.8ex]
       Master of Science & & & \\
       \hline 
       \hline 
       \\[-1.8ex]
     \end{tabular}  
     \end{threeparttable}                         
     \end{adjustwidth} 
    \end{table}
\end{itemize}

\section{Modeling Job Applications}
\label{sec:appendixmodelingapplications}

The difference in the composition of job applications across castes is not economically significant in my setting. However, I show below that the model can be extended to incorporate job application behavior. Therefore, the decision to omit job applications is not a restriction on the generality of my model of the job placement process.

\textbf{Choosing Jobs Instead of Job Portfolios.} The key trick in modeling job application behavior is to convert the student's search from one over potential \emph{job application portfolios} to one over \emph{jobs}. The intuition is simple: for any job a student applied to, the expected marginal benefit from adding the job to his application vector should exceed the cost of applying to the job. 

Let $A_i$ denote the application vector of student $i$. Following the notation in \citet{howell2010}, define

\begin{equation}
\label{eq:modify_app}
    A_{i/k} = \begin{cases}
     \{m | m \in A_i, m \ne k\} \hspace{0.3cm} &\text{if} \hspace{0.2cm}k \in A_i \\    
     \{m | m \in A_i\} \cup \{k\}  \hspace{0.3cm} &\text{if} \hspace{0.2cm}k \notin A_i
    \end{cases}
\end{equation}
Then, it must be true that 
\begin{align}
    MV_{i/k} > 0 \hspace{0.3cm} \forall k \in A_i \\
      MV_{i/p} < 0 \hspace{0.3cm} \forall p \notin A_i 
\end{align}
$MV_{i/k} = V(A_i) - V(A_{i/k})$ denotes the marginal value from modifying the application vector according to Equation~\ref{eq:modify_app} above. To make the computation tractable, one proceeds by reducing the search space by eliminating dominated strategies. Following \citet{howell2010}, we categorize strategies into four main categories: adjacent, non-adjacent, single-swap, and multiple-swap strategies. 

Consider an application vector, $A_i = \{\text{Goldman Sachs}, \text{Microsoft}, \text{Google}\}$. Removing ``Goldman Sachs'' from the application vector is an \emph{adjacent strategy}. Removing both ``Goldman Sachs'' and ``Google'' from the application vector is a \emph{non-adjacent strategy}. Replacing ``Goldman Sachs'' with ``Facebook'' in the application vector is a \emph{single-swap strategy}. Replacing ``Goldman Sachs'' and ``Microsoft'' with ``Facebook'' and ``Uber'' in the application vector is a \emph{multiple-swap strategy}.

\citet{howell2010} shows that if a student's application strategy is preferred to all adjacent and single-swap strategies, then it will also be preferred to all non-adjacent and multiple-swap strategies. Hence, to begin with, a student only needs to examine $J$ applications and find the first job to apply to. Next, \citet{howell2010} shows that the student needs to evaluate $J - 1$ applications and find the second job to apply to and so on. At most, he needs to evaluate a total of $J +(J - 1)+ \dots +2+1 =
\frac{J(J+1)}{2}$
applications. The complexity of the problem is reduced dramatically. When searching over \emph{job portfolios}, the complexity of the problem is $\mathcal{O}(2^J)$, where $J$ is the number of jobs. However, when searching sequentially over \emph{jobs}, the complexity of the problem is only $\mathcal{O}(J)$, where $J$ is the number of jobs. The cost of job applications can then be modeled in a manner similar to \citet{howell2010}. Finally, a logit kernel smoother is used to obtain closed form solutions \citep{train2003}.

\section{Calculation of Job Offer Probabilities}
\label{sec:appendix1}

In this section, I show how to calculate job offer probabilities, which take into account the key features of the job placement process.

Let $A^{k}_{i}$ be a vector of indicators that takes the value 1 if student $i$ applies to a job allotted interview day $k$.\footnote{Recall, a ``job'' means a job designation within a firm.} Similarly, let $Z^{k}_{i}$ be a vector of indicators that takes the value 1 if student $i$ gets accepted from a job allotted interview day $k$. Taking student applications as given, job $j$ accepts student $i$ on interview day $k$ with probability $\pi^{i}_{j}$, which depends on both student and job characteristics.

For a given interview day allotment to firms, define the probability of interview day $k$ job offers given interview day $k$ job applications, conditional on being eligible for an interview day $k$ job offer, by

\begin{align}
\label{eq:jobofferprobdaywise}
&f_{k}(Z^{k}_{i} | A^{k}_{i}) =  \prod_{j = 1}^{J}\bigg(A^{k}_{ij} \bigg[\pi^{i}_{j}Z^{k}_{ij} + (1- \pi^{i}_{j}) (1 - Z^{k}_{ij} )\bigg] + (1 - A^{k}_{ij}) (1 - Z^{k}_{ij}) \bigg).
\end{align}

From Section~\ref{sec:key_unmodeled}, it is reasonable to assume interview day allotments to jobs as exogenous. However, for the purpose of illustrating the formula for job offer probabilities, it will be easier to also assign probabilities to interview day allotments.

Recall that $Z_i$ is the offer vector for student $i$ and $A_i$ is the application vector for student $i$. Let $f(Z_i | A_i)$ denote the probability of realizing $Z_i$ given $A_i$. Then, $f(Z_{i} | A_{i})$ is defined as

\begin{small}
\begin{align}
\label{eq:full}
f(Z_{i} | A_{i}) = 
\begin{cases}
\prod_{l = 0, 1, \dots, K} \widetilde{f}_{l}(\underbrace{(0,0, \dots, 0)}_{J} | A_{i}) \hspace{0.2cm} \hspace{2cm}&\text{if} \hspace{0.1cm} Z_{ij} = 0 \hspace{0.1cm}\forall j  \\
\sum_{k = 1}^{K}\bigg(\prod_{l = 0, 1, \dots, k -1} \widetilde{f}_{l}(\underbrace{(0,0, \dots, 0)}_{J} | A_{i}) \bigg)\widetilde{f}_{k}(Z_i | A_{i}) \hspace{2cm} &\text{else}, 
\end{cases} 
\end{align}
\end{small}

where $\widetilde{f}_{l}(Z_i | A_{i}) = \sum_{\{m : Z_i \times D^{l}_{m} = Z_i\}} \text{Pr}(D^{l}_{m}) f_{l}(Z^{l}_{i} | A^{l}_{i}) $ is the probability of realizing the offer vector $Z_i$ on interview day $l$, $D^{l}_{m}$ is a collection of indicator variables denoting a possible interview day assignment for day $l$, $m = 1, \dots, 2^{|\{1, \dots, J\}|}$, the symbol ``$X$" denotes \emph{elementwise} vector product (also called Hadamard product), and $f_{l}(Z^{l}_{i} | A^{l}_{i})$ is defined by Equation~\ref{eq:jobofferprobdaywise} above. The term $\prod_{l = 0, 1, \dots, k -1} \widetilde{f}_{l}(\underbrace{(0,0, \dots, 0)}_{J} | A_{i})$ in the big brackets denotes the probability that student $i$ is eligible for a job offer on interview day $k$. For completeness,

\begin{itemize}
\item[(1)] Let $\widetilde{f}_{0}(\underbrace{(0,0, \dots, 0)}_{J} | A_{i}) = 1$.
\item[(2)] If, for a given $k$, there is no such $m$ such that $Z_{i} \times D^{k}_{m} = Z_{i}$, then set \\ $\sum_{\{m : Z_{i} \times D^{k}_{m} = Z_{i}\}} \text{Pr}(D^{k}_{m}) f_{k}(Z^{k}_{i} | A^{k}_{i}) = 0$.
\end{itemize}

\section{Jobs Follow Cutoff Hiring Rules}
\label{sec:appendix_prop}

\begin{prop}
\label{prop}
Each job $j$ follows a cutoff hiring rule denoted by $\underline{k^{*}_{j}}$ and hires a student $i$ iff $V_{ij} > \underline{k^{*}_{j}}$.
\end{prop}
\begin{proof}
The proof follows from \citet{kapor2020}. I prove the proposition above by contradiction. Let $\{1, \dots, I\}$ denote the set of all students job $j$ has to choose from. Let \texttt{Hire}$\{j\}:\{1, \dots, I\} \rightarrow [0,1]$ be a hiring rule used by job $j$ that satisfies Equation \ref{eq:eleven}.\footnote{Equation~\ref{eq:eleven} is given by $\sum_{i: V_{ij} > \underline{k^{*}_{j}}, \hspace{0.1cm} j \in A_i} \text{Pr}(C^{*}_{i} = j) \le \overline{M_{j}}$ and denotes the ex-ante hiring constraint of job $j$. The left-hand side of Equation~\ref{eq:eleven} is the expected size of the incoming cohort $C(j)$ for job $j$, where $V_{ij}$ is the utility to job $j$ from student $i$, $\underline{k^{*}_{j}}$ is the job-specific hiring cutoff, $A_i$ is the application vector of student $i$, and $C^{*}_i$ is the optimal job choice by student $i$ at the job choice stage. The right-hand side of Equation~\ref{eq:eleven} is the ex-ante hiring cap of each job $j$, denoted by $\overline{M_{j}}$ (which is not a parameter).} Suppose it is not a cutoff rule. Then there exist two students $i$ and $i'$ such that $V_{ij} > V_{i'j}$ but \texttt{Hire}$\{j\}(i) < 1$ and \texttt{Hire}$\{j\}(i') > 0$. Let $P_{ij}$ and $P_{i'j}$ denote the probabilities that students $i$ and $i'$ accept offers from job $j$. Then, for some $\epsilon > 0$, it is feasible for job $j$ to increase \texttt{Hire}$\{j\}(i)$ by $\frac{\epsilon}{P_{ij}}$, reduce \texttt{Hire}$\{j\}(i')$ by $\frac{\epsilon}{P_{i'j}}$, and increase overall cohort quality. 
\end{proof}

\section{Estimation Details and Standard Errors}
\label{sec:appendix2}

I describe each of the choice probabilities below, the likelihood function to be estimated, and the estimation method.

\textbf{Job Choice by Students.} Conditional on $q_{i} \sim N(0, \sigma^2_q)$ and given the assumption that each element in  the vector of job acceptance shocks, $\epsilon_{i}$, follows independent Type-1 extreme value distributions, the probability of student $i$ choosing job $j$ at the job choice stage is 

\begin{align}
\label{eq:choiceprob}
\text{Pr}(C^*_{i} = j | X_{ij}, w_j, \text{NP}_{j}, q_i) = \frac{\text{exp}(u_{ij})}{\sum_{k \in \mathcal{O}(Z_{i})} \text{exp}(u_{ik})},
\end{align}
where $\mathcal{O}(Z_{i})$ denotes the offer set of student $i$, $X_{ij}$ is the vector of student and firm characteristics, $w_j$ is the (log) salary, $\text{NP}_j$ is the vector of non-pecuniary amenities, and $u_{ij} = X_{ij}'{\beta} + \text{NP}_{j}'{\Psi} + w_j \tau + q_i + q_i \times \sum_{m = 1}^{M} \gamma_{m}\text{NP}_{jm}$.

\textbf{Student Choice by Jobs.} Conditional on $q_{i} \sim N(0, \sigma^2_q)$ and given the assumption that the idiosyncratic match specific term $\mu_{ij}$ between student $i$ and job $j$ follows a standard logistic distribution, the probability of student $i$ getting accepted from job $j$ is

\begin{align}
\pi^{i}_{j}(S_{ij}, w_j, q_{i}, \underline{k^{*}_j}) =  \frac{\text{exp}(S_{ij}'{\alpha} + \text{Disadv. Caste}_i \times  \eta - w_{j}\phi + q_{i}\delta - \underline{k^{*}_{j}} )}{1 + \text{exp}(S_{ij}'{\alpha} + \text{Disadv. Caste}_i \times  \eta - w_{j}\phi + q_{i}\delta - \underline{k^{*}_{j}} ) },
\end{align}
where $S_{ij}$ includes student and job characteristics. Student characteristics include controls for pre-college skills, within-college academic performance, previous labor market experience, and other employer-relevant skills, including indicators for whether the student qualified got past the application reading, written aptitude test, or group debate stage. Job characteristics include indicators for the job sector and the entire set of over 50 non-pecuniary amenities, as applicable. These job characteristics are usually interacted with student characteristics entering $S_{ij}$. The term $w_{j}$ denotes the (log) salary offered by job $j$, $q_i$ is econometrician-unobserved student-level attributes, and $\underline{k^{*}_j}\equiv\underline{k^{*}_j}(w_j, \overline{X_j})$ is a job-specific cutoff that is estimated for each job $j$ with the vector $\overline{X_j}$ denoting features of the job besides wage. Let $f(Z_i | A_i)$ denote the probability of realizing a job offer vector $Z_i$ given an application vector $A_i$. The formula for $f(Z_i | A_i)$ is shown in Appendix Section~\ref{sec:appendix1}. 

\textbf{Likelihood.} Let $\theta $ denote the parameters to be estimated. The complete likelihood contribution of student $i$ with endogenous job offers and job choices, $(Z^{*}_{i}, C^{*}_{i})$, is given by
\begin{align}
\mathcal{L}_{i}(Z^{*}_{i}, C^{*}_{i} | A_{i}, \overline{X_{i}}, \theta) = & \int_{q} f(Z^{*}_{i}| A_{i}, \overline{X_{i}}, q, \theta) \times \text{Pr}(C^{*}_{i} = j | Z^{*}_{i}, \overline{X_{i}}, q, \theta) \hspace{0.05cm} dF(q| \theta),
\end{align} 
where $A_i$ is the application vector for student $i$ and $\overline{X_i}$ is the vector of all other exogenous characteristics entering the likelihood function of student $i$.

Let $\mathcal{L}^{r}_{i}(\theta )$ be the likelihood for individual $i$ in simulation $r$. Define
\begin{align}
\label{eq:hypotheticalsim}
\widehat{\mathcal{L}_{i}}(\theta ) = \frac{1}{R} \sum_{r = 1}^{R} \mathcal{L}^{r}_{i}(\theta ),
\end{align}
where $R$ is the total number of simulation draws. The MSL estimator is then defined by
\begin{align}
\widehat{\theta }_{MSL} = \arg \max_{\theta }\frac{1}{N} \sum_{i = 1}^{N} \text{log} \hspace{0.1cm} \widehat{\mathcal{L}_{i}}(\theta ) = \arg \max_{\theta } \bigg( \frac{1}{N} \sum_{i = 1}^{N} \text{log} \hspace{0.1cm} \bigg[\frac{1}{R} \sum_{r = 1}^{R} \mathcal{L}^{r}_{i}(\theta )\bigg] \bigg).
\end{align}
If $R$ rises at any rate with $N$, the MSL estimator is consistent \citep{train2003}. 

I calculate standard errors using the information identity. By the information identity, the sample hessian, $\widehat{H}$, can be computed by the average outer product of the gradient of simulated likelihood evaluated at $\widehat{\theta }_{MSL}$, i.e.

\begin{align}
\label{eq:hessian} 
\widehat{H} = \frac{1}{N}\sum_{i = 1}^{N} \nabla_{\theta} \text{log} \hspace{0.1cm} \widehat{\mathcal{L}_{i}}(\widehat{\theta }_{MSL}) \nabla_{\theta} \text{log} \hspace{0.1cm} \widehat{\mathcal{L}_{i}}(\widehat{\theta }_{MSL} )'.
\end{align}
Then, $\widehat{H}^{-1}$ is a consistent estimate of the variance of $\sqrt{N}(\widehat{\theta }_{MSL} - \theta^*)$, where $\theta^*$ is the vector of true parameter values.

\section{Modeling Choices and Their Implications (Continued)}
\label{sec:additional_channels_appendix}

In this section, I discuss some other modeling choices and argue that they do not necessarily limit the scope of my analysis for this population. These include:

\begin{enumerate}

    \item \textbf{Firm entry.} The cost of doing business in India is substantial. The World Bank's ``Ease of Doing Business" rankings lists India low, alongside Mexico, Colombia, and Jamaica. Moreover, starting a new business in India takes about 5 times the duration as it does in the U.S. \citep{worldbank2019}. Given these factors, it is unlikely that elite firm entry would meaningfully respond to compensatory policies, such as hiring subsidies, in the short- to medium-term.
    
    \item  \textbf{Student preference measures kept fixed in counterfactuals.} It is possible that compensatory policies for disadvantaged castes may change the willingness of advantaged caste students to accept elite jobs. While modeling these dynamics would be interesting, my model does not capture such ``full equilibrium" effects.

     \item \textbf{Not modeling multiple firm screening stages and job applications.} The choice of modeling firm screening as a one-stage process is guided by the decomposition of the earnings gap. This choice does not necessarily restrict my model's generalizability. The extension of my model of firm screening can be done in a manner similar to the basic model of labor demand laid out in Section~\ref{sec:joboffers}, with the written test, group debate, and interview stage each having its own cutoff. Modeling application behavior of elite Indian college students is not necessarily economically interesting, as streamlined and centralized application systems effectively make students apply for all eligible jobs (Section~\ref{sec:key_unmodeled}). 
     
     \item  \textbf{Not considering either wages, performance, hiring by firms of workers from other universities or job changing.} The job placement process studied in this paper offers a representative window into how elite college graduates transition into elite entry-level jobs, firms offer about the same job-specific wage to students from other universities across all locations in India, and focusing on just initial placements in elite jobs is important (see point 1 above and Section~\ref{sec:key_unmodeled}). Moreover, under the plausible assumption that advantaged castes benefit more from ``job changing" (e.g., by procuring other job offers once the current job starts and using them as leverage), my paper likely underestimates caste disparities in initial placements.\footnote{Participants in the placement processes of elite Indian colleges are barred from offline job search (Section \ref{sec:alternative explanations}).}

    \item \textbf{Same random effect, $q$, entering both student and firm utility.} See the discussion under ``The above specification of the random effect is not necessarily a limitation" in Section~\ref{sec:joboffers}.
     
    \item \textbf{Omitting $q$ from student utility does not affect the main conclusions in the paper.} The main result on the student side---that there are no average caste differences in preferences over non-pecuniary amenities---holds with or without the inclusion of a random effect in student utility (Section~\ref{sec:student preferences over student characteristics}). In addition, the magnitudes of the coefficients on employer utility (crucial for counterfactuals) do not critically depend upon $q$ entering student utility.
      
    \item \textbf{Equity-efficiency tradeoff.} My counterfactual analysis does not directly take this tradeoff into account. My estimates show that the model-implied subsidy equivalent to elite firms of the pre-college intervention policy is about 0.6\% of average annual salary, which is only about 10\% of the caste penalty. Therefore, the efficiency gains from pre-college interventions (omitted from my cost-effectiveness analysis) are likely small. Moreover, leaning toward equity through interventions in later stages, like hiring subsidies, might not necessarily sacrifice efficiency. Recent work has shown that disadvantaged groups are likely to benefit the most from selective education or job opportunities, whereas displaced advantaged groups are likely to be not much worse off \citep{blackdemmingrothstein2020}. Similar legal arguments have been made recently in the U.S. in favor of redistributive policies in later stages, especially in college admissions \citep{fisherutexas2016}.

     \item \textbf{Legally challenging subjective interviews.} My analysis does not consider such counterfactuals. Legally challenging practices such as personal interviews is likely untenable, as employers typically have free rein to value non-group characteristics, like candidate background \citep{lang2020}. Legal challenges are further complicated by the lack of a federal ombudsman to oversee private-sector hiring practices. Moreover, while explicit caste-based discrimination is illegal, the Indian legal system does not recognize or enforce ``disparate impact." Neither is there a systematic legal provision (anywhere) to penalize employers for judging ``cultural fit" based on myriad characteristics correlated with protected status \citep{jodhka2017}. Moreover, recent research has also shown the loss in screening precision due to the removal of subjective screening practices may outweigh equity gains \citep{mocanu2022}. 
     
\end{enumerate}

\section{Tables and Figures}

\begin{table}[!htbp]
\centering 
\small 
\caption{Foreign-Based versus India-Based Firms in the Data}
\label{tab:foreign_based_versus_india_based}
  \begin{threeparttable}
  \begin{tabular}{@{\extracolsep{0pt}} c c c c c c} 
        \\[-1.8ex]
        \hline 
        \hline
        \\[-1.8ex] 
        
        Designation & Count && Proportion &&\\  
        \hline 
        
        \\[-1.8ex]
        \\[-1.8ex]
        
        Foreign-Based Firms & 622 && 96.58 &&\\
        \\[-1.8ex] 
        \\[-1.8ex] 
        
        India-Based Firms & 22 && 3.42  &&\\
        \\[-1.8ex] 
        \\[-1.8ex] 
        
        Total Firms & 644 && 1 &&\\
        \\[-1.8ex] 

        \hline 
        \hline 
        \\[-1.8ex]

        \end{tabular}
          \begin{tablenotes}
          \begin{footnotesize}
            \item Notes: Online Appendix Table~\ref{tab:foreign_based_versus_india_based} shows the proportion of foreign-based MNCs versus India-based firms in the data that spans from e.g., 2012-15 (exact years omitted to preserve anonymity).
          \end{footnotesize}
          \end{tablenotes}
\end{threeparttable}
\end{table}

\vspace{0.7cm}

\begin{table}[p]
\begin{adjustwidth}{-0.4cm}{}
  \centering
  \scriptsize
      \caption{Proportion of Offices and Entry-Level Employees Outside India for Foreign-Based MNCs}
    \label{tab:proportion_offices_outside_india}

  \begin{threeparttable}

                \begin{tabular}{@{\extracolsep{0pt}} c c c c}
       \hline 
       \hline 

        (1) & (2) & (3) & (4) \\

       Firm Name & Headquarters & Offices Outside India (\%) & Entry-Level Employees Outside India (\%) \\ 
       
       \hline 

        Citicorp  & New York, USA &  99.62 & 98.62 \\
         eBay Inc. & San Jose, USA &  97.67 & 98.22 \\
        Rolls Royce  & London, UK &  97.61 & 96.36 \\
        LinkedIn  & Sunnyvale, USA &  94.66 & 96.73 \\
        Google  & Mountain View, USA & 94.28 & 97.50 \\
        Intel  & Santa Clara, USA & 97.11 & 96.48  \\
        Amazon  &  Seattle, USA &  97.45 & 96.13 \\
        Boston Consulting Group & Boston, USA  &  96.27 & 96.23 \\
        Cisco & San Jose, USA &  96.59 & 84.91 \\  
        Schlumberger & Houston, USA &  98.60 & 98.91 \\ 
        NetApp & Sunnyvale, USA &  96.77 & 83.33  \\
        Citrix & Fort Lauderdale, USA &  94.11 & 95.85 \\
        Ronald Berger  & Munich, Germany &  96.15 & 96.83\\
        Applied Materials & Santa Clara, USA &  96.91 & 96.30 \\
        Epic  & Verona, USA &  96.29 & 98.23  \\
        General Electric  & Boston, USA &  97.92 & 94.73 \\
        Analog Devices Pvt. Ltd.  & Norwood, USA &  96.67 & 98.75 \\
        ARM Embedded Technologies  & Cambridge, UK &  92.59 & 97.20  \\
        Microsoft  & Redmond, USA & 98.17 & 95.58   \\
        VISA  & San Francisco, USA & 97.72 & 96.35   \\
        Texas Instruments  & Dallas, USA & 96.71 & 96.56  \\
        Samsung  & Suwon-si, South Korea & 98.69 & 96.34  \\
        J.P. Morgan \& Chase  & New York, USA & 95.59 & 97.42  \\
        Capital One  & McLean, USA & 95.12 & 98.08  \\
         ASEA Brown Boveri (ABB)  & Zurich, Switzerland & 95.60 & 97.14 \\
        Caterpillar  & Deerfield, USA & 97.95 & 94.02  \\
        \hline 
        \hline \\[-1.8ex]
        \end{tabular}
          \begin{tablenotes}
          \begin{scriptsize}
            \item[] Notes: Online Appendix Table~\ref{tab:proportion_offices_outside_india} reports the proportions of offices of foreign-based MNCs that are outside of India for a select sample of firms. In addition, the table also reports the shares of firm-specific entry-level employees that are hired from India.  Column (1) includes the firm name, column (2) denotes the location of the firm headquarters, column (3) reports the proportions of offices of foreign-based MNCs that are outside of India, and column (4) shows the shares of firm-specific entry-level employees that are hired from India. The average of the proportions reported in column (3) is 96.65\%.
            The average of the proportions reported in column (4) is 95.88\%. The numbers reported in columns (3) and (4) are also similar in magnitude across all jobs in the sample, although only a select number of jobs are shown in this table.
          \end{scriptsize}
          \end{tablenotes}
\end{threeparttable}
\end{adjustwidth}
\end{table}

\pagebreak

\begin{table}[bp]
\centering 
\footnotesize
\caption{Distribution of Students by Caste for Each College Degree}
\label{tab:descriptive statistics students appendix}
  \begin{threeparttable}
\begin{tabular}{@{\extracolsep{0pt}} c c c c c c} 
        \\[-1.8ex]
        \hline 
        \hline
        \\[-1.8ex] 
        
        Degree & Adv. Caste & Disadv. Caste & Total \\  
        \hline \\[-1.8ex]
        
        Bachelor of Technology & 579 &  710 &  1289 && \\
        Dual Degree & 622 &  617 &  1239 && \\
        Master of Technology & 616 &  586  & 1202 && \\
        Master of Science & 350 & 127 & 477 && \\

        \hline \\[-1.8ex] 
        
        $N$  & 2167 & 2040 & 4207 \\
        Fraction  & 0.51 & 0.49 & 1 \\
        
        \hline 
        \hline 
        \\[-1.8ex]

        \end{tabular}
          \begin{tablenotes}
          \begin{footnotesize}
            \item Notes: Online Appendix Table  \ref{tab:descriptive statistics students appendix} includes the total number of students belonging to each caste for each college degree. The college degrees included are Bachelor of Technology (B.Tech.), Dual Degree (a five-year integrated bachelor's and master's degree), Master of Technology (M.Tech.) and Master of Science (M.S.). Adv. Caste stands for advantaged caste and Disadv. Caste stands for disadvantaged caste.
           \end{footnotesize}
          \end{tablenotes}
\end{threeparttable}
\end{table}

\begin{table}[!htbp]
\begin{adjustwidth}{-0.5cm}{}
    \centering
    \footnotesize
    \caption{Differences in Pre-College Skills across Castes}
     \label{tab:differences pre-college skills across castes}

  \begin{threeparttable}
    \begin{tabular}{@{\extracolsep{0pt}} c c c c c c c c} 
\\[-1.8ex]
\hline 

\hline \\[-1.8ex]

       & & \underline{B.Tech. Degree}  & & &   &  &  \\

       \\[-1.8ex]

       & & Adv. Caste  & & & Disadv. Caste  &  & Difference (S.D.) \\
       
       \hline \\[-1.8ex]

        Avg. entrance exam score  & \phantom{Salary (log)} & \phantom{-}0.41 & & & -0.37 & & \bf{$\phantom{-}0.78^{***}$} \\
        Avg. 10th grade score & \phantom{Salary (log)} & \phantom{-}0.07 & & & -0.06 & & $\phantom{-}0.13\phantom{^{***}}$ \\
        Avg. 12th grade score & \phantom{Salary (log)} & \phantom{-}0.04 & & & -0.03 & & $\phantom{-}0.07\phantom{^{***}}$ \\
        
        \hline \\[-1.8ex]
        
        & & \underline{Dual Degree}  & & &   &  &  \\

       \\[-1.8ex]

       & & Adv. Caste & & & Disadv. Caste &  & Difference (S.D.) \\
       
       \hline \\[-1.8ex]

        Avg. entrance exam score & \phantom{Salary (log)} & \phantom{-}0.34 & & & -0.38 & & \bf{$\phantom{-}0.72^{***}$} \\
        Avg. 10th grade score & \phantom{Salary (log)} & \phantom{-}0.03 & & & -0.03 & & $\phantom{-}0.06\phantom{^{***}}$ \\
        Avg. 12th grade score & \phantom{Salary (log)} & -0.03 & & & \phantom{-}0.03 & & $-0.06\phantom{^{***}}$ \\

        \hline \\[-1.8ex]
        
     & & \underline{M.Tech. Degree}  & & &   &  &  \\

       \\[-1.8ex]

       & & Adv. Caste & & & Disadv. Caste &  & Difference (S.D.) \\
       
       \hline \\[-1.8ex]

        Avg. entrance exam score  & \phantom{Salary (log)} & \phantom{-}0.26 & & & -0.28 & & \bf{$\phantom{-}0.54^{***}$} \\
        Avg. 10th grade score & \phantom{Salary (log)} & \phantom{-}0.04 & & & -0.04 & & $\phantom{-}0.08\phantom{^{***}}$ \\
        Avg. 12th grade score & \phantom{Salary (log)} & \phantom{-}0.02 & & & -0.02 & & $\phantom{-}0.04\phantom{^{***}}$ \\

        \hline \\[-1.8ex]
        
      & & \underline{M.S. Degree}  & & &   &  &  \\

       \\[-1.8ex]

       & & Adv. Caste & & & Disadv. Caste &  & Difference (S.D.) \\
       
       \hline \\[-1.8ex]

        Avg. entrance exam score  & \phantom{Salary (log)} & -0.02 & & & \phantom{-}0.07 & & \bf{$-0.09\phantom{^{***}}$} \\
        Avg. 10th grade score & \phantom{Salary (log)} & \phantom{-}\phantom{-}0.001 & & & \phantom{-}-0.001 & & $\phantom{-}0.002\phantom{^{**}}$ \\
        Avg. 12th grade score & \phantom{Salary (log)} & \phantom{-}0.01 & & & -0.02 & & $\phantom{-}0.03\phantom{^{***}}$  \\

            \hline 
            \hline \\[-1.8ex]
       \end{tabular}
          \begin{tablenotes}
            \begin{footnotesize}
            \item Notes: Online Appendix Table  \ref{tab:differences pre-college skills across castes} documents differences in pre-college skills across castes. Pre-college skills include scores on 10th and 12th grade national level examinations, and college entrance exam scores. All scores are pooled and normalized to have zero mean and unit standard deviation. College entrance exam scores are originally ranks, which have been renormalized so that higher numbers are better. The difference across castes is reported in standard deviation units. Adv. Caste stands for advantaged caste and Disadv. Caste stands for disadvantaged caste. *$p<0.1$; **$p<0.05$; ***$p<0.01$.
            \end{footnotesize}
          \end{tablenotes}
\end{threeparttable}
\end{adjustwidth}
\end{table}

\pagebreak

\begin{table}[bp]
\begin{adjustwidth}{-1.0cm}{}
\footnotesize
    \centering
    \caption{Differences in Average Overall GPA (Not Adjusted for Major) across Castes}
     \label{tab:differences GPA across castes}

  \begin{threeparttable}
    \begin{tabular}{@{\extracolsep{0pt}} c c c c} 
    \\[-1.8ex]

             \hline 
             \hline \\[-1.8ex]

        & & \hspace{-2.5cm}\underline{B.Tech. Degree}  & \\ 

        \\[-1.8ex]

       & Adv. Caste  & Disadv. Caste  &  Difference (S.D.) \\
       
       \hline \\[-1.8ex]

        Avg. Overall GPA & 0.51 &  -0.42 &  \bf{$0.93^{***}$} \\

        \hline \\[-1.8ex]

       & & \hspace{-2.5cm}\underline{Dual Degree}  & \\

        \\[-1.8ex]

       & Adv. Caste & Disadv. Caste &  Difference (S.D.) \\
       
       \hline \\[-1.8ex]

        Avg. Overall GPA & 0.43 & -0.43 &  \bf{$0.86^{***}$} \\

        \hline \\[-1.8ex]

                       & & \hspace{-2.5cm}\underline{M.Tech. Degree}  & \\

        \\[-1.8ex]

       & Adv. Caste & Disadv. Caste & Difference (S.D.) \\
       
       \hline \\[-1.8ex]

       Avg. Overall GPA & 0.33 & -0.35 & \bf{$0.68^{***}$} \\
       
           \hline \\[-1.8ex]

        & & \hspace{-2.5cm}\underline{M.S. Degree}  & \\ 

        \\ [-1.8ex]

       & Adv. Caste & Disadv. Caste & Difference (S.D.) \\
       
       \hline \\[-1.8ex]

        Avg. Overall GPA & 0.05 & -0.13 &  \bf{$0.18^{**}$} \\

            \hline 
            \hline \\[-1.8ex]
       \end{tabular}
          \begin{tablenotes}
            \begin{footnotesize}
            \item Notes: Online Appendix Table  \ref{tab:differences GPA across castes} documents differences in average college GPA (not adjusted for major) across castes. All scores are pooled and normalized to have zero mean and unit standard deviation. Adv. Caste stands for advantaged caste and Disadv. Caste stands for disadvantaged caste. *$p<0.1$; **$p<0.05$; ***$p<0.01$.
            \end{footnotesize}
          \end{tablenotes}
\end{threeparttable}
\end{adjustwidth}
\end{table}

\begin{table}[!htbp]
\scriptsize
\begin{adjustwidth}{-0.5cm}{}
    \centering
    \caption{Differences in Previous Labor Market Experience across Castes}
     \label{tab:differences pre-college skills across castes previous labor}
  \begin{threeparttable}
    \begin{tabular}{@{\extracolsep{0pt}} c c c c}  \\[-1.8ex]

        \hline 
        \hline \\[-1.8ex]

       & \underline{B.Tech. and Dual Degrees} &  &   \\

       & Adv. Caste  & Disadv. Caste  &  Difference  \\
       
       \hline \\[-1.8ex]

        Avg. Internship Duration (Weeks) & 8.00 (0.06) &  7.81 (0.07) & $\phantom{-}0.19^{**}$ \\
        Fraction Worked in the IT Sector &  0.22 (0.05) & 0.22 (0.04) & $\phantom{-}0.00\phantom{^{**}}$ \\
        Fraction Worked in the Consulting Sector &  0.35 (0.05) & 0.37 (0.05) & $-0.02\phantom{^{**}}$ \\
        Fraction Worked in the Manufacturing Sector &  0.43 (0.05) & 0.41 (0.06) &  $\phantom{-}0.02\phantom{^{**}}$ \\
        Fraction Worked at a Startup &  0.34 (0.05) & 0.30 (0.05) & $\phantom{-}0.04\phantom{^{**}}$ \\
        Total Internship Pay (\$) & \phantom{-}3042.24 (249.40) & 2877.28 (220.89) &  $\phantom{-}164.96\phantom{^{**}}$ \\
        
        \hline \\[-1.8ex]
        
      & \underline{M.Tech. and M.S. Degrees} &  &   \\

               & Adv. Caste  & Disadv. Caste  &  Difference  \\
       
       \hline \\[-1.8ex]

       Avg. Part-Time/Full-Time Employment Duration (Weeks) & 68.48 (4.52) &  \phantom{-}68.93 (6.96) &  $-0.45\phantom{^{***}}$ \\
        Fraction Worked in the IT Sector &  0.36 (0.04) & 0.18 (0.07) & $\phantom{-}0.18^{***}$ \\
        Fraction Worked in the Consulting Sector & 0.19 (0.04) & 0.15 (0.06) &  $\phantom{-}0.04\phantom{^{***}}$ \\
        Fraction Worked in the Manufacturing Sector & 0.45 (0.05) & 0.67 (0.08) &  $-0.12^{***}$ \\
        Total Part-Time/Full-Time Employment Pay (\$) &  22523.80 (1458.03) & 19645.89 (1390.32) &  $2877.91$ \\

            \hline
            \hline \\[-1.8ex]
       \end{tabular}
          \begin{tablenotes}
            \begin{scriptsize}
            \item Notes: Online Appendix Table~\ref{tab:differences pre-college skills across castes previous labor} documents differences in previous labor market experience across castes. Previous labor market experience includes internship duration (weeks), part-time or full-time employment duration (weeks), total pay during internships, total pay during part-time or full-time employment, sectors of employment and employment in startups. Standard errors are reported in parentheses. All dollar amounts are in purchasing power parity units. T-tests are conducted for differences in overall means. *$p<0.1$; **$p<0.05$; ***$p<0.01$.
            \end{scriptsize}
          \end{tablenotes}
\end{threeparttable}
\end{adjustwidth}
\end{table}

\clearpage

\begin{table}[!htbp]
\centering 
\small
\caption{Firm Transition Matrix}
\label{tab:firm_transition_matrix}
  \begin{threeparttable}
\begin{tabular}{@{\extracolsep{0pt}} c c c c c c c c} 
        \\[-1.8ex]
        \hline 
        \hline
        \\[-1.8ex] 
        
        From/To & Elite Private-Sector && Other & &&&\\  
        \hline 
        
        \\[-1.8ex]
        \\[-1.8ex]
        
        Elite Private-Sector & 97.88 && 2.12 & &&&\\
        \\[-1.8ex] 
        \\[-1.8ex] 
        
        Other & 1.21 &&   98.79 & &&&\\
        \\[-1.8ex] 

        \hline 
        \hline 
        \\[-1.8ex]

        \end{tabular}
          \begin{tablenotes}
          \begin{footnotesize}
            \item Notes: Online Appendix Table~\ref{tab:firm_transition_matrix} shows the probability of transitioning from elite private-sector jobs to ``other" jobs. ``Other" jobs include elite public-sector jobs, other private-sector jobs, unemployment, and so on. The dataset is constructed from separate samples of the India Human Development Survey (IHDS) and the Periodic Labor Force Survey (PLFS), which is collected by the National Sample Survey Office (NSSO). Note that the NSSO defines unemployment as a situation in which all those who owing to lack of work are not working, but seek work through employment exchanges, intermediaries, friends or relatives \citep{nssconcepts}. Therefore, being unemployed in the data collected by the NSSO is closer to being actually out of work (e.g., it does not include self-employment). On the other hand, students who are ``unemployed" through job the placement process I study could include those who are self-employed or taking gap years. The sample construction from the PLFS follows \citet{jyotirmoy2021}. The same construction from the IHDS follows \citet{sarkar2017}.
        \end{footnotesize}
          \end{tablenotes}
\end{threeparttable}
\end{table}

\vspace{1cm}

\begin{table}[!htbp] 
\centering 
\small
\caption{Total Number of Firms and Average Salary by Sector}
\label{tab:descriptive statistics firms main}

  \begin{threeparttable}
        \begin{tabular}{@{\extracolsep{0pt}} c c c} 
        \\[-1.8ex]
        \hline 
        \hline 
        \\[-1.8ex]
        
        \\[-1.8ex]
        
         & (1) & (2)  \\ 
       
       \\[-1.8ex]
        
       Sector & Total (Fraction) & Avg. Salary (\$)
        \\
        \hline \\[-1.8ex] 
        Technology & 335 (0.52) &  67302.64 \\
        Consulting & 129 (0.20) &  63544.02 \\
        Manufacturing & 180 (0.28) &  43525.25 \\
        \hline 
        \hline \\[-1.8ex]
        \end{tabular}
          \begin{tablenotes}
            \begin{footnotesize}
            \item[] Notes: Online Appendix Table  \ref{tab:descriptive statistics firms main} shows the distribution of firms by sector and the average salary across all jobs by sector. Column (1) shows the number of firms in each sector with their proportions in parentheses. Column (2) shows the average salary of all jobs in a given sector. All dollar amounts are in purchasing power parity (PPP) terms. 
            \end{footnotesize}
          \end{tablenotes}
\end{threeparttable}
\end{table}

\clearpage

\newgeometry{top=0.5in, bottom=0.5in, left=0.5in, right=0.5in}
\begin{landscape}
\thispagestyle{empty}
\begin{table}[!htbp]
\begin{adjustwidth}{-0.2cm}{}
    \centering
    \caption{Non-Pecuniary Amenities}
     \label{tab:non pecuniary amenities full 1}

  \begin{threeparttable}
        \begin{tabular}{@{\extracolsep{0pt}} c c c}

        \hline
        \hline 
        \\[-1.8ex]

       \underline{Row Number} & \underline{Non-Pecuniary Amenity} & \underline{Additional Details} \gdef\rownumber{\stepcounter{magicrownumbers}\arabic{magicrownumbers}} \\

       \\[-1.8ex]

       1. & Variable annual pay? &  \\
       2. & Is the variable compensation taxable? &  \\
       3. & Restricted stock units? &  \\
       4. & Paid leave? & Non-personal, non-educational purposes leave \\
       5. & Sickness or disability leave? &  \\
       6. & Signing bonus? &  \\
       7. & Bonus for spending 1 year at the firm? &  \\
        8. & Bonus for spending 2 years at the firm? & \\
        9. & Bonus for spending 3 years at the firm? & \\
        10. & Bonus for spending 4 years at the firm? & \\
       11. & Annual bonus? &  \\
       12. & Variable bonus? &  In addition to fixed bonus\\
       13. & Performance bonus? &  Could be project specific\\
       14. & Stakeholder bonus? & \\
      15. & Festival bonus? & \\ 
      16. & Loyalty bonus? & Might vary by job tenure\\
      17. & ELRP bonus? & Also called deferred compensation\\
      18. & Probation completion bonus?  & \\
      19. & Relocation bonus? & \\
      20. & Relocation assistance? & Arranging moving company\\
      21. & Employees' provident fund (EPF)? & Similar to a 401k benefit \\
      22. & Voluntary provident fund? & Voluntary employee contribution over and above EPF\\ 
      23. & Medical insurance? &  \\ 
      24. & Dental insurance? & \\
      25. & Eye insurance? & \\
      26. & Life insurance? & \\
      27. & Food allowance? & \\
      28. & Temporary accommodation? & \\
      29. & Stipend during temporary accommodation? & \\
      30. & Travel allowance? & Air, rail and road travel\\
      31. & Leave travel concession (LTC)? & Non-work-related travel \\ 
            \hline 
            \hline 
            \\[-1.8ex]
       \end{tabular}
          \begin{tablenotes}
            \begin{footnotesize}
            \item Notes: Online Appendix Table  \ref{tab:non pecuniary amenities full 1} shows the complete list of unique non-pecuniary amenities offered by each job (job designation within a firm) along with an added description of the perks, unless self-explanatory.
            \end{footnotesize}
          \end{tablenotes}
\end{threeparttable}
\end{adjustwidth}
\end{table}
\end{landscape}
\restoregeometry

\newgeometry{top=0.5in, bottom=0.5in, left=0.5in, right=0.5in}
\begin{landscape}
\thispagestyle{empty}
\begin{table}[!htbp]
\begin{adjustwidth}{-0.2cm}{}
    \centering
    \caption*{\textbf{Table OA.9:} Non-Pecuniary Amenities (Continued)}
  \begin{threeparttable}
        \begin{tabular}{@{\extracolsep{0pt}} c c c}

        \hline
        \hline 
        \\[-1.8ex]

       \underline{Row Number} & \underline{Non-Pecuniary Amenity} & \underline{Additional Details} \gdef\rownumber{\stepcounter{magicrownumbers}\arabic{magicrownumbers}} \\

       \\[-1.8ex]
       
       32. & House rent allowance (HRA)? & \\
               33. & Telephone/mobile phone allowance? & \\
        34. & Conveyance allowance? & Covers travel between work and residence\\
              35. & Night shift allowance? & \\
       36. & Counseling services? &  \\
       37. & Option to work from home? &  \\
       38. & Paid maternity Leave? &  \\
       39. & Sodexo Coupons? & Tax-free vouchers for restaurants, grocery stores, etc.  \\ 
       40. & Flexible working hours? &  \\ 
       41. & Paid day care for kids? &  \\
       42. & Happy fridays? &  \\
       43. & Gym subsidies? & \\
       44. & Lunch on company campus? \\
        45. & Child psychology services?  & \\ 
        46. & Personal development classes?  & Yoga, cooking, dancing, etc. \\
              47. & Family days? & \\
                    48. & Smoking zones? & \\
    49. & Telemedicine? &  \\
      50. & Parental day care? &  \\ 
       51. & Financial literacy classes? &   \\
       52. & Employee assistance program? &  \\
       53. & Subsidized personal leave? & Usually up to 6 months\\ 
       54. & Subsidized educational leave?  & \\
        55. & Subsidized high-school education for kids? & \\
        56. & Subsidized housing? & \\
        57. & Gratuity? &  Lump sum payment after 4 years and 8 months of service\\
      58. & Leave encashments? & Unused paid leave reimbursed as part of salary\\
      59. & Option to return after sabbatical? & \\
            \hline 
            \hline 
            \\[-1.8ex]
       \end{tabular}
          \begin{tablenotes}
            \begin{footnotesize}
            \item Notes: Online Appendix Table \ref{tab:non pecuniary amenities full 1} shows the complete list of unique non-pecuniary amenities offered by each job (job designation within a firm) along with an added description of the perks, unless self-explanatory. 
            \end{footnotesize}
          \end{tablenotes}
\end{threeparttable}
\end{adjustwidth}
\end{table}
\end{landscape}
\restoregeometry

\clearpage

\thispagestyle{empty}
\begin{table}[p]
  \centering
  \footnotesize
        \caption{Earnings Gap}
    \label{tab:earnings gap main}
  \begin{threeparttable}
    \begin{tabular}{@{\extracolsep{0pt}} c c c c c}
           \hline 
        \hline \\[-1.8ex]
       
               \multicolumn{4}{c}{\hspace{2.3cm}{ \underline{Baseline Log Earnings (USD PPP))}}} \\
                      \\[-1.8ex]
        \multicolumn{4}{c}{\hspace{2.3cm}{ \underline{\textbf{Baseline Specification}}}} \\
        
                 \\[-1.8ex]
        
        \vphantom{Coefficient} & (1) & (2) & (3) & (4)  \\ 
       
       \\[-1.8ex]

       Coefficient & Linear & Quadratic & Cubic & Splines \\ 
       
        \hline 
         \\[-1.8ex]
        
        \text{Disadv. Caste} & $-0.113^{***}$ (0.014) & $-0.105^{***}$ (0.017) & $-0.104^{***}$ (0.019) & $-0.104^{***}$ (0.024) \\
         
        \hline 
        $N$  & 2927 & 2927 & 2927 & 2927 \\
        $R^{2}$ & 0.452 & 0.532 & 0.553 & 0.578 \\
        Adjusted \hspace{0.05cm}$R^{2}$ & 0.447 & 0.486 & 0.490 & 0.497 \\

        \\[-1.8ex]
\\[-1.8ex]
       
       & \multicolumn{4}{c}{\hspace{-2.9cm}{ \underline{\textbf{Manufacturing Sector}}}} \\
        
                         \\[-1.8ex]

       \\[-1.8ex]
       
             Coefficient & Linear & Quadratic & Cubic & Splines \\ 
       
        \hline 
         \\[-1.8ex]

        \text{Disadv. Caste} & $-0.084^{***}$ (0.022) & $-0.070^{**}$ (0.027) & $-0.091^{***}$ (0.032) & $-0.091^{***}$ (0.032) \\

        \hline 

        $N$  & 789 & 789 & 789 & 789 \\

        $R^{2}$ & 0.344 & 0.547 & 0.604 &  0.619 \\
        Adjusted \hspace{0.05cm}$R^{2}$ &  0.323  & 0.402 & 0.408 & 0.431 \\

                         \\[-1.8ex]

       \\[-1.8ex]
        
        & \multicolumn{4}{c}{\hspace{-2.9cm}{ \underline{\textbf{Technology Sector}}}} \\
        
                         \\[-1.8ex]

       \\[-1.8ex]

       Coefficient & Linear & Quadratic & Cubic & Splines  \\ 
       
       \hline 
        
        \text{Disadv. Caste} & $-0.080^{***}$ (0.022) & $-0.077^{***}$ (0.028) & $-0.061^{*}$ (0.033) & $-0.071^{**}$ (0.033) \\

        \hline 

        $N$  & 1435 & 1435 & 1435 & 1435 \\
        $R^{2}$ & 0.418 & 0.535 & 0.574 & 0.575 \\
        Adjusted \hspace{0.05cm}$R^{2}$ &  0.406  & 0.438 & 0.443 & 0.446 \\
        
\\[-1.8ex]
\\[-1.8ex]
        
& \multicolumn{4}{c}{\hspace{-2.9cm}{ \underline{\textbf{Consulting Sector }}}} \\
        
                         \\[-1.8ex]

       \\[-1.8ex]

       Coefficient & Linear & Quadratic & Cubic & Splines  \\ 
       
       \hline

        \text{Disadv. Caste} & $-0.119^{***}$ (0.032) & $-0.104^{**}$ (0.041) & $-0.087^{*}$ (0.048) & $-0.109^{**}$ (0.054) \\
         
        \hline 
        $N$  & 703 & 703 & 703 & 703 \\

        $R^{2}$ & 0.494 & 0.636 & 0.688 & 0.689 \\
        Adjusted \hspace{0.05cm}$R^{2}$ & 0.473 & 0.502 & 0.528 & 0.528 \\

                            \\[-1.8ex]

       \\[-1.8ex]
       
        & \multicolumn{4}{c}{\hspace{-2.9cm}{ \underline{\textbf{Client-Facing Jobs}}}} \\
        
                         \\[-1.8ex]

       \\[-1.8ex]

       Coefficient & Linear & Quadratic & Cubic & Splines  \\ 
       
       \hline

        \text{Disadv. Caste} & $-0.117^{***}$ (0.029) & $-0.129^{***}$ (0.037) & $-0.115^{***}$ (0.044) & $-0.119^{***}$ (0.045) \\

        \hline 
        
        $N$  & 822 & 822 & 822 & 822 \\

        $R^{2}$ & 0.437 & 0.568 & 0.614 & 0.616 \\
        Adjusted \hspace{0.05cm}$R^{2}$ & 0.418 & 0.434 & 0.454 & 0.456 \\

      \\[-1.8ex]
       \\[-1.8ex]
        
        & \multicolumn{4}{c}{\hspace{-2.9cm}{ \underline{\textbf{Non-Client-Facing Jobs}}}} \\
        
      \\[-1.8ex]
       \\[-1.8ex]

       Coefficient & Linear & Quadratic & Cubic & Splines  \\ 
       
       \hline 
        
        \text{Disadv. Caste} & $-0.080^{***}$ (0.016) & $-0.070^{***}$ (0.020) & $-0.074^{***}$ (0.022) & $-0.071^{***}$ (0.022) \\

        \hline
        $N$  & 2105 & 2105 & 2105 & 2105 \\

        $R^{2}$ & 0.499 & 0.581 & 0.609 & 0.610 \\
        Adjusted \hspace{0.05cm}$R^{2}$ & 0.492 & 0.522 & 0.528 & 0.529 \\

        \hline 
        \hline \\[-1.8ex]
        \end{tabular}
          \begin{tablenotes}
            \begin{scriptsize}
            \item[] Notes: Online Appendix Table  \ref{tab:earnings gap main} includes estimates from an earnings regression run on the sample of all students who graduated with jobs. The dependent variable is log earnings. I include detailed controls including measures of pre-college skills, within-college academic performance, previous labor market experience, and other employer-relevant skills (Section~\ref{sec:data overview}).  Each column is a separate regression and includes all the controls mentioned above. In column (1),  all controls enter linearly. In column (2), GPA and entrance exam scores enter as quadratic polynomials, while other controls enter linearly. In column (3), GPA and entrance exam scores enter as cubic polynomials, while other controls enter linearly. In column (4), estimates are reported from a fully flexible quadratic polynomial regression with all possible interactions between controls. PPP stands for purchasing power parity. *$p<0.1$; **$p<0.05$; ***$p<0.01$. 
            \end{scriptsize}
          \end{tablenotes}
\end{threeparttable}
\end{table}

\begin{table}[p]
 \begin{adjustwidth}{-0.5cm}{}
    \centering
    \footnotesize
    \caption{GPA and Entrance Exam Score Comparisons of All Students vs. Those Without Jobs}
     \label{tab:comparison gpa overall and without jobs}
  \begin{threeparttable}

      \begin{tabular}{@{\extracolsep{0pt}} c c c c c c c} 
      \\[-1.8ex]
       \hline 
       \hline 
       
                       & \multicolumn{6}{c}{\hspace{-3.6cm}{\textbf{\underline{GPA Comparisons}}}}\\

                        \\[-1.8ex]
                         \\[-1.8ex]

     & \multicolumn{6}{c}{\hspace{-3.6cm}{\underline{B.Tech. Degree}}}\\
       
                                \\[-1.8ex]
        
    (1) & & (2) & & (3)  & & (4)
       
       \\[-1.8ex]
            \\[-1.8ex]
         \\[-1.8ex]

      \phantom{Adv. Caste} & \textbf{Overall} & & &
      \phantom{Adv. Caste} & \textbf{Students without Jobs} & \\
      
     \hline

       Adv. Caste & \phantom{Overall} & Disadv. Caste & & Adv. Caste & \phantom{Students without Jobs} & Disadv. Caste \\

        8.08 & \phantom{Top 25\%}  & 7.00 & \phantom{Top 25\%} & 7.97 & \phantom{Top 25\%} & $6.58^{***}$  \\

        \\[-1.8ex]

                                & \multicolumn{6}{c}{\hspace{-3.6cm}{\underline{Dual Degree}}}\\

      \phantom{Adv. Caste} & Overall & & & \phantom{Adv. Caste} & Students without Jobs & \\
      
      \hline 
       
       Adv. Caste & \phantom{Overall} & Disadv. Caste & & Adv. Caste & \phantom{Students without Jobs} & Disadv. Caste \\

        8.05 & \phantom{Top 25\%}  & 7.15 & \phantom{Top 25\%} & 8.02 & \phantom{Top 25\%} & $6.86^{**}$  \\

        \\[-1.8ex]

                                    & \multicolumn{6}{c}{\hspace{-3.6cm}{\underline{M.Tech. Degree}}}\\
                
      \phantom{Adv. Caste} & Overall & & & \phantom{Adv. Caste} & Students without Jobs & \\
      
      \hline 
       
       Adv. Caste & \phantom{Overall} & Disadv. Caste & & Adv. Caste & \phantom{Students without Jobs} & Disadv. Caste \\

        8.33 & \phantom{Top 25\%}  & 7.62 & \phantom{Top 25\%} & $8.00^{***}$ & \phantom{Top 25\%} & $7.35^{***}$  \\
        
        \hline \\[-1.8ex]

                                    & \multicolumn{6}{c}{\hspace{-3.6cm}{\underline{M.S. Degree}}}\\

      \phantom{Adv. Caste} & Overall & & & \phantom{Adv. Caste} & Students Without Jobs & \\
      
      \hline 
       
       Adv. Caste & \phantom{Overall} & Disadv. Caste & & Adv. Caste & \phantom{Students without Jobs} & Disadv. Caste \\

        8.49 & \phantom{Top 25\%}  & 8.42 & \phantom{Top 25\%} & 8.46 & \phantom{Top 25\%} & $8.23^{*}$ \\

        \hline

         \\[-1.8ex]
                         \\[-1.8ex]
                         
                       & \multicolumn{6}{c}{\hspace{-3.6cm}{\textbf{\underline{Entrance Exam Score Comparisons}}}}\\

                        \\[-1.8ex]
                         \\[-1.8ex]
                         
                         & \multicolumn{6}{c}{\hspace{-3.6cm}{\underline{B.Tech. Degree}}}\\

      \phantom{Adv. Caste} & \textbf{Overall} & & & \phantom{Adv. Caste} & \textbf{Students without Jobs} & \\
      
      \hline 
       
       Adv. Caste & \phantom{Overall} & Disadv. Caste & & Adv. Caste & \phantom{Students without Jobs} & Disadv. Caste \\

        -1617.89 & \phantom{Top 25\%}  & -3707.45 & \phantom{Top 25\%} & $-1879.32^{*}$ & \phantom{Top 25\%} & $-4315.18^{**}$  \\

        \\[-1.8ex]
        
                & \multicolumn{6}{c}{\hspace{-3.6cm}{\underline{Dual Degree}}}\\

      \phantom{Adv. Caste} & Overall & & & \phantom{Adv. Caste} & Students without Jobs & \\
      
      \hline 
       
       Adv. Caste & \phantom{Overall} & Disadv. Caste & & Adv. Caste & \phantom{Students without Jobs} & Disadv. Caste \\

        -2096.60 & \phantom{Top 25\%}  & -4067.13 & \phantom{Top 25\%} & $-2602.79^{***}$ & \phantom{Top 25\%} & $-5743.80^{***}$  \\

        \\[-1.8ex]
        
        & \multicolumn{6}{c}{\hspace{-3.6cm}{\underline{M.Tech. Degree}}}\\

      \phantom{Adv. Caste} & Overall & & & \phantom{Adv. Caste} & Students without Jobs & \\
      
      \hline 
       
       Adv. Caste & \phantom{Overall} & Disadv. Caste & & Adv. Caste & \phantom{Students without Jobs} & Disadv. Caste \\

        -653.94 & \phantom{Top 25\%}  &  -2445.64 & \phantom{Top 25\%} & $-1052.61^{***}$ & \phantom{Top 25\%} & $-3310.677^{**}$ \\

        \\[-1.8ex]
 
        & \multicolumn{6}{c}{\hspace{-3.6cm}{\underline{M.S. Degree}}}\\

      \phantom{Adv. Caste} & Overall & & & \phantom{Adv. Caste} & Students without Jobs & \\
      
      \hline 
       
       Adv. Caste & \phantom{Overall} & Disadv. Caste & & Adv. Caste & \phantom{Students without Jobs} & Disadv. Caste \\

        -558.94 & \phantom{Top 25\%}  & -1416.09 & \phantom{Top 25\%} & -642.18 & \phantom{Top 25\%} & -1411.26  \\

            \hline 
            \hline \\[-1.8ex]
       \end{tabular}
          \begin{tablenotes}
            \begin{footnotesize}
            \item Notes: Online Appendix Table  \ref{tab:comparison gpa overall and without jobs} compares the average GPA and entrance exam scores of all students versus those of students without jobs. T-tests are conducted for differences in overall means versus means of students without jobs \emph{within} each caste. Significance denoted by asterisks are shown in the third and fourth columns. Adv. Caste stands for advantaged caste and Disadv. Caste stands for disadvantaged caste. *$p<0.1$; **$p<0.05$; ***$p<0.01$.
            \end{footnotesize}
          \end{tablenotes}
\end{threeparttable}
\end{adjustwidth}
\end{table}

\clearpage

\newgeometry{top=0.2in, bottom=0.2in, left=0.2in, right=0.2in}
\begin{landscape}
\thispagestyle{empty}
\begin{table}[bp]
\begin{adjustwidth}{-0.4cm}{}
  \centering
  \small
      \caption{Job-Specific Salary Comparisons between Salaries Reported on the Major U.S. Platform Versus those in my Sample}
    \label{tab:Salaries for Select Firms in the Same Location}

  \begin{threeparttable}

                \begin{tabular}{@{\extracolsep{0pt}} c c c c c}
       \hline 
       \hline 

        (1) & (2) & (3) & (4) & (5) \\

       Firm Name & Job Designation & Job Location & Headquartered in & $\Delta (\%)$  \\ 
       
       \hline 

        Citicorp  & Analyst & India & New York, USA & 0.07  \\
         eBay Inc. & Software Engineer  & India & San Jose, USA  & 1.29 \\
        Indian Register of Shipping  & Assistant Surveyor & India & Powai, India & 1.45 \\
        Rolls Royce  & Engineering Graduate & India & London, UK & 1.46 \\
        LinkedIn  & Software Engineer  & India & Sunnyvale, USA & 1.86 \\
        Google  & Software Engineer  & India & Mountain View, USA & 2.76 \\
        Intel  & Component Design Engineer  & India  & Santa Clara, USA & 2.44 \\
        Amazon  & Area Manager & India & Seattle, USA & 0.47 \\
        Boston Consulting Group & Associate & India & Boston, USA & 1.42 \\
        Cisco & Software Engineer & India & San Jose, USA & 1.80 \\ 
        Schlumberger & Software Engineer & India & Houston, USA & 3.08 \\
        NetApp & Member Technical Staff & India & Sunnyvale, USA & 1.48  \\
        Citrix & Software Engineer & India & Fort Lauderdale, USA & 2.66 \\
        Ronald Berger  & Business Analyst & India & Munich, Germany & 2.80\\
        Applied Materials  & Application Engineer & India & Santa Clara, USA & 0.85 \\
        Epic  & Software Developer & India & Verona, USA & 1.64  \\
        General Electric  & Edison Engineer & India  & Boston, USA & 2.33 \\
        Analog Devices Pvt. Ltd.  & Software Engineer & India & Norwood, USA & 1.88 \\
        ARM Embedded Technologies  & Graduate Engineer & India & Cambridge, UK & 4.01  \\
        Microsoft  & Software Engineer & India & Redmond, USA & 2.31  \\
        VISA  & Software Engineer & India  & San Francisco, USA & 1.74  \\
        Texas Instruments  & Analog Engineer & India  & Dallas, USA & 4.85 \\
        Samsung  & Software Engineer & India & Suwon-si, South Korea & 4.92 \\
        J.P. Morgan \& Chase  & Associate & India & New York, USA & 1.38 \\
        Capital One  & Associate & India & McLean, USA & 4.35 \\
         ASEA Brown Boveri (ABB)  & Software Engineer & India &  Zurich, Switzerland &  1.38\\
        Caterpillar  & Associate Engineer & India & Deerfield, USA & 2.88 \\
        \hline 
        \hline \\[-1.8ex]
        \end{tabular}
          \begin{tablenotes}
            \item[] \scriptsize Notes: Online Appendix Table~\ref{tab:Salaries for Select Firms in the Same Location} includes comparisons between job-specific salaries offered on reported on the major U.S. platform versus those in my sample. The table reports these comparisons for select firms in the sample. Column (1) includes the firm name, column (2) includes job designation, column (3) includes the job location, column (4) reports the location of the firm headquarters, and column (5) reports the absolute percentage difference between the salaries from the U.S. platform (in PPP)  versus those in my sample (in PPP). \\ The denominator is the average salary across all jobs in the sample (\$56,767.29 PPP). The average of $\Delta (\%)$ reported in column (5) is 2.21\%. This average difference is also similar in magnitude across all jobs in the sample, although only a select number of jobs are shown in Online Appendix Table~\ref{tab:Salaries for Select Firms in the Same Location}. The PPP conversion factor is taken from the \href{www.data.oecd.org/conversion/purchasing-power-parities-ppp.htm}{OECD website}. 
          \end{tablenotes}
\end{threeparttable}
\end{adjustwidth}
\end{table}
\end{landscape}

\clearpage

\begin{table}[p]
  \centering
  \large 
      \caption{Predicting Interview Days with Job Characteristics and ``Firm Identity"}
    \label{tab:hedonic regression firm characteristics only 1}

  \begin{threeparttable}

        \begin{tabular}{@{\extracolsep{0pt}} c c c c}
       \hline 
       \hline 
               & \multicolumn{3}{c}{\hspace{-2cm}{\underline{Dependent Variable: Assigned a Particular Interview Day}}} \\
                       \\[-1.8ex]
        \\[-1.8ex]
        & \multicolumn{3}{c}{\hspace{-2cm}{ \underline{\textbf{Job Characteristics Only}}}} \\

                                            \\[-1.8ex]
        
        \vphantom{Coefficient} & (1) & (2) & (3)  \\ 
       
       \\[-1.8ex]

       Coefficient & Logistic & Random Forest & Decision Tree \\ 
       
       \hline 

        Accuracy  & 0.734 & 0.759 & 0.721\\
        95\% CI & [0.690, 0.7745] & [0.716, 0.798] & [0.676, 0.762] \\
        Kappa& 0.304 & 0.366 & 0.356\\
        
        \\[-1.8ex]
        \\[-1.8ex]

& \multicolumn{3}{c}{\hspace{-2cm}{\underline{\textbf{Job Characteristics and ``Firm Identity"}}}} \\
        
            \\[-1.8ex]

       \\[-1.8ex]

       Coefficient & Logistic & Random Forest & Decision Tree \\ 
       
       \hline 

        Accuracy  & 0.948 & 0.951 & 0.952\\
        95\% CI & [0.923, 0.967] & [0.926, 0.969] & [0.929, 0.971] \\
        Kappa & 0.879 & 0.884 & 0.890 \\

        \hline
        \hline \\[-1.8ex]

        \end{tabular}
          \begin{tablenotes}
            \item[] \scriptsize Notes: Online Appendix Table  \ref{tab:hedonic regression firm characteristics only 1} includes measures of predictive accuracy of interview day assignments given job characteristics and measures of ``firm identity.'' ``Firm identity'' is proxied by previous interview day assignment of the same firm. The dependent variable is the interview day assigned to a firm. Controls include job salaries, job sectors, and job titles. In column (1), an ordered logistic model is estimated. In column (2), a random forest model is estimated. In column (3), a decision tree model is estimated. Accuracy is the total number of correct predictions divided by the total number of observations. The Kappa statistic, which lies between 0 and 1, measures how classification results compare to values assigned by chance. A higher Kappa statistic is better. Full regression results are available on request.  
          \end{tablenotes}
\end{threeparttable}
\end{table}

\clearpage

\vspace{1cm}

\begin{table}[p]
    \centering
    \caption{Model Fit: Job Offer, Job Choice, Unemployment, and Earnings Gap}
     \label{tab:model fit job choice and job offers main 1}

  \begin{threeparttable}
 \begin{tabular}{@{\extracolsep{0pt}} c c c c c c c}
       \hline 
       \hline 
        & \multicolumn{6}{c}{\hspace{-3cm}Model Fit} \\

       \hline \\[-1.8ex]
       
       \\[-1.8ex]
        
        \vphantom{Coefficient} & (1) & &  & (2)  \\ 
       
       \\[-1.8ex]
       \\[-1.8ex]

        & \multicolumn{6}{c}{\hspace{-2.5cm}{\underline{\textbf{Job Offer}}}} \\

       & & \hspace{-2cm}Data & & Model &  & \\
       
       \hline \\[-1.8ex]

        Consulting & \phantom{Salary (log)} & \hspace{-2cm}0.25 & \phantom{Salary (log)} & \hspace{0.2cm}0.23 & &  \\

        Technology & \phantom{Salary (log)} & \hspace{-2cm}0.48 & \phantom{Salary (log)} & \hspace{0.2cm}0.51 & &   \\
 
        Manufacturing & \phantom{Salary (log)} & \hspace{-2cm}0.27 & \phantom{Salary (log)} & \hspace{0.2cm}0.26 & &   \\

        \hline \\[-1.8ex]
        
        & \multicolumn{6}{c}{\hspace{-2.5cm}{\underline{\textbf{Job Choice}}}} \\

       & & \hspace{-2cm}Data & & Model &  & \\
       
       \hline \\[-1.8ex]

        Consulting & \phantom{Salary (log)} & \hspace{-2cm}0.24 & \phantom{Salary (log)} & \hspace{0.2cm}0.22 & &   \\

        Technology & \phantom{Salary (log)} & \hspace{-2cm}0.49 & \phantom{Salary (log)} & \hspace{0.2cm}0.51 & &   \\

        Manufacturing & \phantom{Salary (log)} & \hspace{-2cm}0.27 & \phantom{Salary (log)} & \hspace{0.2cm}0.27 & &   \\

        \hline \\[-1.8ex]

        & \multicolumn{6}{c}{\hspace{-2.5cm}{\underline{\textbf{Unemployed}}}} \\

       & & \hspace{-2cm}Data & & Model &  & \\
       
       \hline \\[-1.8ex]

       --- & \phantom{Salary (log)} & \hspace{-2cm}0.30 & \phantom{Salary (log)} & \hspace{0.2cm}0.31 & &   \\
       
       \hline \\[-1.8ex]

        & \multicolumn{6}{c}{\hspace{-2.5cm}{\underline{\textbf{Earnings Gap}}}} \\

       & & \hspace{-2cm}Data & & Model &  & \\
       
       \hline \\[-1.8ex]

       --- & \phantom{Salary (log)} & \hspace{-2cm}-11.3\% & \phantom{Salary (log)} & \hspace{0.2cm} -10.6\%& &   \\
       
            \hline 
            \hline \\[-1.8ex]
       \end{tabular}
          \begin{tablenotes}
            \begin{footnotesize}
            \item Notes: Online Appendix Table  \ref{tab:model fit job choice and job offers main 1} compares the moments in the data to the corresponding model-simulated moments. Earnings gap reported in the first column corresponds to the regression specification where all controls enter linearly. Model-simulated
            moments are computed by simulating the model 300 times for each observation in the sample and then averaging over the number of observations and the number of simulation draws.
            \end{footnotesize}
          \end{tablenotes}
\end{threeparttable}
\end{table}

\clearpage

\vspace{2cm}

\begin{table}[p]
    \centering
    \caption{Select Job Cutoffs by Pay Category, Job Sector, and Job Title}
     \label{tab:firm cutoffs main}

  \begin{threeparttable}

    \begin{tabular}{@{\extracolsep{0pt}} c c c}
       \hline 
       \hline 
    & Job Cutoffs (Job Utility) & \\

       \hline \\[-1.8ex]

       & \underline{\textbf{Pay Category}} & \\

       Parameter & Estimate & Std. Error \\
       
       \hline \\[-1.8ex]

        Top 25\% & $-16.300^{***}$ & 0.749  \\

         50\%-75\% & $-16.487^{***}$ & 0.765 \\

         25\%-50\% & $-16.779^{***}$ & 0.762 \\

         Bottom 25\% & $-17.138^{***}$ & 0.767 \\
         
         \hline \\[-1.8ex]

        & \underline{\textbf{Job Sector}} & \\

       Parameter & Estimate & Std. Error \\
       
       \hline \\[-1.8ex]

        Technology  & $-17.031^{***}$ & 0.788  \\

        Consulting & $-16.165^{***}$ & 0.734 \\
        
       Manufacturing & $-16.274^{***}$ & 0.724 \\
       
       \hline \\[-1.8ex]

        & \underline{\textbf{Job Title}} & \\

       Parameter & Estimate & Std. Error \\
       
       \hline \\[-1.8ex]

        Engineer & $-16.643^{***}$ & 0.760  \\

        Consultant & $ -16.415^{***}$ & 0.751 \\

     Manager & $-17.253^{***}$ & 0.782 \\

            \hline 
            \hline \\[-1.8ex]
       \end{tabular}
          \begin{tablenotes}
            \begin{footnotesize}
            \item Notes: Online Appendix Table  \ref{tab:firm cutoffs main} includes estimates of the job cutoffs by pay category, job sector, and job title for aggregate firms. An ``aggregate'' firm in a given category (e.g., sector) has the hiring cutoff averaged over all firms in that category. Note that the job cutoff estimates are not structural parameters, as they are allowed to change under counterfactual policies. Full estimation tables are available upon request. Average Salary = \$56,767.29 (PPP), $N$ = 4207 (no. of students), $J$ = 644 (no. of jobs). PPP stands for purchasing power parity.\\  * significant at 10\%, ** significant at 5\%, *** significant at 1\%.
            \end{footnotesize}
          \end{tablenotes}
\end{threeparttable}
\end{table}

\clearpage

\newgeometry{top=0.2in, bottom=0.2in, left=0.2in, right=0.2in}
\begin{landscape}
\thispagestyle{empty}
\begin{table}[bp]
\begin{adjustwidth}{-0.4cm}{}
  \centering
  \small
      \caption{Differences in Job-Specific Salaries between Indian Locations versus those in Establishments Located in the ``MNC headquarters region"}
    \label{tab:differences in nominal salaries}

  \begin{threeparttable}

                \begin{tabular}{@{\extracolsep{0pt}} c c c c c}
       \hline 
       \hline 

        (1) & (2) & (3) & (4) & (5) \\

       Firm Name & Job Designation & Job Location & Headquartered in & $\Delta (\%)$  \\ 
       
       \hline 

        Citicorp  & Analyst & India & New York, USA & 2.12  \\
         eBay Inc. & Software Engineer  & India & San Jose, USA & 0.45 \\
        Rolls Royce  & Engineering Graduate & India & London, UK & 2.67 \\
        LinkedIn  & Software Engineer  & India & Sunnyvale, USA & 3.52 \\
        Google  & Software Engineer  & India & Mountain View, USA & 2.34 \\
        Intel  & Component Design Engineer  & India & Santa Clara, USA & 1.74 \\
        Amazon  & Area Manager & India & Seattle, USA & 2.82 \\
        Boston Consulting Group & Associate & India & Boston, USA & 0.66 \\
        Cisco & Software Engineer & India & San Jose, USA & 2.91 \\ 
        Schlumberger & Software Engineer & India & Houston, USA & 3.22 \\
        NetApp & Member Technical Staff & India & Sunnyvale, USA & 1.32  \\
        Citrix & Software Engineer & India & Fort Lauderdale, USA & 3.67 \\
        Ronald Berger  & Business Analyst & India & Munich, Germany & 2.49\\
        Applied Materials  & Application Engineer & India & Santa Clara, USA & 3.56 \\
        Epic  & Software Developer & India & Verona, USA & 3.47  \\
        General Electric  & Edison Engineer & India & Boston, USA & 3.04 \\
        Analog Devices Pvt. Ltd.  & Software Engineer & India & Norwood, USA & 1.76 \\
        ARM Embedded Technologies  & Graduate Engineer & India & Cambridge, UK & 3.94  \\
        Microsoft  & Software Engineer & India & Redmond, USA & 2.91  \\
        VISA  & Software Engineer & India & San Francisco, USA & 2.84  \\
        Texas Instruments  & Analog Engineer & India  & Dallas, USA & 3.11 \\
        Samsung  & Software Engineer & India & Suwom-si, South Korea & 2.75 \\
        J.P. Morgan \& Chase  & Associate & India & New York, USA & 2.74 \\
        Capital One  & Associate & India & McLean, USA & 3.67 \\
         ASEA Brown Boveri (ABB)  & Software Engineer & India & Zurich, Switzerland & 3.61\\
        Caterpillar  & Associate Engineer & India & Deerfield, USA & 3.42 \\
        \hline 
        \hline \\[-1.8ex]
        \end{tabular}
          \begin{tablenotes}
            \item[] \scriptsize Notes: Online Appendix Table~\ref{tab:differences in nominal salaries} includes select comparisons between real job-specific salaries offered by MNCs at Indian locations versus locations in the ``MNC headquarters region" (typically North America and Europe). Column (1) includes the firm name, column (2) includes job designation, column (3) includes the job location, column (4) denotes the location of the firm headquarters, and column (5) reports the absolute percentage difference between real job-specific salaries at Indian locations versus locations in the ``MNC headquarter region or similar," where the denominator is the average salary across all jobs in my sample (\$56,767.29 PPP). The average of $\Delta (\%)$ reported in column (5) is 2.72\%. This average difference is  also similar in magnitude across all jobs in the sample, although only a select number of jobs are shown in Online Appendix Table~\ref{tab:differences in nominal salaries}. Firm salaries for Indian locations are taken from my sample. Firm salaries at locations in the ``MNC headquarters region" (typically North America and Europe) are taken from major salary reporting platforms in the U.S.    
          \end{tablenotes}
\end{threeparttable}
\end{adjustwidth}
\end{table}
\end{landscape}

\clearpage

\begin{table}[p]
  \centering
  \small
     \caption{Negative Correlation between College GPA and Entrance Exam Score}
    \label{tab:gpa_and_test_score_negative_correlation}
  \begin{threeparttable}
        \begin{tabular}{@{\extracolsep{0pt}} c c c c} 
        \\[-1.8ex]
        \hline 
        \hline \\[-1.8ex]
                & \multicolumn{3}{c}{\hspace{-3cm}{ \underline{Dependent Variable: log  GPA}}} \\
        
                 \\[-1.8ex]

       \\[-1.8ex]
        & \multicolumn{3}{c}{\hspace{-3cm}{ \underline{\textbf{B.Tech. Degree Students}}}} \\
        
         \\[-1.8ex]
        
        \vphantom{Coefficient} & (1) & (2) & (3)  \\ 
       
       \\[-1.8ex]

       Coefficient & All & Non-Selective Majors & Selective Majors  \\ 
       
       \hline \\[-1.8ex]

        \text{Disadv. Caste} & $-0.171^{***}$ (0.010) & $-0.162^{***}$ (0.011) & $-0.187^{***}$ (0.020) \\
        
        \text{Entrance Exam Score} & $-0.025^{***}$ (0.006) & $-0.008\phantom{^{***}}$ (0.007) & $-0.060^{***}$ (0.010) \\
         
        \hline \\[-1.8ex]
        $N$  & 1289 & 902 & 387\\
        $R^{2}$ & 0.237 & 0.232 & 0.264 \\
        Adjusted \hspace{0.05cm}$R^{2}$ & 0.230 & 0.225 & 0.249 \\
        
        \\[-1.8ex]
        \\[-1.8ex]

        & \multicolumn{3}{c}{\hspace{-3cm}{ \underline{\textbf{Dual Degree Students}}}} \\
        
       \\[-1.8ex]
        \\[-1.8ex]
        
               Coefficient & All & Non-Selective Majors & Selective Majors  \\ 
       
       \hline \\[-1.8ex]
       
        \text{Disadv. Caste} & $-0.147^{***}$ (0.009) & $-0.140^{***}$ (0.012) & $-0.155^{***}$ (0.014) \\
        
        \text{Entrance Exam Score} & $-0.029^{***}$ (0.006) & $-0.021^{**\phantom{*}}$ (0.010) & $-0.036^{***}$ (0.007) \\
         
        \hline
        $N$  & 1239 & 780 & 459\\
        $R^{2}$ & 0.221 & 0.190 & 0.276 \\
        Adjusted \hspace{0.05cm}$R^{2}$ & 0.212 & 0.182 & 0.262 \\
        
       \\[-1.8ex]

        \\[-1.8ex]
        
        & \multicolumn{3}{c}{\hspace{-3cm}{ \underline{\textbf{M.Tech. Degree Students}}}} \\
        
      \\[-1.8ex]
       \\[-1.8ex]

       Coefficient & All & Non-Selective Majors & Selective Majors  \\ 
       
       \hline \\[-1.8ex]
       
        \text{Disadv. Caste} & $-0.071^{***}$ (0.007) & $-0.078^{***}$ (0.010) & $-0.048^{***}$ (0.013) \\
        
        \text{Entrance Exam Score} & $-0.033^{***}$ (0.006) & $-0.042^{***}$ (0.011) & $-0.022^{***}$ (0.004) \\
         
        \hline 
        $N$  & 1202 & 840 & 362\\
        $R^{2}$ & 0.245 & 0.271 & 0.206 \\
        Adjusted \hspace{0.05cm}$R^{2}$ & 0.236 & 0.264 & 0.183 \\
        
           \\[-1.8ex]

        \\[-1.8ex]
        
        & \multicolumn{3}{c}{\hspace{-3cm}{ \underline{\textbf{M.S. Degree Students}}}} \\
        
                 \\[-1.8ex]

       \\[-1.8ex]

       Coefficient & All & Non-Selective Majors & Selective Majors  \\ 
       
       \hline \\[-1.8ex]
       
        \text{Disadv. Caste} & $-0.011^{*\phantom{**}}$ (0.056) & $-0.019^{**\phantom{*}}$ (0.008) & $\phantom{-}0.003^{\phantom{***}}$ (0.011) \\
        
        \text{Entrance Exam Score} & $-0.004^{***}$ (0.001) & $-0.004\phantom{^{***}}$ (0.010) & $-0.005^{***}\phantom{}$ (0.001) \\
         
        \hline 
        $N$  & 477 & 322 & 155\\
        $R^{2}$ & 0.076 & 0.055 & 0.157 \\
        Adjusted \hspace{0.05cm}$R^{2}$ & 0.046 & 0.031 & 0.098 \\
                \hline 
        \hline \\[-1.8ex]
        \end{tabular}
          \begin{tablenotes}
            \begin{scriptsize}
            \item[] \scriptsize Notes: Online Appendix Table  \ref{tab:gpa_and_test_score_negative_correlation} includes estimates from a regression of grade point averages of B.Tech., Dual, M.Tech., and M.S. degree holders on student characteristics. The dependent variable is log GPA. Controls include college major, entrance exam score (standardized), grades on 10th and 12th grade national level examinations (standardized), and caste. College major includes indicators for each major. College entrance exam scores (ranks) have been renormalized so that higher numbers are better. In column (1),  I report results for all students. In column (2), I report results only for students in non-selective majors. In column (3), I report results only for students in selective majors. *$p<0.1$; **$p<0.05$; ***$p<0.01$.
            \end{scriptsize}
          \end{tablenotes}
\end{threeparttable}
\end{table}

\clearpage

\begin{table}[bp]
\begin{adjustwidth}{-0.7cm}{}
    \centering
    \small 
    \caption{Earnings Gap under Subsidies versus PCI}
        \label{tab:earnings gap baseline and cf}
  \begin{threeparttable}
       \begin{tabular}{@{\extracolsep{5pt}} c c c}
       \hline 
       \hline \\[-1.8ex]
        &  Earnings Gap (\%) &
       \\
       
       \hline \\[-1.8ex]
               &  \underline{Perfectly Elastic Labor Demand} &
       \\ 
       \\[-1.8ex]
        \underline{Hiring Subsidy} & &\underline{PCI} \\
       \\[-1.8ex]
       -5.5\% & &-8.9\% \\
              \hline \\[-1.8ex]
       \\[-1.8ex]
                     &  \underline{Perfectly Inelastic Labor Demand} &
       \\ 
       \\[-1.8ex]
       \underline{Hiring Subsidy} & &\underline{PCI} \\
        \\[-1.8ex]
        -7.6\% & & -9.5\% \\
        \hline 
        \hline \\[-1.8ex]
       \end{tabular}
          \begin{tablenotes}
            \begin{footnotesize}
            \item Notes: \scriptsize Appendix Table  \ref{tab:earnings gap baseline and cf} shows the earnings gap under hiring subsidies and the pre-college intervention policy. ``PCI'' stands for the pre-college intervention (PCI) policy. 
            \end{footnotesize}
          \end{tablenotes}
\end{threeparttable}
\end{adjustwidth}
\end{table}

\begin{table}[p]
\begin{adjustwidth}{-0.5cm}{}
    \centering
    \small 
    \caption{Displacement Effects (Unemployment)}
        \label{tab:unemployment baseline and cf}

  \begin{threeparttable}

       \begin{tabular}{@{\extracolsep{0pt}} c c c c c c}
       \hline 
       \hline \\[-1.8ex]
       
       & \multicolumn{3}{c}{\underline{\textbf{Subsidies and Pre-College Intervention}}} \\

       \\[-1.8ex]
       \\[-1.8ex]
       
        \vphantom{blahblah} & \underline{$\%$ Unemployed} & \underline{\hspace{2.2cm}$\Delta$ \text{Unemployed} ($\%$)} & \phantom{---} \\

        & \underline{Adv. Caste} \hspace{0.2cm} \underline{Disadv. Caste} \hspace{0.2cm} \underline{Overall} & \underline{Adv. Caste} \hspace{0.2cm} \underline{Disadv. Caste} & \hspace{-0.5cm}\underline{Overall}  \\
        \\[-1.8ex]
        
        Baseline & \hspace{0.3cm}25\% \hspace{1.7cm} 36\% \hspace{1.7cm} 31\% & --- \hspace{1.5cm} --- & \hspace{-0.5cm}--- \\
        \hline \\[-1.8ex]
        
        &  \hspace{3cm}\underline{Perfectly Elastic Labor Demand} &  & \\ 
        \\[-1.8ex]
        
        Subsidy &  \hspace{0.3cm}25\% \hspace{1.7cm} 24\% \hspace{1.7cm} 28\% & \phantom{0}-0\% \hspace{1cm} -35\% & \hspace{-0.5cm}-20\%\\
        PCI &  \hspace{0.3cm}25\% \hspace{1.7cm} 31\% \hspace{1.7cm} 25\% & \phantom{0}-0\% \hspace{1cm} -15\% & \hspace{-0.5cm}-9\%\\
       \hline \\[-1.8ex]
       
        &  \hspace{3cm}\underline{Perfectly Inelastic Labor Demand}&  & \\ 
        \\[-1.8ex]
        
        Subsidy &  \hspace{0.3cm}33\% \hspace{1.7cm} 28\% \hspace{1.7cm} 31\% & \phantom{1}+31\% \hspace{1cm} -23\% & \hspace{-0.5cm}-0\%\\
        PCI &  \hspace{0.3cm}28\% \hspace{1.7cm} 33\% \hspace{1.7cm} 31\% & +12\% \hspace{1cm} -9\% & \hspace{-0.5cm}-0\%\\
        \hline \\[-1.8ex]
        
               & \multicolumn{3}{c}{\underline{\textbf{Hiring Quotas}}} \\

       \\[-1.8ex]
       \\[-1.8ex]
        
            \vphantom{blahblah} & \underline{$\%$ Unemployed} & \hspace{2.2cm}\underline{$\Delta$ \text{Unemployed} ($\%$)} & \phantom{---}\\

        &  Adv. Caste \hspace{0.2cm} Disadv. Caste \hspace{0.2cm}\underline{Overall}& \underline{Adv. Caste} \hspace{0.2cm} \underline{Disadv. Caste} & \hspace{-0.5cm}\underline{Overall}\\

        \\[-1.8ex]

        Baseline &  \hspace{0.5cm}25\% \hspace{1.5cm} 36\% \hspace{1.7cm} 31\% & --- \hspace{1.5cm} --- & \hspace{-0.5cm}---\\
        \hline \\[-1.8ex]
        Hiring Quotas   & \hspace{0.5cm}35\% \hspace{1.5cm} 31\% \hspace{1.7cm} 33\% & +37\% \hspace{1cm} -16\% & \hspace{-0.5cm}+7\%\\
        \hline 
        \hline \\[-1.8ex]
       \end{tabular}
          \begin{tablenotes}
            \begin{footnotesize}
            \item Notes: \scriptsize Online Appendix Table  \ref{tab:unemployment baseline and cf} shows unemployment by caste under the baseline, hiring subsidies, PCI and quotas. ``PCI'' stands for the pre-college intervention (PCI) policy. 
            \end{footnotesize}
          \end{tablenotes}
\end{threeparttable}
\end{adjustwidth}
\end{table}

\clearpage

\begin{figure}[p]
\captionsetup{justification=centering}
  \centering\captionsetup[subfloat]{labelfont=bf}
   \label{fig:common support rank}
  \includegraphics[width=0.8\textwidth]{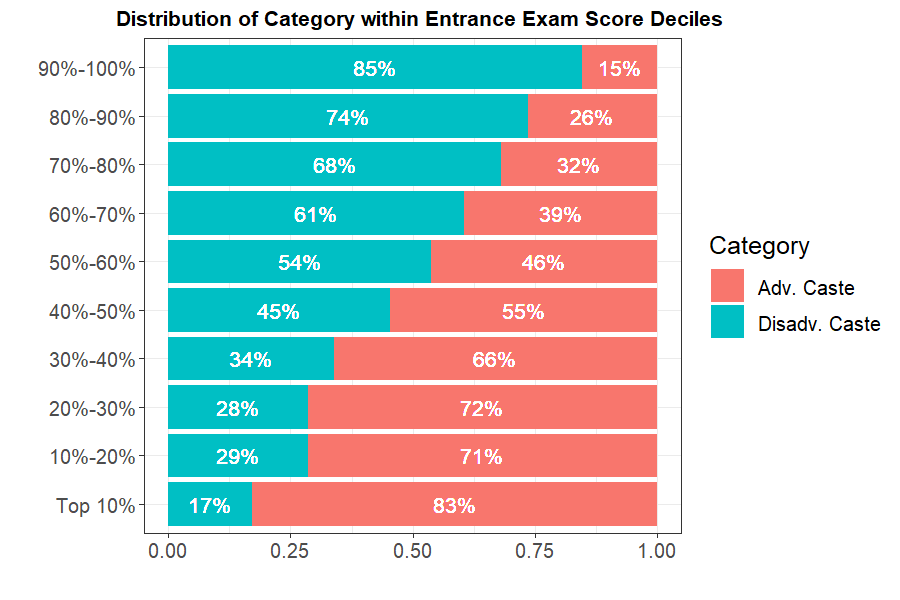}
  \vspace*{0.5cm}%
   \label{fig:common support gpa}
   \includegraphics[width=0.8\textwidth]{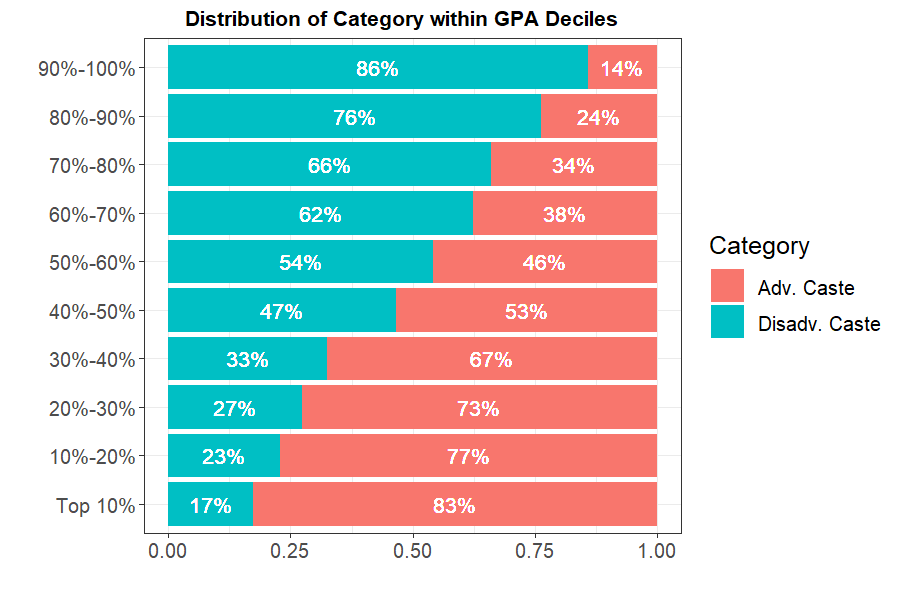}
        \caption{This figure shows that students from both disadvantaged \\ and advantaged castes belong within each entrance exam score or GPA decile.}\label{fig:common support}
\end{figure}

\clearpage

{\eject \pdfpagewidth=11.5in \pdfpageheight=11.5in
\begin{landscape}
\thispagestyle{empty}
\begin{figure}[htb]
\captionsetup{justification=centering}
\centering
  \begin{subfigure}[b]{.43\linewidth}
    \centering
    \includegraphics[width=.99\textwidth]{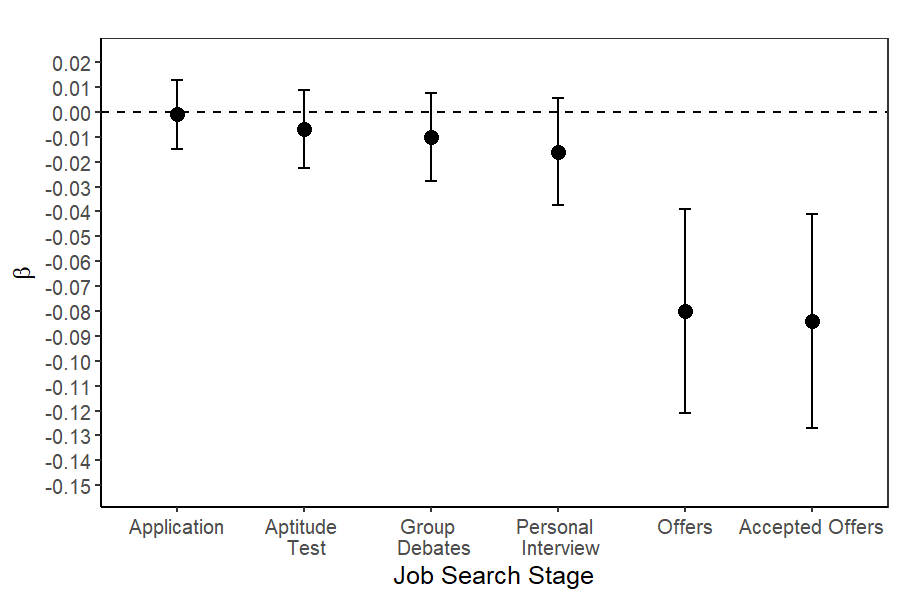}
            \caption{Manufacturing}\label{fig:1a}
  \end{subfigure}%
  \begin{subfigure}[b]{.43\linewidth}
    \centering
    \includegraphics[width=.99\textwidth]{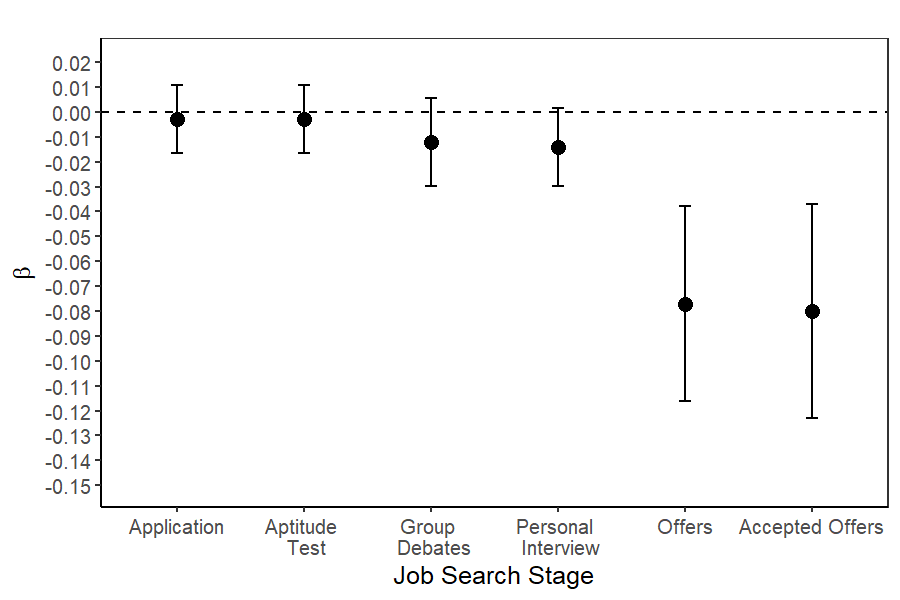}
    \caption{Technology}\label{fig:1b}
  \end{subfigure}\\%
\vspace{1cm}
  \begin{subfigure}[b]{.43\linewidth}
    \centering
    \includegraphics[width=.99\textwidth]{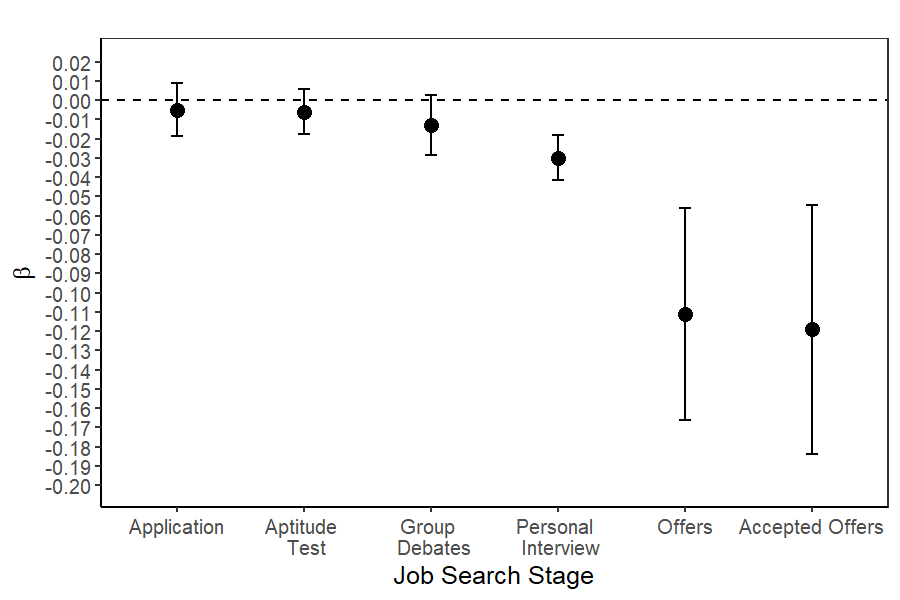}
    \caption{Consulting}\label{fig:1c}
  \end{subfigure}%
  \begin{subfigure}[b]{.43\linewidth}
    \centering
    \includegraphics[width=.99\textwidth]{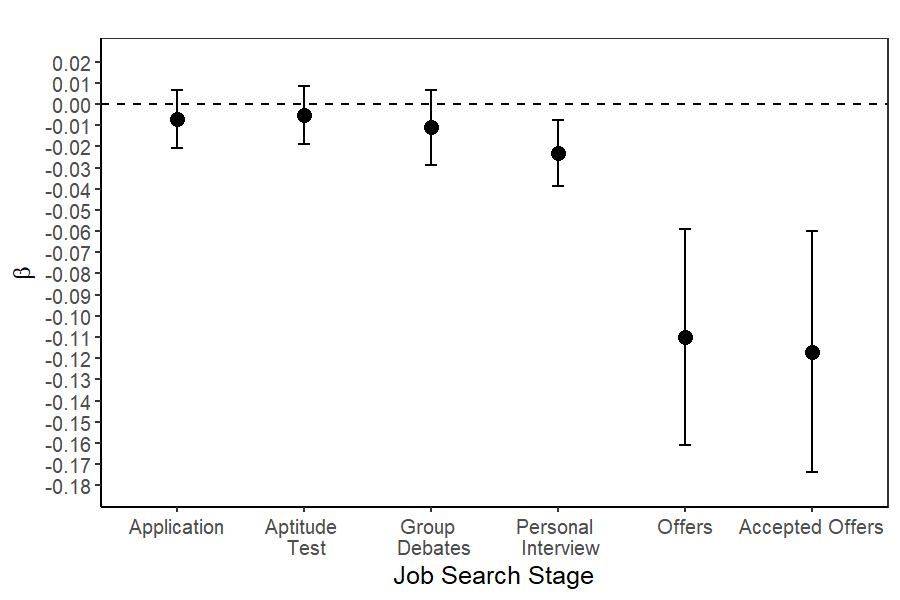}
    \caption{Client Facing}\label{fig:1d}
  \end{subfigure}  \\%
  \vspace{1cm}
  \begin{subfigure}[b]{.43\linewidth}
    \centering
    \includegraphics[width=.99\textwidth]{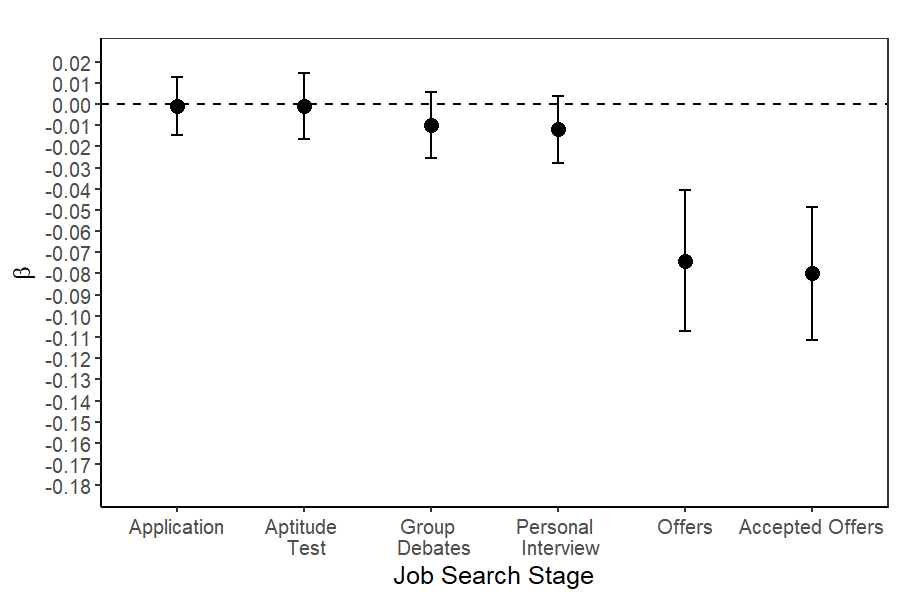}
    \caption{Non-Client Facing}\label{fig:1e}
  \end{subfigure}%
  \vspace{0.5cm}
  \caption{This figure shows the coefficient $\beta$ corresponding to the regression in Equation~\ref{eq:app behavior main} across job sectors and job types. $\beta$ represents the percentage difference in the average salary at each job search stage between advantaged and disadvantaged castes. Each dot is the coefficient $\beta$ from a separate regression. The vertical bars are 95\% confidence intervals. These regressions include controls. This figure shows that the increase in the earnings gap is even \emph{more} concentrated for technology, manufacturing, and non-client-facing jobs. The entire earnings gap among these jobs occurs after personal interviews.}\label{fig:earnings gap hiring stage sectors and job types}
\end{figure}
\end{landscape}}

\clearpage

{\eject \pdfpagewidth=8.5in \pdfpageheight=11in
\begin{landscape}
\thispagestyle{empty}
\begin{figure}[htb]
\captionsetup{justification=centering}
\centering
  \begin{subfigure}[b]{.50\linewidth}
    \centering
\includegraphics[width=.99\textwidth]{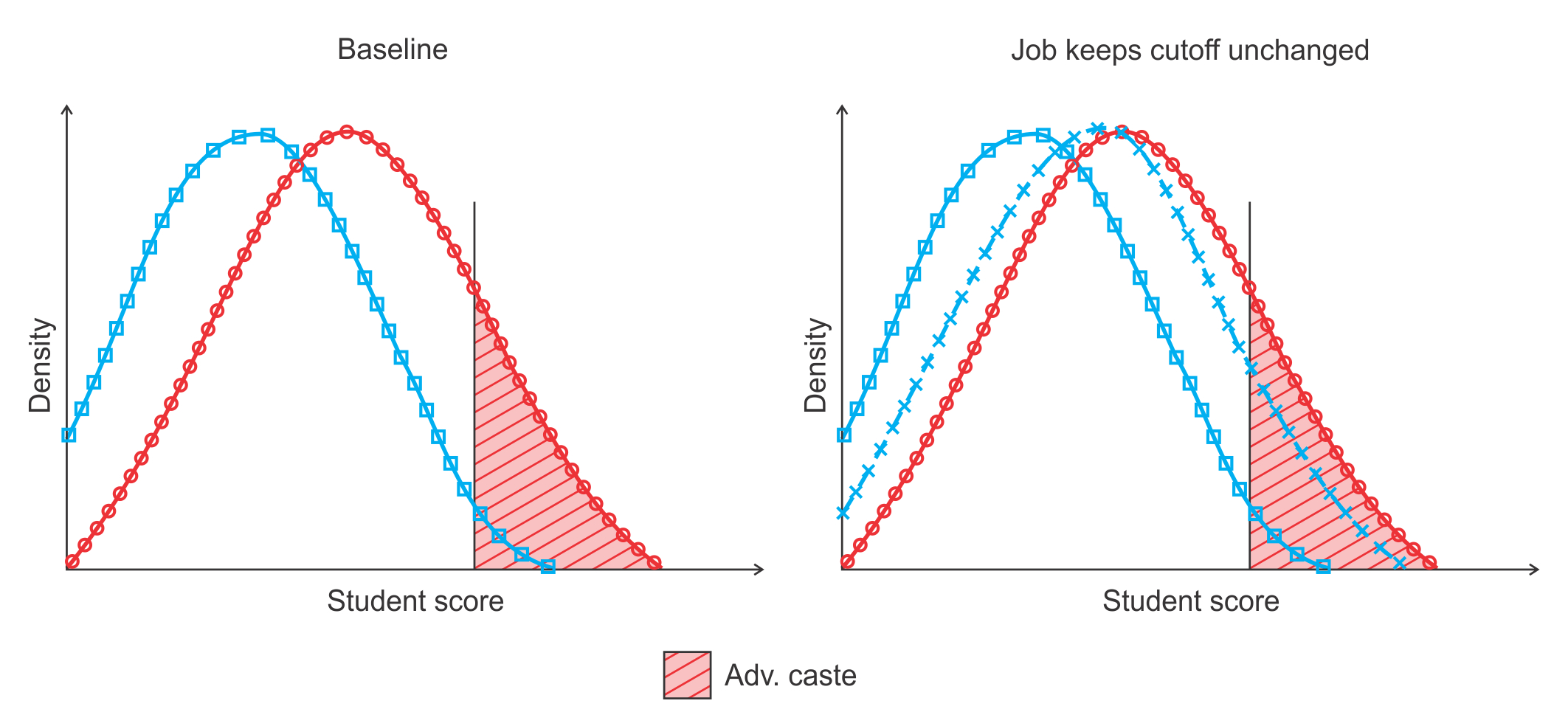}
            \caption{Advantaged Caste Hires Under Perfectly Elastic Demand}\label{fig:2a}
  \end{subfigure}%
  \begin{subfigure}[b]{.50\linewidth}
    \centering
    \includegraphics[width=.99\textwidth]{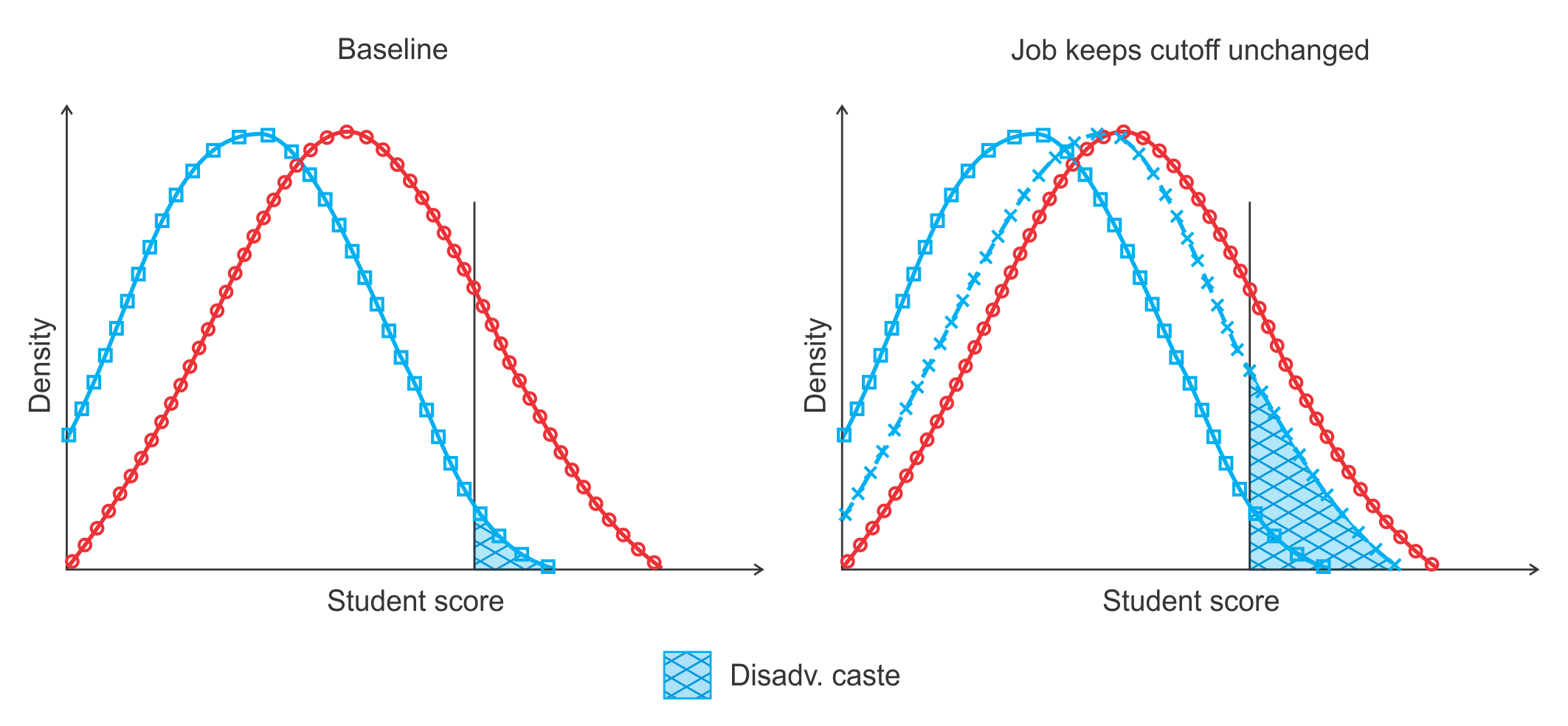}
    \caption{Disadvantaged Caste Hires Under Perfectly Elastic Demand}\label{fig:2b}
  \end{subfigure}\\%
\vspace{1cm}
  \begin{subfigure}[b]{.50\linewidth}
    \centering
    \includegraphics[width=.99\textwidth]{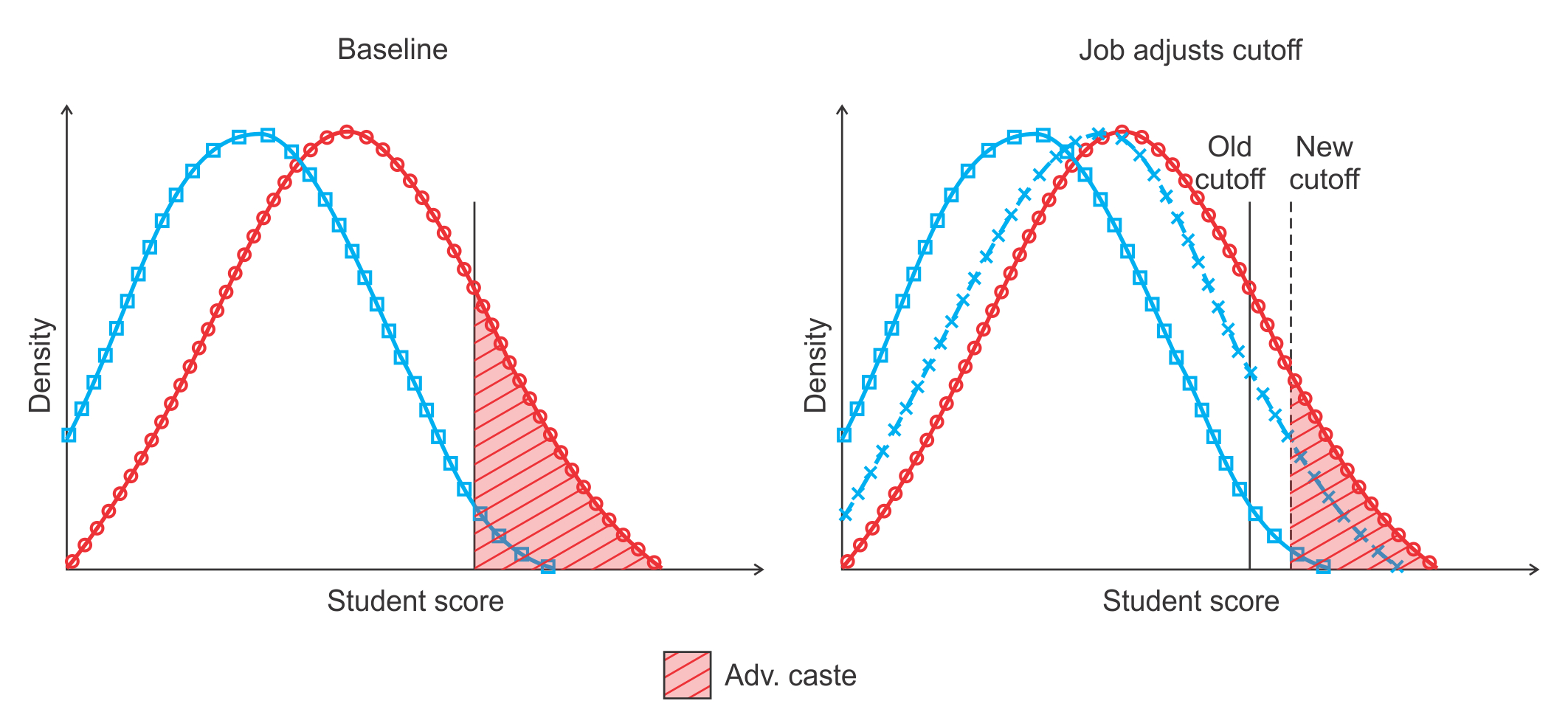}
    \caption{Advantaged Caste Hires Under Perfectly Inelastic Demand}\label{fig:2c}
  \end{subfigure}%
  \begin{subfigure}[b]{.50\linewidth}
    \centering
    \includegraphics[width=.99\textwidth]{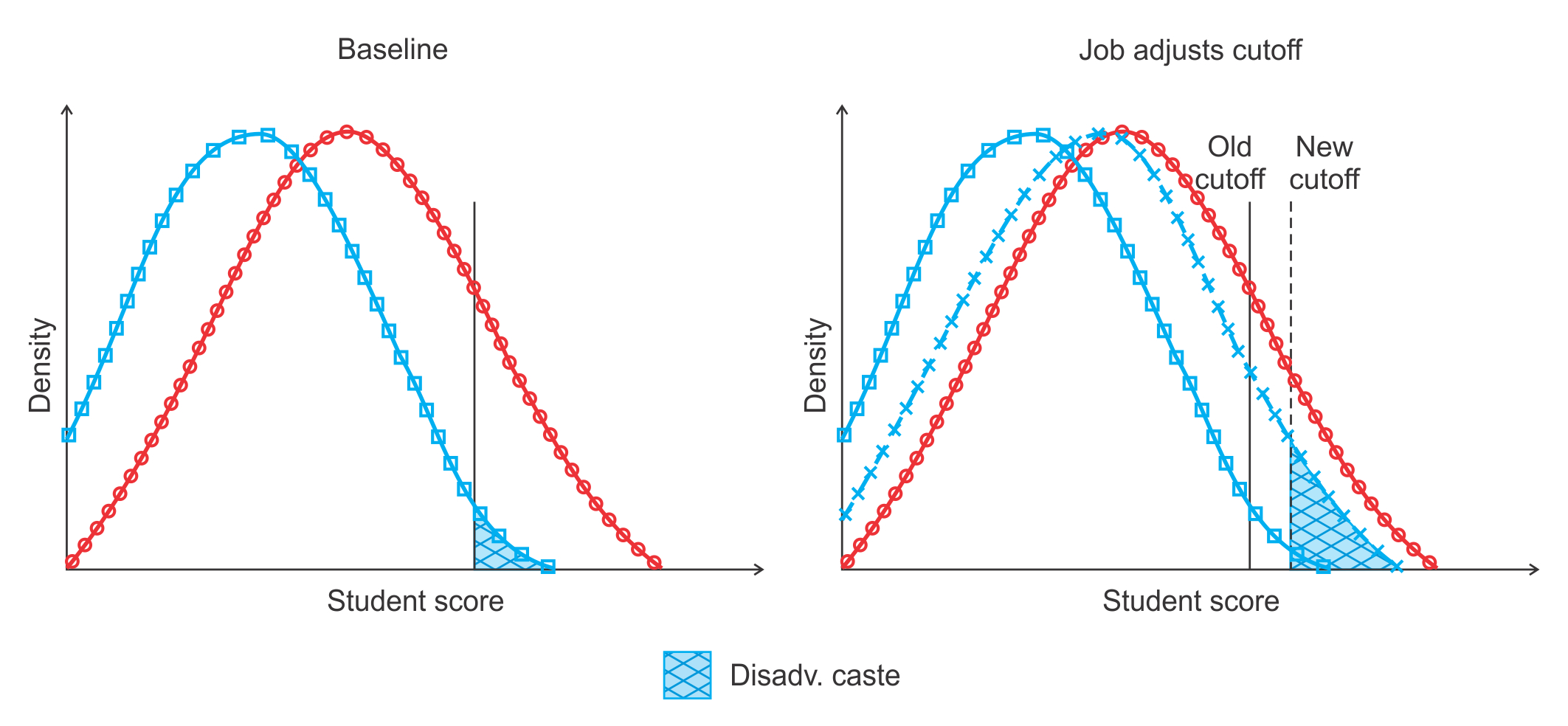}
    \caption{Disadvantaged Caste Hires Under Perfectly Inelastic Demand}\label{fig:2d}
  \end{subfigure}  \\%
  \vspace{0.5cm}
  \caption{This figure shows how the model bounds both the negative and positive employment effects on advantaged and disadvantaged castes, respectively, under hiring subsidies and the pre-college intervention policy. The distribution of advantaged caste ``scores'' are shown in red. These scores are to the right of the distribution of disadvantaged caste scores, shown in blue. Scores can be calculated from Equation~\ref{eq:8}. Under both hiring subsidies and the pre-college intervention policy, the distribution of disadvantaged caste scores shifts to the right. The top panel represents a scenario where labor demand is perfectly elastic. In this scenario, there is no displacement of advantaged castes and the number of disadvantaged caste hires is at least as large as in the baseline. The bottom panel represents a scenario where labor demand is perfectly inelastic. In this scenario, the number of disadvantaged caste hires is at least as large as in the baseline but not as large as when labor demand is perfectly elastic. On the other hand, the displacement of advantaged castes is larger than when demand is perfectly elastic (where it is zero).}\label{fig:elastic_inelastic_labor_demand}
\end{figure}
\end{landscape}}

\end{document}